\documentclass[a4paper]{article}
\usepackage{color}

\usepackage{geometry}
\usepackage[figuresright]{rotating}
\usepackage{graphicx}
\usepackage{subcaption}
\usepackage{longtable}
\usepackage{array}

\usepackage{booktabs}
\usepackage{multirow}
\usepackage{makecell}

\usepackage{tensor}
\usepackage{mathtools}
\usepackage{amsmath}
\usepackage{upgreek}
\usepackage{mathrsfs}

\usepackage{appendix}

\usepackage{bm}

\usepackage[noblocks]{authblk}

\usepackage[numbers,sort&compress]{natbib}

\usepackage{CJK}
\usepackage{txfonts}
\usepackage{times}
\usepackage{ragged2e}
\usepackage{indentfirst}

\usepackage{verbatim}

\usepackage{ntheorem}

\usepackage{lineno}

\usepackage[colorlinks=true,linkcolor=blue,citecolor=blue,urlcolor=green]{hyperref}

\newtheorem{theorem}{Theorem}
\newtheorem{lemma}{Lemma}
\newtheorem*{proof}{Proof}

\renewcommand{\raggedright}{\leftskip=0pt \rightskip=0pt plus 0cm}

\allowdisplaybreaks[4]
\begin{document}

\title{Entanglement swapping theory and beyond}

\author{Zhaoxu Ji$^{\dag}$, Huanguo Zhang
\\
{\small 
Key Laboratory of Aerospace Information Security and Trusted Computing,
Ministry of Education, School of Cyber Science and Engineering, Wuhan University, Wuhan 430072, China	\\
$^{\dag}$jizhaoxu@whu.edu.cn}}
\date{}

\maketitle

\begin{abstract}

We focus on the general theory of entanglement swapping, including entanglement swapping of pure and mixed states.
We also study the theory of entangled swapping chains, which can be regarded as an application of entangled swapping.
For maximally entangled states, we consider the entanglement swapping of 2-level maximally entangled states
without limiting each subsystem to be in the same basis.
We further consider the entanglement swapping between d-level maximally entangled states,
which is realized by performing a joint measurement on the particles that contain the first particle of
the selected entangled states and the particles without involving the first particle in other entangled states.
For the entanglement swapping of general pure state,
we generalize the case of two bipartite general pure states to multi-state cases.
Besides, we propose the entanglement swapping between bipartite general pure states and maximally entangled states.
For entanglement swapping chains, we propose the entanglement swapping chains for maximally entangled states,
which is realized by performing joint measurements on multiple particles in each state,
we use mathematical induction to prove the results of entanglement swapping chains of
maximally entangled states and that of general pure states.
Moreover, we study the entanglement swapping and entanglement swapping chains of mixed states,
including $X$ states and mixed maximally entangled states.
Finally, we provide a new proof for our previous work [2022, Physica A, 585, 126400].

\end{abstract}




\tableofcontents


\section{Introduction}

\noindent
Quantum mechanics phenomena, discovered at the beginning of the last century, 
have been widely regarded as bizarre because there is no existing theory that can explain them \cite{JiZX12442025}.
Shockingly, it is precisely the peculiar properties of quantum mechanics that make it closer to the truth of 
the universe than any other theory, as the origin of the universe is absolutely inexplicable \cite{JiZX12442025}.
Quantum superposition, entanglement, and entanglement swapping are three typical
quantum mechanics phenomena, among which entanglement swapping is built upon
the first two \cite{JiZX5852022,HorodeckiR8122009,RahmanAU5522022}.
Suppose that there are several independent entangled states in which each state contains at least two particles,
then two new entangled states can be created simultaneously by performing a joint measurement on the particles
selected from all the entangled states. This process is called entanglement swapping,
which has attracted extensive attention since it was discovered \cite{JiZX12442025,JiZX5852022}.

The original idea of entanglement swapping appeared in the quantum teleportation scheme proposed by 
Bennett et al. in 1992 \cite{BennettCH70131993}. Soon after, Zeilinger et al. formally proposed entanglement swapping 
and completed its experimental realization using parametric down-converters \cite{ZukowskiM711993}.
The basic idea of entanglement swapping is that Alice and Bob share an entangled state,
and Bob shares another one with Charlie beforehand;
Then, Bob performs a Bell measurement on the two particles he holds,
which eventually enables Alice and Charlie to share a new entangled state.
Afterwards, Bose et al. \cite{BoseS5721998} and Bouda et al. \cite{BoudaJ34202001}
generalized entanglement swapping to multi-qubit systems and multi-qudit systems, respectively.
Hardy and Song proposed the entanglement swapping chains of
d-level bipartite general pure states \cite{HardyL6252000}.
Later, Karimipour et al. considered the entanglement swapping between
a multi-particle maximally entangled state and Bell states in d-level systems \cite{KarimipourV652002}.
Entanglement swapping is not only an important way to verify the nonlocality of quantum mechanics,
but also plays an important role in quantum computing
and quantum information security \cite{JiZX5852022,GalindoA7422002,ZhangHG16102019}.
A particularly attractive application is that it is the key to construct quantum repeaters,
making it possible to construct large-scale quantum networks \cite{JiZX5852022}.

This paper mainly focuses on the theoretical research of entanglement swapping, 
including entanglement swapping and entanglement swapping chains for pure and mixed states.
Firstly, we study entanglement swapping between 2-level maximally entangled states
without limiting all the subsystems to be in the natural basis,
which is enlightening for studying the entanglement swapping between graph states.
Secondly, we study entanglement swapping between d-level maximally entangled states,
including the case of performing a joint measurement on the selected particles
containing the first particle in some entangled states,
and the ones without the first particle in the remaining entangled states.
Further, we give the formulas for two extreme entanglement swapping cases, that is,
measuring the first part of the particles in all states and measuring the last part of the particles in all states.
For the entanglement swapping of general pure states,
we consider the generalization of the entanglement swapping between
two bipartite general pure states to multi-state case.
Moreover, we propose the entanglement swapping between general pure states and maximally entangled states.
On the entanglement swapping chains, we employ mathematical induction to 
provide a proof for the results of the entanglement swapping chains of general pure states.
We propose entanglement swapping chains of multi-particle maximally entangled states,
and also give a proof for the results using mathematical induction.
In addition, we consider the entanglement swapping chains for mixed Bell states.
Finally, we review and provide a neat proof for our work in Ref. \cite{JiZX5852022}.

The structure of the rest of this paper is as follows.
Sec. 2 presents some basic theory of entanglement swapping.
Sec. 3 introduces commonly used measurement operators for realizing entanglement swapping.
Sec. 4 shows the entanglement swapping of pure states.
Firstly, we describe the entanglement swapping between maximally entangled states for 2-level
systems and d-level systems, respectively. Then, we describe the cases for general pure states.
In Sec. 5, we depict the entanglement swapping of mixed states, including $X$ states and mixed maximally entangled states.
Sec. 6 shows the entanglement swapping chains for pure states and mixed maximally entangled states.
Sec. 7 shows the new proof for the conclusion in Ref. \cite{JiZX5852022}.
This paper is summarized in Sec. 8.


\section{Basic theory of entanglement swapping}

\noindent
There are two ways to formulate entanglement swapping.
One way is to formulate entanglement swapping directly from the definition of quantum measurements;
Indeed, entanglement swapping is realized by measurements.
Another way is a purely mathematical process, which can be simply described as the expansion of 
polynomials composed of vectors, combining like terms, permutation of vectors, 
re-expansion of polynomials, and finally combining like terms again.
In what follows, we would first like to give the basic principles of these two ways for pure states,
and give a general theory of deriving entanglement swapping results through density matrix.
Then, we further give the general theory of entanglement swapping of mixed states.
We will show that entanglement swapping results are determined by the permutation of particle order, 
that is, the entanglement swapping results vary with the permutation of particle order.
Meanwhile, we will consider, without loss of generality, the cases under the assumption that all subsystems in
entanglement swapping are in the same dimension
(we always assume that all subsystems are in d-level systems, unless we make a special statement),
because the Hilbert space of any subsystem can be embedded in a larger one by local actions \cite{BennettCH5451996};
Indeed, a low dimensional system is a special form of a higher one.

\subsection{Basics of pure states entanglement swapping}

Before starting the topic, let us introduce two lemmas and provide brief proofs,

\begin{lemma}
\label{lemma-1}

Suppose that there are $n$ arbitrary pure states containing $m_r$ particles each,
denoted as $\left\{ \left\vert \psi_r \right\rangle \right\}_{r=1}^n$,
then their tensor product, $\otimes_{r=1}^n \left\vert \psi_r \right\rangle$, can always be expressed in the following form:
\begin{gather}
\label{Lemma-1}
\sum_i p_i \bigotimes_{j=1}^M \left\vert l_{ij} \right\rangle,											\\
\sum_i \left\lvert p_i \right\rvert^2 = 1,
\phantom{i}
M = \sum_{r=1}^n m_r,																	\notag
\end{gather}
where $\left\vert l_{ij} \right\rangle$ are single-particle states.

\end{lemma}

\begin{proof}

Suppose that $\left\{ \left\vert e_1^r e_2^r \cdots e_{m_r}^r \right\rangle_{1,2,\dots,m_r} \right\}$,
where the subscripts $1,2,\dots,m_r$ denote the particles respectively,
is an arbitrary orthonormal basis in the Hilbert space $\mathcal{H}^{\otimes m_r}$ corresponding to $\left\vert \psi_r \right\rangle$,
for example, the natural basis
\begin{gather}
\left\{ 
\left\vert e_1^r e_2^r \cdots e_{m_r}^r \right\rangle_{1,2,\dots,m_r}
\middle\vert
e_i^r = 0,1,\dots,d-1, \phantom{i} i = 1,2,\dots,m_r
\right\}.
\end{gather}
One can mark $\left\vert \psi_r \right\rangle$ by
\begin{gather}
\left\vert \psi_r \right\rangle
:=
\sum_{e_1^r, e_2^r, \dots, e_{m_r}^r} \lambda_{e_1^r e_2^r \dotsm e_{m_r}^r} \left\vert e_1^r e_2^r \dotsm e_{m_r}^r \right\rangle, \\
\sum_{e_1^r, e_2^r, \dots, e_{m_r}^r} \left\lvert \lambda_{e_1^r e_2^r \dotsm e_{m_r}^r} \right\rvert^2 = 1,				\notag
\end{gather}
such that one can arrive at
\begin{gather}
\label{composition-proof}
\bigotimes_{r=1}^{n} \left\vert \psi_r \right\rangle		
= \mathcal{P}
\sum_{e_1^1, e_2^1, \dots, e_{m_1}^1}
\sum_{e_1^2, e_2^2, \dots, e_{m_2}^2}
\cdots
\sum_{e_1^n, e_2^n, \dots, e_{m_n}^n}
\prod_{r=1}^n \lambda_{e_1^r e_2^r \dotsm e_{m_r}^r}
\bigotimes_{r=1}^n \left\vert e_1^r e_2^r \dotsm e_{m_r}^r \right\rangle,		\\
\mathcal{P} = 
\frac{1}{\sqrt{
\sum_{e_1^1, e_2^1, \dots, e_{m_1}^1}
\sum_{e_1^2, e_2^2, \dots, e_{m_2}^2}
\cdots
\sum_{e_1^n, e_2^n, \dots, e_{m_n}^n}
\prod_{r=1}^n \left\vert \lambda_{e_1^r e_2^r \dotsm e_{m_r}^r} \right\vert^2}}. \notag
\end{gather}
Here, the symbol $\mathcal{P}$ is the added probability normalization factor (this symbol will be frequently used in this paper).
It can be seen that Eq. \ref{composition-proof} is consistent with Eq. \ref{Lemma-1} in form.
QED.

\end{proof}

\begin{lemma}
\label{lemma-2}

Suppose that there is a composite pure state $\left\vert \ell_1 \ell_2 \cdots \ell_m \right\rangle$ composed of
$m$ single-particle states $\left\vert \ell_1 \right\rangle$, $\left\vert \ell_2 \right\rangle,\dots,\left\vert \ell_m \right\rangle$,
then $\left\vert S_k \left( \ell_1 \ell_2 \cdots \ell_m \right) \right\rangle$ 
are all the same iff $\ell_i \equiv \ell_j \forall i,j \in \left\{ 1,2,\dots,m \right\}$,
where $k=1,2,\dots,m!$ and $S_k \left( \ell_1 \ell_2 \cdots \ell_m \right)$ refers to 
an arbitrary permutation of the sequence $\ell_1 \ell_2 \cdots \ell_m$.

\end{lemma}

\begin{proof}

The necessity of this lemma is obvious; it remains only to prove its sufficiency.
Since $\left\vert S_k \left( \ell_1 \ell_2 \cdots \ell_m \right) \right\rangle$, $k=1,2,\dots,m!$, are all the same implies that
\begin{gather}
\left\vert \ell_i S \left( \left\{\ell_2 \ell_3 \cdots \ell_m \right\} \setminus \left\{ \ell_i \right\} \right) \right\rangle = 
\left\vert \ell_j S \left( \left\{\ell_2 \ell_3 \cdots \ell_m \right\} \setminus \left\{ \ell_j \right\} \right) \right\rangle,
\forall i,j \in \left\{ 1,2,\dots,m \right\},
\end{gather}
in which $S \left( \cdot \right)$ represents all permutations, one can get $\ell_i \equiv \ell_j \forall i,j \in \left\{ 1,2,\dots,m \right\}$.
QED.

\end{proof}

Let us now present the first basic principle for characterizing entanglement swapping.

\begin{theorem}
\label{theorem-1}

Suppose that there are $n$ pure states $\left\{ \left\vert \psi_r \right\rangle \right\}_{r=1}^n$
containing $m_r$ particles each,
and that $k_r$ particles in $\left\vert \psi_r \right\rangle$ are selected.
According to Lemma \ref{lemma-1}, one can mark the joint state by
\begin{gather}
\label{decomposition-assumption}
\left\vert \Psi \right\rangle :=
\sum_{i} \lambda_i \left\vert \mathcal{S} \left( \imath_{i,1} \imath_{i,2} \cdots \imath_{i,K} \right) \right\rangle 
\left\vert \mathcal{R} \left( \jmath_{i,1} \jmath_{i,2} \cdots \jmath_{i,M-K} \right) \right\rangle,		
\\
\sum_i \left\lvert \lambda_i \right\rvert^2 = 1,
\phantom{i}
K = \sum_{r=1}^n k_r,
\phantom{i}
M = \sum_{r=1}^n m_r,						\notag												
\end{gather}
where $\left\vert \mathcal{S} \left( \imath_{i,1} \imath_{i,2} \cdots \imath_{i,K} \right) \right\rangle $
and $\left\vert \mathcal{R} \left( \jmath_{i,1} \jmath_{i,2} \cdots \jmath_{i,M-K} \right) \right\rangle$
are a permutation of the order of all the selected particles and remaining ones, respectively.
Suppose that a measurement operator, selected from the measurement operator set
$\left\{ \left\vert \upphi_j \right\rangle \left\langle \upphi_j \right\vert \right\}$
constructed by the orthonormal basis $\left\{ \left\vert \upphi_j \right\rangle \right\}$ in $\mathcal{H}^{\otimes K}$,
is performed on the selected particles, such that a state in the basis can be obtained,
that is, the measurement result is in one of $\left\{ \left\vert \upphi_j \right\rangle \right\}$.
With the basis, $\left\vert \mathcal{S} \left( \imath_{i,1} \imath_{i,2} \cdots \imath_{i,K} \right) \right\rangle$ can be expressed as
\begin{gather}
\label{decomposition-bases}
\left\vert \mathcal{S} \left( \imath_{i,1} \imath_{i,2} \cdots \imath_{i,K} \right) \right\rangle = 
\sum_j \lambda_{ij} \left\vert \upphi_j \right\rangle,				\\
\sum_j \left\lvert \lambda_{ij} \right\rvert^2 = 1,												\notag
\end{gather}
such that entanglement swapping results, that is, the joint states of the measurement result and the state
that the remaining particles collapse onto, are given by
\begin{gather}
\label{ES-basics-1}
\left\vert \widehat{\Psi} \right\rangle = \mathcal{P}
\sum_i \sum_j \lambda_i \lambda_{ij} \left\vert \upphi_j \right\rangle 
\left\vert \mathcal{R} \left( \jmath_{i,1} \jmath_{i,2} \cdots \jmath_{i,M-K} \right) \right\rangle, 		\\
\mathcal{P} = 
\frac{1}{\sqrt{\sum_i \sum_j \left\vert \lambda_i \lambda_{ij} \right\vert^2}}. 	\notag
\end{gather}
Suppose that the selected measurement operator is
$\left\vert \upphi_{\hat{j}} \right\rangle \left\langle \upphi_{\hat{j}} \right\vert$,
then the remaining particles collapse onto the state
\begin{gather}
\label{ES-basics-2}
\left\vert \widetilde{\Psi} \right\rangle =
\mathcal{P}
\sum_i \lambda_i \lambda_{i\hat{j}} \left\vert \mathcal{R} \left( \jmath_{i,1} \jmath_{i,2} \cdots \jmath_{i,M-K} \right) \right\rangle, 
\\
\mathcal{P} = \frac{1}{\sqrt{\sum_i \left\vert \lambda_i \lambda_{i\hat{j}} \right\vert^2}}. 	\notag
\end{gather}

\end{theorem}

The entanglement swapping result shown in Eq. \ref{ES-basics-1} is the superposition of all possible states, 
while the entanglement swapping result
$\left\vert \upphi_{\hat{j}} \right\rangle \otimes \left\vert \widetilde{\Psi} \right\rangle$ is one of these states.
Lemma \ref{lemma-1} shows that it is feasible to make such an assumption as Eq. \ref{decomposition-assumption}.
It is also feasible to make the hypothesis shown in Eq. \ref{decomposition-bases},
since the measurement operator set $\left\{ \left\vert \upphi_j \right\rangle \left\langle \upphi_j \right\vert \right\}$
is constructed by the orthonormal basis $\left\{ \left\vert \upphi_j \right\rangle \right\}$.
It can be seen from Lemma \ref{lemma-2} that the entanglement swapping results caused by different permutations of
particles are different in a large probability, because the permutation only remains unchanged when
$\imath_{i,1} \equiv \imath_{i,2} \equiv \cdots \equiv \imath_{i,K}$
and $\jmath_{i,1} \equiv \jmath_{i,2} \equiv \cdots \equiv \jmath_{i,K}$ for given $i$.

From the definition of quantum measurement \cite{NielsenMA2000}, entanglement swapping results can also be obtained by

\begin{theorem}
\label{theorem-2}

Suppose the joint state of the selected particles and the remaining particles in entangled states is $\left\vert \uppsi \right\rangle$,
and that a measurement operator, denoted as $M_m$, is performed on the selected particles in 
$\left\vert \Psi \right\rangle$ (see Eq. \ref{decomposition-assumption}).
Denote the measurement result as $\mathrm{A}$,
and the state that the remaining particles collapse onto as $\mathrm{B}$, 
then the joint state after entanglement swapping is given by
\begin{align}
\label{theorem-2-Eq}
\mathrm{A} \otimes \mathrm{B} = 
\frac{
\left( M_m \otimes I \right) \left\vert \uppsi \right\rangle
}
{
\sqrt{\left\langle \uppsi \right\vert \left( M_m^{\dag} M_m \otimes I \right) \left\vert \uppsi \right\rangle}
}.
\end{align}

\end{theorem}

Let us provide more details about Eq. \ref{theorem-2-Eq}.
Before doing this, let us introduce the following lemma:

\begin{lemma}
\label{lemma-3}

Suppose that there are m arbitrary orthonormal basis in $\mathcal{H}$,
denoted as $\left\{ \left\vert \epsilon_i^1 \right\rangle \right\},\left\{ \left\vert \epsilon_i^2 \right\rangle \right\},\dots,
\left\{ \left\vert \epsilon_i^m \right\rangle \right\}$ respectively,
then the $m$-particle state set
\begin{gather}
\left\{
\left\vert \hat{\epsilon}_{i_1} \hat{\epsilon}_{i_2} \cdots \hat{\epsilon}_{i_m} \right\rangle
\phantom{i}
\middle\vert
\phantom{i}
\left\vert \hat{\epsilon}_{i_r} \right\rangle \in \left\{ \left\vert \epsilon_i^r \right\rangle \right\},
\phantom{i}
r = 1,2,\dots,m
\right\},															 	
\end{gather}
construct an orthonormal basis in $\mathcal{H}^{\otimes m}$.

\end{lemma}

\begin{proof}

$\left\{ \left\vert \hat{\epsilon}_{i_1} \hat{\epsilon}_{i_2} \cdots \hat{\epsilon}_{i_m} \right\rangle \right\}$ 
is complete and normalized since
\begin{align}
\sum_{\hat{\epsilon}_{i_1}, \hat{\epsilon}_{i_2}, \dots, \hat{\epsilon}_{i_m}}	
\left\vert \hat{\epsilon}_{i_1} \hat{\epsilon}_{i_2} \cdots \hat{\epsilon}_{i_m} \right\rangle
\left\langle \hat{\epsilon}_{i_1} \hat{\epsilon}_{i_2} \cdots \hat{\epsilon}_{i_m} \right\vert 			
= \sum_{\hat{\epsilon}_{i_1}, \hat{\epsilon}_{i_2}, \dots, \hat{\epsilon}_{i_m}}	
\bigotimes_{r=1}^m
\left\vert \hat{\epsilon}_{i_r} \right\rangle \left\langle \hat{\epsilon}_{i_r} \right\vert			
= \bigotimes_{i} I = \hat{I},
\end{align}
where $I$ and $\hat{I}$ are the identity matrix in $\mathcal{H}$ and $\mathcal{H}^{\otimes m}$, respectively.
Due to
\begin{align}
\left\langle \hat{\epsilon}_{i_1} \hat{\epsilon}_{i_2} \cdots \hat{\epsilon}_{i_m}
\middle\vert 
\hat{\epsilon}_{i_1'} \hat{\epsilon}_{i_2'} \cdots \hat{\epsilon}_{i_m'} \right\rangle
=
\prod_{r=1}^m \left\langle \hat{\epsilon}_{i_r} \middle\vert \hat{\epsilon}_{i_r'} \right\rangle
= \prod_{r=1}^m \delta_{i_r,i_r'},
\end{align}
$\left\vert \hat{\epsilon}_{i_1} \hat{\epsilon}_{i_2} \cdots \hat{\epsilon}_{i_m} \right\rangle$ and 
$\left\vert \hat{\epsilon}_{i_1'} \hat{\epsilon}_{i_2'} \cdots \hat{\epsilon}_{i_m'} \right\rangle$
are the same or orthogonal. QED.

\end{proof}

According to Lemma \ref{lemma-1}, the state $\left\vert \uppsi \right\rangle$ can be wrote 
in the same form as Eq. \ref{decomposition-assumption}. For example, let us suppose that
$\left\vert \uppsi \right\rangle =
\sum_i \lambda_i \left\vert \mathcal{A}_i \right\rangle \left\vert \mathcal{B}_i \right\rangle$,
such that the probability of getting the measurement result $m$ is
\begin{align}
p(m) 
& = 
\sqrt{\left\langle \uppsi \right\vert \left( M_m^{\dag} M_m \otimes I \right) \left\vert \uppsi \right\rangle}						\notag \\
& = \sqrt{\left( \sum_i \lambda_i \left\vert \mathcal{A}_i \right\rangle \left\vert \mathcal{B}_i \right\rangle \right)^{\dag}
\left( M_m^{\dag} M_m \otimes I \right)
\left( \sum_{i'} \lambda_{i'} \left\vert \mathcal{A}_{i'} \right\rangle \left\vert \mathcal{B}_{i'} \right\rangle \right)} 					\notag \\
& = \sqrt{\sum_i \lvert \lambda_i \rvert^2 \left\langle \mathcal{A}_i \right\vert M_m^{\dag} M_m \left\vert \mathcal{A}_i \right\rangle
\left\langle \mathcal{B}_{i} \middle\vert \mathcal{B}_{i} \right\rangle}												\notag \\
&  = \sqrt{\sum_i \lvert \lambda_i \rvert^2 \left\langle \mathcal{A}_i \right\vert M_m^{\dag} M_m \left\vert \mathcal{A}_i \right\rangle}.
\end{align}
Here, Lemma \ref{lemma-2} is used in the derivation of the above equation.
We can now arrive at a clearer form than Eq. \ref{theorem-2-Eq} as follow:
\begin{align}
\label{theorem-2-clear-Eq}
\mathrm{A} \otimes \mathrm{B} = 
\frac{
\sum_{i} \lambda_i M_m \left\vert \mathcal{A}_i \right\rangle \left\vert \mathcal{B}_i \right\rangle
}
{
\sqrt{\sum_i \lvert \lambda_i \rvert^2 \left\langle \mathcal{A}_i \right\vert M_m^{\dag} M_m \left\vert \mathcal{A}_i \right\rangle}
}.
\end{align}

Theorem \ref{theorem-2} can also be described with density matrix,
both of which are essentially the same.

\begin{theorem}
\label{theorem-3}

Let $\rho = \left\vert \uppsi \right\rangle \left\langle \uppsi \right\vert$, and mark the measurement result as $\rho_A$
and the state that the remaining particles collapses onto as $\rho_B$,
then entanglement swapping results are given by
\begin{align}
\label{theorem-3-ES-results}
\rho_A \otimes \rho_B
=
\frac{
\left( M_m \otimes I \right) \rho \left( M_m \otimes I \right)^{\dag}
}
{
\textnormal{Tr}
\left[ \left( M_m \otimes I \right)^{\dag} \left( M_m \otimes I \right) \rho \right]
}
\equiv
\frac{
\left( M_m \otimes I \right) \rho \left( M_m \otimes I \right)^{\dag}
}
{
\textnormal{Tr}
\left[ \left( M_m^{\dag} M_m \otimes I \right) \rho \right]
}.													
\end{align}

\end{theorem}

\subsection{Basics of mixed state entanglement swapping}

In what follows we study the general theory for the entanglement swapping of mixed states.
Suppose that $\rho_{\mathrm{mix}}$ is the mixture of the pure states $\rho_i$ with the probabilities $\lambda_i$,
where $\rho_i = \left\vert \varphi_i \right\rangle \left\langle \varphi_i \right\vert$, such that
\begin{gather}
\label{mixed-state}
\rho_{\mathrm{mix}} = \sum_i \lambda_i \left\vert \varphi_i \right\rangle \left\langle \varphi_i \right\vert,				\\
\sum_i \lambda_i = 1.																				\notag
\end{gather}
As Theorem \ref{theorem-1}, let us mark $\left\vert \varphi_i \right\rangle$ after swapping particles by
\begin{gather}
\left\vert \hat{\varphi}_i \right\rangle =
\sum_j \lambda_{ij} \left\vert \mathcal{S}_{ij} \right\rangle \left\vert \mathcal{R}_{ij} \right\rangle,
\end{gather}
such that Eq. \ref{mixed-state} can be transformed into
\begin{gather}
\hat{\rho}_{\mathrm{mix}} = \sum_{i,j,j'} \lambda_i \lambda_{ij} \lambda_{ij'}
\left\vert \mathcal{S}_{ij} \right\rangle \left\langle \mathcal{S}_{ij'} \right\vert 
\otimes
\left\vert \mathcal{R}_{ij} \right\rangle \left\langle \mathcal{R}_{ij'} \right\vert.
\end{gather}
As Eq. \ref{decomposition-bases}, 
let $\left\vert \mathcal{S}_{ij} \right\rangle = \sum_k \lambda_{ijk} \left\vert \upphi_k \right\rangle$, 
then we can get the entanglement swapping results as follows:
\begin{gather}
\hat{\rho}_{\mathrm{mix}} = \mathcal{P} \sum_{i,j,j',k} \lambda_i \lambda_{ij} \lambda_{ij'} \lambda_{ijk} \lambda_{ij'k}
\left\vert \upphi_k \right\rangle \left\langle \upphi_{k} \right\vert 
\otimes
\left\vert \mathcal{R}_{ij} \right\rangle \left\langle \mathcal{R}_{ij'} \right\vert,							\\
\mathcal{P} = \frac{1}{\sum_{i,j,j',k} \lambda_i \lambda_{ij} \lambda_{ij'} \lambda_{ijk} \lambda_{ij'k}}, 		\notag
\end{gather}
which means that if the measurement result is assumed to be $\left\vert \upphi_{\hat{k}} \right\rangle$,
then the remaining particles collapse onto
\begin{gather}
\mathcal{P} \sum_{i,j,j'} \lambda_i \lambda_{ij} \lambda_{ij'} \lambda_{ij\hat{k}} \lambda_{ij'\hat{k}}
\left\vert \mathcal{R}_{ij} \right\rangle \left\langle \mathcal{R}_{ij'} \right\vert,							\\
\mathcal{P} = \frac{1}{\sum_{i,j,j'} \lambda_i \lambda_{ij} \lambda_{ij'} \lambda_{ij\hat{k}} \lambda_{ij'\hat{k}}}, 		\notag
\end{gather}

The mixed states entanglement swapping results can also be derived from Theorem \ref{theorem-3}.
Substituting $\hat{\rho}_{\mathrm{mix}}$ into Eq. \ref{theorem-3-ES-results}, we can arrive at
\begin{align}
\rho_A \otimes \rho_B
=
\frac{
\sum_{i,j,j'} \lambda_i \lambda_{ij} \lambda_{ij'}
M_m \left\vert \mathcal{S}_{ij} \right\rangle \left\langle \mathcal{S}_{ij'} \right\vert M_m^{\dag}
\otimes
\left\vert \mathcal{R}_{ij} \right\rangle \left\langle \mathcal{R}_{ij'} \right\vert
}
{
\sum_{i,j} \lambda_i \left\vert \lambda_{ij} \right\vert^2
\textnormal{Tr}
\left( 
M_m^{\dag} M_m \left\vert \mathcal{S}_{ij} \right\rangle \left\langle \mathcal{S}_{ij} \right\vert
\right)
}.												
\end{align}


\section{Measurement operators}

\noindent
In this section, we introduce the measurement operators that will be used frequently to realize the entanglement swapping
cases considered in this paper. Before doing this, let us introduce the d-level $m$-particle maximally entangled states
\cite{BoudaJ34202001,KarimipourV652002},
\begin{gather}
\label{d-level-maximally-entangled-states}
\left\vert\phi\left( u_1,u_2,\dots,u_m \right)\right\rangle =
\frac{1}{\sqrt d} \sum_{l=0}^{d-1} \zeta^{l u_1} \left\vert l, l \oplus u_2, l \oplus u_3,\dots,l \oplus u_m \right\rangle,	\\
u_1,u_2,\dots,u_m = 0,1,\dots,d-1,					\notag
\end{gather}
where $\zeta = e^{2\pi i/d}$ and the symbol $\oplus$ denotes addition modulo $d$ throughout this paper.
These states form an orthonormal basis in the d-level Hilbert space
$\mathcal{H}^{\otimes m}$ since
\begin{gather}
\phantom{i} \left\langle\phi \left(u_1,u_2,\dots,u_m \right) \middle\vert \phi \left(u_1',u_2',\dots,u_{m}' \right)\right\rangle = 
\prod_{i=1}^{m} \delta_{u_i,u_i'},
\notag \\
\sum_{u_1,u_2,\dots,u_m=0}^{d-1} 
\left\vert \phi \left( u_1,u_2,\dots,u_m \right) \right\rangle \left\langle \phi \left( u_1,u_2,\dots,u_m \right) \right\vert = I,
\end{gather}
where $\delta$ is Kronecker delta and $I$ is the identity matrix in $\mathcal{H}^{\otimes m}$. 
From Eq. \ref{d-level-maximally-entangled-states}, one can get the following relations:
\begin{gather}
\left\vert l_1, l_2,\dots,l_m \right\rangle = \frac{1}{\sqrt d} \sum_{l'=0}^{d-1} \zeta^{-l' l_1} 
\left\vert\phi\left( l',l_2 \ominus l_1,l_3 \ominus l_1,\dots,l_m \ominus l_1 \right)\right\rangle,
\end{gather}
where the symbol $\ominus$ denotes subtraction modulo $d$ throughout this paper.

A particular case is that when $m=2$, the above maximally entangled states reduce to the familiar d-level Bell states
\begin{gather}
\left\vert\phi\left( u_1,u_2 \right)\right\rangle =
\frac{1}{\sqrt d} \sum_{l=0}^{d-1} \zeta^{l u_1} \left\vert l, l \oplus u_2 \right\rangle,
\label{d-level-Bell-states}										
\end{gather}
which form an orthonormal basis in $\mathcal{H}^{\otimes 2}$.
Another special case is provided by limiting $u_1,u_2,\dots,u_m = 0,1$
in $\left\vert\phi\left( u_1,u_2,\dots,u_m \right)\right\rangle$, which is given by
\begin{gather}
\label{2-level-maximally-entangled-states}
\left\vert \tilde{\phi} \left( u_1,u_2,\dots,u_m \right)\right\rangle =
\frac{1}{\sqrt 2} \left( \left\vert 0, u_2, u_3,\dots, u_m \right\rangle + (-1)^{u_1}
\left\vert 1, \bar{u}_2, \bar{u}_3,\dots, \bar{u}_m \right\rangle \right),	\\
u_1,u_2,\dots,u_m = 0,1,   									\notag
\end{gather}
where a bar over a bit indicates the logical negation of the bit throughout this paper 
(e.g., if $u_2 = 0$, then $\bar{u}_2 = 1$, otherwise $\bar{u}_2 = 0$).
These states are often called multi-particle GHZ states.
In this case, one can get
\begin{gather}
\left\vert u_1,u_2,\dots,u_m \right\rangle =
\frac{1}{\sqrt 2} \left( \left\vert \tilde{\phi} \left( 0,u_2,u_3,\dots,u_m \right)\right\rangle + (-1)^{u_1}
\left\vert \tilde{\phi} \left( 1,u_2,u_3,\dots,u_m \right)\right\rangle \right).
\end{gather}
It is easy to check that
\begin{gather}
\left\vert \tilde{\phi} \left( 0,0 \right)\right\rangle \equiv \left\vert\phi\left( 0,0 \right)\right\rangle =
\frac{1}{\sqrt 2} \left( \left\vert 0,0 \right\rangle + \left\vert 1,1 \right\rangle \right),
\phantom{i}
\left\vert \tilde{\phi} \left( 0,1 \right)\right\rangle \equiv \left\vert\phi\left( 0,1 \right)\right\rangle =
\frac{1}{\sqrt 2} \left( \left\vert 0,1 \right\rangle + \left\vert 1,0 \right\rangle \right),
\notag \\
\left\vert \tilde{\phi} \left( 1,0 \right)\right\rangle \equiv \left\vert\phi\left( 1,0 \right)\right\rangle =
\frac{1}{\sqrt 2} \left( \left\vert 0,0 \right\rangle - \left\vert 1,1 \right\rangle \right),
\phantom{i}
\left\vert \tilde{\phi} \left( 1,1 \right)\right\rangle \equiv \left\vert\phi\left( 1,1 \right)\right\rangle =
\frac{1}{\sqrt 2} \left( \left\vert 0,1 \right\rangle - \left\vert 1,0 \right\rangle \right),
\label{2-level-Bell-states}
\end{gather}
and that
\begin{align}
\label{2-level-GHZ-states}
\left\vert \tilde{\phi} \left( 0,0,0 \right)\right\rangle \equiv \left\vert\phi\left( 0,0,0 \right)\right\rangle = 
\frac 1{\sqrt{2}} \left(   \left|000\right\rangle + \left|111\right\rangle \right),
\quad
\left\vert \tilde{\phi} \left( 0,0,1 \right)\right\rangle \equiv \left\vert\phi\left( 0,0,1 \right)\right\rangle = 
\frac 1{\sqrt{2}} \left(   \left|001\right\rangle + \left|110\right\rangle \right), 		\notag\\
\left\vert \tilde{\phi} \left( 0,1,0 \right)\right\rangle \equiv \left\vert\phi\left( 0,1,0 \right)\right\rangle = 
\frac 1{\sqrt{2}} \left(   \left|010\right\rangle + \left|101\right\rangle \right),
\quad
\left\vert \tilde{\phi} \left( 0,1,1 \right)\right\rangle \equiv \left\vert\phi\left( 0,1,1 \right)\right\rangle = 
\frac 1{\sqrt{2}} \left(   \left|011\right\rangle + \left|100\right\rangle \right), 		\notag\\
\left\vert \tilde{\phi} \left( 1,0,0 \right)\right\rangle \equiv \left\vert\phi\left( 1,0,0 \right)\right\rangle = 
\frac 1{\sqrt{2}} \left(   \left|000\right\rangle - \left|111\right\rangle \right),
\quad
\left\vert \tilde{\phi} \left( 1,0,1 \right)\right\rangle \equiv \left\vert\phi\left( 1,0,1 \right)\right\rangle = 
\frac 1{\sqrt{2}} \left(   \left|001\right\rangle - \left|110\right\rangle  \right),		\notag\\
\left\vert \tilde{\phi} \left( 1,1,0 \right)\right\rangle \equiv \left\vert\phi\left( 1,1,0 \right)\right\rangle = 
\frac 1{\sqrt{2}} \left(   \left|010\right\rangle - \left|101\right\rangle  \right),
\quad
\left\vert \tilde{\phi} \left( 1,1,1 \right)\right\rangle \equiv \left\vert\phi\left( 1,1,1 \right)\right\rangle = 
\frac 1{\sqrt{2}} \left(   \left|011\right\rangle - \left|100\right\rangle  \right),
\end{align}
which are the familiar Bell states and GHZ states, respectively.

Let us now introduce the measurement operators.
With the d-level maximally entangled states introduced above, one can construct the measurement operators
\begin{gather}
\mathrm{\widehat{M}} = \left\vert \widehat{\upphi} \right\rangle \left\langle \widehat{\upphi} \right\vert,
\phantom{i}
\left\vert \widehat{\upphi} \right\rangle \in \left\{ \left\vert \phi \left( u_1,u_2 \right) \right\rangle \right\},				
\notag \\
\mathrm{\widetilde{M}} = \left\vert \widetilde{\upphi} \right\rangle \left\langle \widetilde{\upphi} \right\vert,
\phantom{i} 
\left\vert \widetilde{\upphi} \right\rangle \in 
\left\{ \left\vert \phi \left( u_1,u_2,\dots,u_m \right) \right\rangle \right\},
\end{gather}
and
\begin{gather}
\mathcal{\widehat{M}} = \left\vert \widehat{\uppsi} \right\rangle \left\langle \widehat{\uppsi} \right\vert,
\phantom{i}
\left\vert \widehat{\uppsi} \right\rangle \in \left\{ \left\vert \tilde{\phi} \left( u_1,u_2 \right) \right\rangle \right\},	\notag \\
\mathcal{\widetilde{M}} = \left\vert \widetilde{\uppsi} \right\rangle \left\langle \widetilde{\uppsi} \right\vert,
\phantom{i} 
\left\vert \widetilde{\uppsi} \right\rangle \in
\left\{ \left\vert \tilde{\phi} \left( u_1,u_2,\dots,u_m \right) \right\rangle \right\},
\label{2-level-Bell-GHZ-measurement-operators}
\end{gather}
which can be used to realize the entanglement swapping of d-level entangled states and 
that of 2-level entangled states, respectively.
Here, $\mathcal{\widehat{M}}$ and $\mathcal{\widetilde{M}}$ 
can be called a Bell measurement operator and a multi-particle GHZ measurement operator, respectively.

Let us then introduce anther class of maximally entangled states, which has the form
\begin{gather}
\label{CAT-state}		
\left\vert \varphi \left( w_1,w_2,\dots,w_m \right) \right\rangle = \frac {1}{\sqrt{2}}
\left( \left\vert e_1^1,w_2,w_3,\dots,w_m \right\rangle
+ (-1)^{w_1}
\left\vert e_2^1,\widehat{w}_2,\widehat{w}_3,\dots,\widehat{w}_m \right\rangle \right),								\\
w_1 \in \left\{ 0,1 \right\},
\phantom{i}
w_k \in \left\{ \left\vert e_1^j \right\rangle,\left\vert e_2^j \right\rangle \right\}_{j=1}^m, 
\phantom{i}
\widehat{w}_k = \left\{ \left\vert e_1^j \right\rangle,\left\vert e_2^j \right\rangle \right\}_{j=1}^m \setminus \left\{ w_k \right\},
\phantom{i}
k = 2,3,\dots,m,    								
\notag																	
\end{gather}
where $\left\{ \left\vert e_1^j \right\rangle,\left\vert e_2^j \right\rangle \right\}_{j=1}^m$ 
is an arbitrary orthonormal basis in $\mathcal{H}_2$. 
For simplicity, we call these states CAT states hereinafter.
It can be seen that they are complete and orthonormal since we have
\begin{align}
& \left\langle \varphi \left( w_1,w_2,\dots,w_m \right) \middle\vert \varphi \left( w_1',w_2',\dots,w_m' \right) \right\rangle \notag \\
= & \frac{1}{2} 
\left(
\left\langle e_1^1 \middle\vert e_1^1 \right\rangle
\prod_{i=2}^{m} \left\langle w_i \middle\vert w_i^{\prime} \right\rangle 
+ (-1)^{w_1 + w_1'}
\left\langle e_2^1 \middle\vert e_2^1 \right\rangle
\prod_{i=2}^{m} \left\langle \widehat{w}_i \middle\vert \widehat{w}_i^{\prime} \right\rangle
\right)																				
= \prod_{i=1}^{m} \delta_{w_i,w_i^{\prime}},
\end{align}
and
\begin{align}
& \sum_{w_1,w_2,\dots,w_m}
\left\vert \varphi \left( w_1,w_2,\dots,w_m \right) \right\rangle 
\left\langle \varphi \left( w_1,w_1,w_2,\dots,w_m \right) \right\vert 									\notag \\
= & \sum_{w_2,w_3,\dots,w_m}
\left(
\left\vert \varphi \left( 0,w_2,\dots,w_m \right) \right\rangle 
\left\langle \varphi \left( 0,w_1,w_2,\dots,w_m \right) \right\vert
+
\left\vert \varphi \left( 1,w_2,\dots,w_m \right) \right\rangle 
\left\langle \varphi \left( 1,w_1,w_2,\dots,w_m \right) \right\vert
\right)																				\notag \\
= & \sum_{w_2,w_3,\dots,w_m}
\left(
\left\vert e_1^1 \right\rangle \left\langle e_1^1 \right\vert  
\bigotimes_{i=2}^{m}
\left\vert w_i \right\rangle \left\langle w_i \right\vert  
+
\left\vert e_2^1 \right\rangle \left\langle e_2^1 \right\vert  
\bigotimes_{i=2}^m
\left\vert \widehat{w}_i \right\rangle \left\langle \widehat{w}_i \right\vert
\right)																				
= I,
\end{align}
thus we can construct the measurement operators
\begin{gather}
\mathbb{\widehat{M}} = \left\vert \widehat{\upvarphi} \right\rangle \left\langle \widehat{\upvarphi} \right\vert,
\phantom{i} 
\left\vert \widehat{\upvarphi} \right\rangle \in
\left\{ \left\vert \varphi \left( w_1,w_2,\dots,w_m \right) \right\rangle \right\}.
\end{gather}
In particular, one can construct the measurement operators
\begin{gather}
\mathbb{\widetilde{M}} = \left\vert \widetilde{\upvarphi} \right\rangle \left\langle \widetilde{\upvarphi} \right\vert,
\phantom{i} 
\left\vert \widetilde{\upvarphi} \right\rangle \in
\left\{ \left\vert \varphi \left( v_1,v_2,\dots,v_m \right) \right\rangle \right\},
\\
v_1 \in \left\{ 0,1 \right\},
\phantom{i}
v_i \in \left\{ \left\vert + \right\rangle,\left\vert - \right\rangle \right\}
\vee
\left\{ \left\vert 0 \right\rangle,\left\vert 1 \right\rangle \right\},
\phantom{i}
v_j \in \left\{ \left\vert + \right\rangle,\left\vert - \right\rangle \right\} \setminus \left\{ v_i \right\}
\vee
\left\{ \left\vert 0 \right\rangle,\left\vert 1 \right\rangle \right\} \setminus \left\{ v_i \right\},
\phantom{i}
i,j = 2,3,\dots,m,
\notag
\end{gather}
where $\left\vert + \right\rangle = \frac{1}{\sqrt 2} \left( \left\vert 0 \right\rangle + \left\vert 1 \right\rangle \right)$
and $\left\vert - \right\rangle = \frac{1}{\sqrt 2} \left( \left\vert 0 \right\rangle - \left\vert 1 \right\rangle \right)$.


\section{Pure state entanglement swapping}

\noindent
From this section, we begin to study entanglement swapping theory.
As shown above, due to entanglement swapping results caused by different permutations of particle order are not all the same,
we would only like to consider one kind of permutation in a entanglement swapping scheme in this paper
(unless a special statement is made).
We will start with the entanglement swapping of generalized pure states for any number of quantum systems,
and then consider special cases including the entanglement swapping of general pure states, which was studied by Hardy et al.
by proposing entanglement swapping chains \cite{HardyL6252000},
and the entanglement swapping of some maximally entangled states such as Bell states.
As mentioned before, since a low dimensional system can be embedded into a higher one,
we assume that all subsystems are in a natural basis with same dimensions.
At the same time, we would only like to consider the entanglement swapping of bipartite generalized pure states, 
which belong to the most basic and representative cases.
Considering that the 2-level system is a special form of the d-level system 
and its importance in quantum information processing, we will devote a lot of attention to the formulation of 
the entanglement swapping in 2-level systems.

\subsection{Entanglement swapping for generalized pure states}

\noindent
Before characterizing the entanglement swapping between pure states, 
let us introduce the following d-level pure states:
\begin{gather}
\label{generalized-pure-states}
\left\vert \mathscr{P}_{pure} \right\rangle = \sum_{l_1,l_2=0}^{d-1} \lambda_{l_1,l_2} \left\vert l_1,l_2 \right\rangle,	\\
\sum_{l_1,l_2=0}^{d-1} \left\vert \lambda_{l_1,l_2} \right\vert^2 = 1, \notag
\end{gather}
For the convenience of expression, we refer to this type of quantum states as generalized pure states in this paper.
Suppose that there are $n$ generalized pure states, denoted as $\left\{ \left\vert \mathscr{P}_{pure}^r \right\rangle = 
\sum_{l_1,l_2=0}^{d-1} \lambda_{l_1^r,l_2^r} \left\vert l_1^r,l_2^r \right\rangle_{2r-1,2r} \right\}_{r=1}^n$ where
the subscripts $(2r-1,2r)$ represent two particles in $\left\vert \mathscr{P}_{pure}^r \right\rangle$, respectively,
then the entanglement swapping realized by performing the measurement operator $\mathrm{\widehat{M}}$
on the first particle in each state, is given by
\begin{align}
\label{generalized-pure-states-ES}
\bigotimes_{r=1}^{n} \left\vert \mathscr{P}_{pure}^r \right\rangle_{2r-1,2r}
= & 	\sum_{l_1^1,l_2^1,l_1^2,l_2^2,\dots,l_1^n,l_2^n=0}^{d-1} \prod_{r=1}^{n} \lambda_{l_1^r,l_2^r}
	\bigotimes_{r=1}^{n} \left\vert l_1^r,l_2^r \right\rangle_{2r-1,2r} \notag \\
\Rightarrow &
	\sum_{l_1^1,l_2^1,l_1^2,l_2^2,\dots,l_1^n,l_2^n=0}^{d-1} \prod_{r=1}^{n} \lambda_{l_1^r,l_2^r}
	\bigotimes_{r=1}^{n} \left\vert l_1^r \right\rangle_{2r-1} \bigotimes_{r=1}^{n} \left\vert l_2^r \right\rangle_{2r}		\notag \\
= &
	\frac{\mathcal{P}}{\sqrt d}
	\sum_{l_1^1,l_2^1,l_2^2,\dots,l_2^n=0}^{d-1}
	\sum_{v_1,v_2,\dots,v_n=0}^{d-1}
	\lambda_{l_1^1,l_2^1} \prod_{r=2}^{n} \lambda_{l_1^1 \oplus v_r,l_2^r} 
	\zeta^{-l_1^1 v_1}
	\left\vert \phi \left( v_1,v_2,\dots,v_n \right) \right\rangle
	\left\vert l_2^1,l_2^2,\dots,l_2^n \right\rangle,
\end{align}
where
\begin{gather}
\mathcal{P} = \frac{1}{\sqrt{
	\sum_{l_1^1,l_2^1,l_2^2,\dots,l_2^n=0}^{d-1}
	\sum_{v_2,v_3,\dots,v_n=0}^{d-1}
	\left\vert \lambda_{l_1^1,l_2^1} \prod_{r=2}^{n} \lambda_{l_1^1 \oplus v_r,l_2^r} \right\vert^2
	}}, 				\notag
\end{gather}
and the symbol $\Rightarrow$, which will be frequently used in this paper, represents the swapping of the particles.
It can be seen that, suppose the measurement result is $\left\vert \phi \left( \tilde{v}_1,\tilde{v}_2,\dots,\tilde{v}_n \right) \right\rangle$, 
the remaining particle collapse onto 
\begin{gather}
	\frac{\mathcal{P}}{\sqrt d}
	\sum_{l_1^1,l_2^1,l_2^2,\dots,l_2^n=0}^{d-1}
	\lambda_{l_1^1,l_2^1} \prod_{r=2}^{n} \lambda_{l_1^1 \oplus \tilde{v}_r,l_2^r} 
	\zeta^{-l_1^1 \tilde{v}_1}
	\left\vert l_2^1,l_2^2,\dots,l_2^n \right\rangle,
\\
\mathcal{P} = \sqrt{ \frac{d}{
	\sum_{l_1^1,l_2^1,l_2^2,\dots,l_2^n=0}^{d-1}
	\left\vert \lambda_{l_1^1,l_2^1} \prod_{r=2}^{n} \lambda_{l_1^1 \oplus \tilde{v}_r,l_2^r} \right\vert^2
	}}. 				\notag
\end{gather}
When $n = 2$, Eq. \ref{generalized-pure-states-ES} reduces to the entanglement swapping 
between two generalized pure states, which is given by
\begin{gather}
\label{two-generalized-pure-states-ES}
	\bigotimes_{r=1}^{2} \left\vert \mathscr{P}_{pure}^r \right\rangle_{2r-1,2r}
\rightarrow
	\frac{\mathcal{P}}{\sqrt d}
	\sum_{l_1^1,l_2^1,l_2^2=0}^{d-1}
	\sum_{v_1,v_2=0}^{d-1}
	\lambda_{l_1^1,l_2^1} \lambda_{l_1^1 \oplus v_2,l_2^2} 
	\zeta^{-l_1^1 v_1}
	\left\vert \phi \left( v_1,v_2 \right) \right\rangle
	\left\vert l_2^1,l_2^2 \right\rangle,
\\
\mathcal{P} = \frac{1}{\sqrt{
	\sum_{l_1^1,l_2^1,l_2^2,v_2=0}^{d-1}
	\left\vert \lambda_{l_1^1,l_2^1} \lambda_{l_1^1 \oplus v_2,l_2^2}  \right\vert^2
	}}. 				\notag
\end{gather}
When $l_1 \equiv l_2 \forall l_1,l_2 \in \left\{ 0,1,\dots,d-1 \right\}$, $\left\vert \mathscr{P}_{pure} \right\rangle$
reduces to the general pure state
\begin{gather}
\label{general-pure-state}
\left\vert \mathscr{P}_{\uppercase\expandafter{\romannumeral1}} \right\rangle 
= \sum_{l=0}^{d-1} \lambda_l \left\vert l,l \right\rangle,
\\
\sum_{l=0}^{d-1} \left\vert \lambda_{l} \right\vert^2 = 1,
\notag
\end{gather}
such that the entanglement swapping between 
$n$ general pure states marked by
$\left\{ \left\vert \mathscr{P}_{\uppercase\expandafter{\romannumeral1}}^r \right\rangle = 
\sum_{l_r=0}^{d-1} \lambda_{l_r} \left\vert l_r,l_r \right\rangle \right\}_{r=1}^n$
in which $\sum_{l_r = 0}^{d-1} \left\vert \lambda_{l_r} \right\vert^2 = 1$, is given by
\begin{gather}
\label{general-pure-states-ES}
	\bigotimes_{r=1}^{n} \left\vert \mathscr{P}_{\uppercase\expandafter{\romannumeral1}}^r \right\rangle						
\rightarrow
	\frac{\mathcal{P}}{\sqrt d}
	\sum_{v_1,v_2,\dotsm,v_n,l_1=0}^{d-1}
	\lambda_{l_1} \prod_{r=2}^{n}  \lambda_{l_1 \oplus v_r} \zeta^{-l_1 v_1}
	\left\vert \phi \left( v_1,v_2,\dots,v_n \right) \right\rangle
	\left\vert l_1,l_1 \oplus v_2,l_1 \oplus v_3,\dots,l_1 \oplus v_n \right\rangle,
\\
\mathcal{P} = \frac{1}{\sqrt{\sum_{v_2,\dotsm,v_n,l_1=0}^{d-1} 
\left\vert \lambda_{l_1} \prod_{r=2}^{n}  \lambda_{l_1 \oplus v_r} \right\vert^2}}. 				\notag
\end{gather}
As a simple special case of the entanglement swapping shown above,
the entanglement swapping between two general pure states,
which was considered in Ref. \cite{HardyL6252000}, can be expressed as
\begin{gather}
\left\vert \mathscr{P}_{\uppercase\expandafter{\romannumeral1}}^1 \right\rangle_{1,2} 
\otimes 
\left\vert \mathscr{P}_{\uppercase\expandafter{\romannumeral1}}^2 \right\rangle_{3,4}
\rightarrow
\frac{\mathcal{P}}{\sqrt{d}} \sum_{v_1,v_2,l_1=0}^{d-1} \lambda_{l_1} \lambda_{l_1 \oplus v_2} \zeta^{- l_1 v_1} 
\left\vert \phi \left( v_1,v_2 \right) \right\rangle_{1,3} \left\vert l_1,l_1 \oplus v_2 \right\rangle_{2,4},	
\label{two-general-pure-states-ES}
\\
\mathcal{P} =
\frac{1}{\sqrt{\sum_{v_2,l_1=0}^{d-1} \left\vert \lambda_{l_1} \lambda_{l_1 \oplus v_2} \right\vert^2}}. 	\notag
\end{gather}

Let us now turn our attention to the entanglement swapping between 2-level systems.
Indeed, by limiting $d=2$ in Eq. \ref{generalized-pure-states-ES} and Eq. \ref{general-pure-states-ES},
one can get, in a direct way, the formulas for the entanglement swapping between 2-level generalized pure states
and the one between 2-level general pure states, respectively.
In what follows we would like to take the entanglement swapping of 2-level general pure states as an example,
and adopt the derivation process shown in Ref. \cite{JiZX5852022}, to derive the entanglement swapping results
through the direct expansion and combination of polynomials. Such a derivation process is more complicated than 
Eq. \ref{general-pure-states-ES}, but it can more specifically show the entangled states generated by entanglement swapping.
Let us assume that there are $n$ 2-level general pure states and mark them by
\begin{gather}
\left\vert \mathscr{P}_{\uppercase\expandafter{\romannumeral2}}^r \right\rangle = 
\lambda_1^r \left\vert 00 \right\rangle_{2r-1,2r} + \lambda_2^r \left\vert 11 \right\rangle_{2r-1,2r},
\\
\left\vert \lambda_1^r \right\vert^2 + \left\vert \lambda_2^r \right\vert^2 = 1, r=1,2,\dots,n,					\notag
\end{gather}
such that the entanglement swapping between them, realized by performing $\mathcal{\widetilde{M}}$ on
the first particle in each state, can be expressed as
\begin{align}
\label{pure-states-ES}
\bigotimes_{r=1}^{n} \left\vert \mathscr{P}_{\uppercase\expandafter{\romannumeral2}}^r \right\rangle_{2r-1,2r}		
= & \bigotimes_{r=1}^{n} \left( \lambda_1^r \left\vert 00 \right\rangle + \lambda_2^r \left\vert 11 \right\rangle \right)_{2r-1,2r} 	\notag \\
= &	\prod_{r=1}^{n} \lambda_1^r \bigotimes_{r=1}^{n} \left\vert 00 \right\rangle_{2r-1,2r}
	+
	\lambda_2^n \prod_{r=1}^{n-1} \lambda_1^r \bigotimes_{r=1}^{n-1} 
	\left\vert 00 \right\rangle_{2r-1,2r} \left\vert 11 \right\rangle_{2n-1,2n}
	\notag	\\
&	+
	\lambda_2^{n-1} \lambda_1^n \prod_{r=1}^{n-2} \lambda_1^r \bigotimes_{r=1}^{n-2} 
	\left\vert 00 \right\rangle_{2r-1,2r} \left\vert 11 \right\rangle_{2n-3,2n-2} \left\vert 00 \right\rangle_{2n-1,2n}
	+
	\dotsm
	+
	\prod_{r=1}^{n} \lambda_2^r \bigotimes_{r=1}^{n} \left\vert 11 \right\rangle_{2r-1,2r}
	\notag	\\
\Rightarrow &
	\prod_{r=1}^{n} \lambda_1^r 
	\bigotimes_{r=1}^{n} \left\vert 0 \right\rangle_{2r-1} \bigotimes_{r=1}^{n} \left\vert 0 \right\rangle_{2r}
	+
	\lambda_2^n \prod_{r=1}^{n-1} \lambda_1^r 
	\bigotimes_{r=1}^{n-1} \left\vert 0 \right\rangle_{2r-1} \left\vert 1 \right\rangle_{2n-1}
	\bigotimes_{r=1}^{n-1} \left\vert 0 \right\rangle_{2r} \left\vert 1 \right\rangle_{2n}
	\notag	\\
&	+
	\lambda_2^{n-1} \lambda_1^n \prod_{r=1}^{n-2} \lambda_1^r
	\bigotimes_{r=1}^{n-2} \left\vert 0 \right\rangle_{2r-1} \left\vert 1 \right\rangle_{2n-3} \left\vert 0 \right\rangle_{2n-1}
	\bigotimes_{r=1}^{n-2} \left\vert 0 \right\rangle_{2r} \left\vert 1 \right\rangle_{2n-2} \left\vert 0 \right\rangle_{2n}
	+
	\dotsm
	+
	\prod_{r=1}^{n} \lambda_2^r 
	\bigotimes_{r=1}^{n} \left\vert 1 \right\rangle_{2r-1} \bigotimes_{r=1}^{n} \left\vert 1 \right\rangle_{2r}
	\notag	\\
= &
	\frac{\mathcal{P}}{\sqrt 2}
	\sum_{u_2,u_3,\dotsm,u_n=0}^1
	\left[
	\left\vert \tilde{\phi} \left( 0,u_2,u_3,\dots,u_n \right)\right\rangle
	\otimes
	\left(			
	\widehat{\uplambda}_{\varSigma} \left\vert 0,u_2,u_3,\dots,u_n \right\rangle
	+
	\widehat{\uplambda}_{2^n - \varSigma + 1} \left\vert 1,\hat{u}_2,\hat{u}_3,\dots,\hat{u}_n \right\rangle
	\right)
	\right.
	\notag	\\
&	\left.
	+
	\left\vert \tilde{\phi} \left( 1,u_2,u_3,\dots,u_n \right)\right\rangle 
	\otimes
	\left(			
	\widehat{\uplambda}_{\varSigma} \left\vert 0,u_2,u_3,\dots,u_n \right\rangle
	-
	\widehat{\uplambda}_{2^n - \varSigma + 1} \left\vert 1,\hat{u}_2,\hat{u}_3,\dots,\hat{u}_n \right\rangle
	\right)
	\right],
\end{align}
where $\varSigma=1,2,\dots,2^{n-1}$ and
\begin{gather}
\mathcal{P} = 
\frac{1}
{\sqrt{\sum_{\varSigma=1}^{2^{n-1}} 
\left(
\left\vert \widehat{\uplambda}_{\varSigma} \right\vert^2 + \left\vert \widehat{\uplambda}_{2^n - \varSigma + 1} \right\vert^2 
\right)}},
\notag \\
\widehat{\uplambda}_1 = \prod_{r=1}^{n} \lambda_1^r,
\phantom{i}
\widehat{\uplambda}_2 = \lambda_2^n \prod_{r=1}^{n-1} \lambda_1^r,
\phantom{i}
\widehat{\uplambda}_3 = \lambda_2^{n-1} \lambda_1^n \prod_{r=1}^{n-2} \lambda_1^r,
\dots,
\widehat{\uplambda}_{2^n} = \prod_{r=1}^{n} \lambda_2^r.
\end{gather}
Then, the entanglement swapping between two 2-level general pure states,
presented in Ref. \cite{HardyL6252000}, can be easily derived and described by
\begin{align}
\left\vert \mathscr{P}_{\uppercase\expandafter{\romannumeral2}}^1 \right\rangle 
\otimes 
\left\vert \mathscr{P}_{\uppercase\expandafter{\romannumeral2}}^2 \right\rangle 
\rightarrow
& 
\left\vert\tilde{\phi}\left( 0,0 \right)\right\rangle_{1,3} \otimes \frac{\mathcal{P}}{\sqrt 2}
\left(
\lambda_1^1 \lambda_1^2 \left\vert 00 \right\rangle + \lambda_2^1 \lambda_2^2 \left\vert 11 \right\rangle
\right)_{2,4}
+
\left\vert\tilde{\phi}\left( 0,1 \right)\right\rangle_{1,3} \otimes \frac{\mathcal{P}}{\sqrt 2}
\left(
\lambda_1^1 \lambda_1^2 \left\vert 00 \right\rangle - \lambda_2^1 \lambda_2^2 \left\vert 11 \right\rangle
\right)_{2,4}
\notag \\
+
& \left\vert\tilde{\phi}\left( 1,0 \right)\right\rangle_{1,3} \otimes \frac{\mathcal{P}}{\sqrt 2}
\left(
\lambda_1^1 \lambda_2^2 \left\vert 01 \right\rangle + \lambda_2^1 \lambda_1^2 \left\vert 10 \right\rangle
\right)_{2,4}
+
\left\vert\tilde{\phi}\left( 1,1 \right)\right\rangle_{1,3} \otimes \frac{\mathcal{P}}{\sqrt 2}
\left(
\lambda_1^1 \lambda_2^2 \left\vert 01 \right\rangle - \lambda_2^1 \lambda_1^2 \left\vert 10 \right\rangle
\right)_{2,4},
\end{align}
where 
\begin{align}
\mathcal{P} = 
\frac{1}{\sqrt{\left\vert \lambda_1^1 \lambda_1^2 \right\vert^2 + \left\vert \lambda_2^1 \lambda_2^2 \right\vert^2
+ \left\vert \lambda_1^1 \lambda_2^2 \right\vert^2 + \left\vert \lambda_2^1 \lambda_1^2 \right\vert^2}}. \notag
\end{align}
Note here that $\left\vert \tilde{\phi} \left( 0,0 \right)\right\rangle,\left\vert \tilde{\phi} \left( 0,1 \right)\right\rangle,
\left\vert \tilde{\phi} \left( 1,0 \right)\right\rangle,\left\vert \tilde{\phi} \left( 1,1 \right)\right\rangle$
are Bell states (see Eq. \ref{2-level-Bell-states}).

\subsection{Entanglement swapping for maximally entangled states}

\noindent
In what follows, we focus on the research of the entanglement swapping between d-level maximally entangled states. 
As before, we will first consider the entanglement swapping between d-level systems, 
and then transition to the cases of 2-level systems.

For d-level systems, we will consider the entanglement swapping realized by performing joint measurement on the particles 
containing the first particle in some entangled states and the particles without the first particles in the remaining entangled states.
Then we turn our attention to special cases, including the entanglement swapping schemes achieved by
measuring the particles containing the first particle in all states, and the scheme achieved by
measuring the particles without containing the first particle of all states, which was studied in Ref. \cite{BoudaJ34202001}.
Several other special cases, such as the entanglement swapping between a
d-level maximally entangled state and a d-level Bell state \cite{KarimipourV652002},
and the entanglement swapping between d-level Bell states \cite{BoudaJ34202001},
will also be introduced. Finally, the entanglement swapping between generalized d-level GHZ states is presented.

For 2-level systems, we will consider the entanglement swapping between CAT states.
These states are a variant of the multi-particle GHZ states. The main feature that distinguishes them from GHZ states 
is that not all subsystems are confined to the natural basis.
In other words, the states of all subsystems are in different basis, 
such as natural basis $\{ \left\vert 0 \right\rangle,\left\vert 1 \right\rangle\}$
and physical basis $\{ \left\vert + \right\rangle,\left\vert - \right\rangle\}$.
We will then present the entanglement swapping for some other special 2-level maximally entangled states,
such as cluster states.

\subsubsection{Entanglement swapping of d-level maximally entangled states}

\noindent
Let us now start considering the entanglement swapping between the multi-particle maximally entangled states
(see Eq. \ref{d-level-maximally-entangled-states}).
As before, let us assume that there are $n$ multi-particle maximally entangled states containing 
$m_1,m_2,\dots,m_n$ particles each, and denote them as
$\left\{ \left\vert \phi \left( u_1^r,u_2^r,\dots,u_{m_r}^r \right) \right\rangle \right\}_{r=1}^{n}$.
Without losing generality, let us select the first $k_r$ particles in
$\left\vert \phi \left( u_1^r,u_2^r,\dots,u_{m_r}^r \right) \right\rangle$ for $r =1,2,\dots,t$,
and the last $k_r$ particles in $\left\vert \phi \left(u_1^r,u_2^r,\dots,u_{m_r}^r \right) \right\rangle$ for $r =t+1,t+2,\dots,n$.
Assuming that the measurement operator $\mathrm{\widetilde{M}}$ is performed on the selected particles, we have
\begin{align}
\label{maximally-ES}
& \bigotimes_{r=1}^{n} \left\vert \phi \left(u_1^r,u_2^r,\dots,u_{m_r}^r \right) \right\rangle \notag \\
= & \frac {1}{d^{n/2}} \sum_{l_1,l_2,\dots,l_n=0}^{d-1} \zeta^{\sum_{r=1}^{n} l_ru_1^r} 
\bigotimes_{r=1}^{n} \left\vert l_r, l_r \oplus u_2^r,l_r \oplus u_3^r,\dots,l_r \oplus u_{m_r}^r \right\rangle_{1,2,\dots,m_r}		\notag \\
\Rightarrow 
& \frac {1}{d^{n/2}} \sum_{l_1,l_2,\dots,l_n=0}^{d-1} \zeta^{\sum_{r=1}^{n} l_r u_1^r}
\bigotimes_{r=1}^{t}
\left\vert l_r, l_r \oplus u_2^r,l_r \oplus u_3^r,\dots,l_r \oplus u_{k_r}^r \right\rangle								
\bigotimes_{r=t+1}^n
\left\vert l_r \oplus u_{m_r-k_r+1}^r, l_r \oplus u_{m_r-k_r+2}^r,\dots,l_r \oplus u_{m_r}^r \right\rangle					\notag \\
&\bigotimes_{r=1}^t
\left\vert l_r \oplus u_{k_r+1}^r,l_r \oplus u_{k_r+2}^r,\dots,l_r \oplus u_{m_r}^r \right\rangle
\bigotimes_{r=t+1}^n
\left\vert l_r,l_r \oplus u_2^r,l_r \oplus u_3^r,\dots,l_r \oplus u_{m_r-k_r}^r \right\rangle								\notag \\
= & \frac {1}{d^{n/2}} \sum_{v_1^1,v_1^2,\dots,v_1^n=0}^{d-1} 
\zeta^{\sum_{r=2}^{n} v_1^r u_1^r - \sum_{r=t+1}^{n} u_{m_r-k_r+1}^r u_1^r - \left( \sum_{r=1}^n u_1^r -v_1^1 \right) u_{{k_1}+1}^1}
\left\vert \phi \left(v_1^1,v_2^1,\dots,v_{k_1}^1,v_1^2,v_2^2,\dots,v_{k_2}^2,\dots,v_1^n,v_2^n,\dots,v_{k_n}^n \right)\right\rangle \notag \\
& \otimes
\left\vert 
\phi \left(
\tilde{v}_1^1,\tilde{v}_2^1,\dots,\tilde{v}_{m_1 - k_1}^1,\tilde{v}_1^2,\tilde{v}_2^2,\dots,\tilde{v}_{m_2 - k_2}^2,\dots,
\tilde{v}_1^n,\tilde{v}_2^n,\dots,\tilde{v}_{m_n - k_n}^n 
\right)
\right\rangle
\end{align}
where
\begin{gather}
v_j^1 = u_j^1, \phantom{i} j =2,3,\dots,k_1,
\notag \\
v_j^i = v_1^i \oplus u_j^i, \phantom{i} i =2,3,\dots,t, \phantom{i} j =2,3,\dots,k_i,
\notag \\
v_j^i = v_1^i \ominus u_{m_i-k_i+1}^i \oplus u_{m_i - k_i + j}^i, \phantom{i} i =t+1,t+2,\dots,n, \phantom{i} j =2,3,\dots,k_i,
\notag \\
\tilde{v}_1^1 = \oplus_{r=1}^{n} u_1^r \ominus v_1^1,
\notag \\ 
\tilde{v}_j^1 = u_{k_1+j}^1 \ominus u_{k_1+1}^1,\phantom{i} j = 2,3,\dots,m_1 - k_1,
\notag \\
\tilde{v}_j^i = v_1^i \ominus u_{k_1+1}^1 \oplus u_{k_i+j}^1, \phantom{i} i =1,2,\dots,t, \phantom{i} j = 1,2,\dots,m_i - k_i,
\notag \\
\tilde{v}_1^i = v_1^i \ominus u_{k_1+1}^1 \ominus u_{m_i - k_i + 1}^i, \phantom{i} i =t+1,t+2,\dots,n,
\notag \\
\tilde{v}_j^i = v_1^i \ominus u_{k_1+1}^1 \ominus u_{m_i - k_i+1}^i \oplus u_j^i, 
\phantom{i} i =t+1,t+2,\dots,n, \phantom{i} j =2,3,\dots,m_i - k_i.
\label{general-maximally-ES-conditions}
\end{gather}
When $m=2$, the above entanglement swapping reduce to the case of d-level Bell states, which is given by
\begin{align}
\label{maximally-Bell-ES}
& \bigotimes_{r=1}^{n} \left\vert \phi \left(u_1^r,u_2^r \right) \right\rangle 										\notag \\
= & \frac {1}{d^{n/2}} \sum_{l_1,l_2,\dots,l_n=0}^{d-1} \zeta^{\sum_{r=1}^{n} l_r u_1^r}
\bigotimes_{r=1}^{n} \left\vert l_r, l_r \oplus u_2^r \right\rangle												\notag \\
\Rightarrow
& \frac {1}{d^{n/2}} \sum_{l_1,l_2,\dots,l_n=0}^{d-1} \zeta^{\sum_{r=1}^{n} l_r u_1^r}
\left\vert l_1, l_2,\dots,l_t,l_{t+1} \oplus u_2^{t+1},l_{t+2} \oplus u_2^{t+2},\dots,l_n \oplus u_2^n \right\rangle
\left\vert l_{t+1},l_{t+2},\dots,l_n,l_1 \oplus u_2^1,l_2 \oplus u_2^2,\dots,l_t \oplus u_2^t \right\rangle					\notag \\
= & \frac {1}{d^{n/2}} \sum_{v_1^1,v_1^2,\dots,v_1^n=0}^{d-1} 
\zeta^{\sum_{r=2}^n v_r u_1^r - \sum_{r=t+1}^n u_1^r u_2^r - \left( \sum_{r=1}^n u_1^r - v_1 \right) \left( v_{t+1} - u_2^{t+1} \right)}
\left\vert \phi \left(v_1,v_2,\dots,v_n \right)\right\rangle														
\otimes
\left\vert \phi \left( \oplus_{r=1}^{n} u_1^r \ominus v_1,v_{t+2} \ominus v_{t+1} \oplus u_2^{t+1} \ominus u_2^{t+2},
\right.\right.																						\notag \\
&
v_{t+3} \ominus v_{t+1} \oplus u_2^{t+1} \ominus u_2^{t+3},\dots,
v_n \ominus v_{t+1} \oplus u_2^{t+1} \ominus u_2^n,u_2^{t+1} \ominus v_{t +1} \oplus u_2^1,
v_2 \ominus v_{t +1} \oplus u_2^{t+1} \oplus u_2^2,v_3 \ominus v_{t +2} \oplus u_2^{t+1} \oplus u_2^3, 			\notag \\
& \left.\left.
\dots,v_t \ominus v_{t +1} \oplus u_2^{t+1} \oplus u_2^t
\right)\right\rangle.
\end{align}

There are two extreme cases included in the entanglement swapping case presented in Eq. \ref{maximally-ES}.
One is to measure the first $k_r$ particles in $\left\vert \phi \left(u_1^r,u_2^r,\dots,u_{m_r}^r \right) \right\rangle$,
which has not been considered in existing literature.
The other is to measure the last $k_r$ particles in $\left\vert \phi \left(u_1^r,u_2^r,\dots,u_{m_r}^r \right) \right\rangle$,
which was proposed in Ref. \cite{BoudaJ34202001} and organized in Ref. \cite{JiZX5852022}.
Let us first characterize the former. Suppose that the measurement operator $\mathrm{\widetilde{M}}$
is performed on the first $k_r$ particles in each state, then the entanglement swapping can be expressed as
\begin{align}
\label{maximally-ES-contain-first-particle}
& \bigotimes_{r=1}^{n} \left\vert \phi \left(u_1^r,u_2^r,\dots,u_{m_r}^r \right) \right\rangle 							\notag \\
= & \frac {1}{d^{n/2}} \sum_{l_1,l_2,\dots,l_n=0}^{d-1} \zeta^{\sum_{r=1}^{n} l_ru_1^r} 
\bigotimes_{r=1}^{n} \left\vert l_r, l_r \oplus u_2^r,l_r \oplus u_3^r,\dots,l_r \oplus u_{m_r}^r \right\rangle		\notag \\
\Rightarrow 
& \frac {1}{d^{n/2}} \sum_{l_1,l_2,\dots,l_n=0}^{d-1} \zeta^{\sum_{r=1}^{n} l_r u_1^r}
\bigotimes_{r=1}^{n}
\left\vert l_r, l_r \oplus u_2^r,l_r \oplus u_3^r,\dots,l_r \oplus u_{k_r}^r \right\rangle
\bigotimes_{r=1}^{n}
\left\vert l_r \oplus u_{k_r+1}^r,l_r \oplus u_{k_r+2}^r,\dots,l_r \oplus u_{m_r}^r \right\rangle					\notag \\
= & \frac {1}{d^{n/2}} \sum_{v_1^1,v_1^2,\dots,v_1^n=0}^{d-1} 
\zeta^{\sum_{r=2}^{n} u_1^r \left( v_1^r - u_{k_1+1}^1 \right) - u_{k_1+1}^1 \left( u_1^1-v_1^1 \right)}
\left\vert \phi \left(v_1^1,v_2^1,\dots,v_{k_1}^1,v_1^2,v_2^2,\dots,v_{k_2}^2,\dots,
v_1^n,v_2^n,\dots,v_{k_n}^n \right)\right\rangle												\notag \\
& \otimes
\left\vert \phi \left( \oplus_{r=1}^{n} u_1^r \ominus v_1^1,
u_{k_2+2}^1 \ominus u_{k_1+1}^1,u_{k_3+3}^1 \ominus u_{k_1+1}^1,\dots,u_{m_1}^1 \ominus u_{k_1+1}^1,
u_{k_2+1}^2 \ominus u_{k_1+1}^1 \oplus v_1^2,u_{k_2+2}^2 \ominus u_{k_1+1}^1 \oplus v_1^2,
\right.\right.																						\notag \\
& \qquad \left.\left.
\dots,u_{m_2}^2 \ominus u_{k_1+1}^1 \oplus v_1^2,\dots,
u_{k_n+1}^n \ominus u_{k_1+1}^1 \oplus v_1^n,u_{k_n+2}^n \ominus u_{k_1+1}^1 \oplus v_1^n,\dots,				
u_{k_n+n}^n \ominus u_{k_1+1}^1 \oplus v_1^n
\right)\right\rangle,
\end{align}
where
\begin{gather}
v_j^1 = u_j^1, \phantom{i} j =2,3,\dots,k_1,
\notag \\
v_{j}^i = v_1^i \oplus u_j^i, \phantom{i} i =2,3,\dots,n, \phantom{i} j =2,3,\dots,k_i.
\label{first-extreme-case-conditions}
\end{gather}
The latter, achieved by measuring the last $k_r$ particles in each state, is given by
\begin{align}
\label{maximally-ES-without-first-particle}
& \bigotimes_{r=1}^{n} \left\vert \phi \left(u_1^r,u_2^r,\dots,u_{m_r}^r \right) \right\rangle \notag \\
= & \frac {1}{d^{n/2}} \sum_{l_1,l_2,\dots,l_n=0}^{d-1} \zeta^{\sum_{r=1}^{n} l_ru_1^r} 
\bigotimes_{r=1}^{n} \left\vert l_r, l_r \oplus u_2^r,l_r \oplus u_3^r,\dots,l_r \oplus u_{m_r}^r \right\rangle		\notag \\
\Rightarrow
& \frac {1}{d^{n/2}} \sum_{l_1,l_2,\dots,l_n=0}^{d-1} \zeta^{\sum_{r=1}^{n} l_r u_1^r}
\bigotimes_{r=1}^{n}
\left\vert l_r, l_r \oplus u_2^r,l_r \oplus u_3^r,\dots,l_r \oplus u_{m_r-k_r}^r \right\rangle
\bigotimes_{r=1}^{n}
\left\vert l_r \oplus u_{m_r-k_r+1}^r,l_r \oplus u_{m_r-k_r+2}^r,\dots,l_r \oplus u_{m_r}^r \right\rangle			\notag \\
= & \frac {1}{d^{n/2}} \sum_{v_1^1,v_1^2,\dots,v_1^n=0}^{d-1}
\zeta^{\sum_{r=2}^{n} \left( v_1^r - u_{m_r-k_r+1} + u_{m_1-k_1+1}^1 \right) u_1^r}
\left\vert \phi \left(\oplus_{r=1}^{n} u_1^r \ominus v_1^1,u_2^1,u_3^1,\dots,u_{m_1-k_1}^1,					
v_1^2 \oplus u_{m_1-k_1+1}^1 \ominus u_{m_2-k_2+1}^2,
\right.\right.																					\notag \\
& \qquad
v_1^2 \oplus u_{m_1-k_1+1}^1 \ominus u_{m_2-k_2+1}^2 \oplus u_2^2,
\dots, 									
v_1^2 \oplus u_{m_1-k_1+1}^1 \ominus u_{m_2-k_2+1}^2 \oplus u_{m_2-k_2}^2,							
\dots,																			
v_1^n \oplus u_{m_1-k_1+1}^1 \ominus u_{m_n-k_n+1}^n,													\notag \\
& \phantom{f} \quad \left.\left.
v_1^n \oplus u_{m_1-k_1+1}^1 \ominus u_{m_n-k_n+1}^n \oplus u_2^n,									
\dots,																					
v_1^n \oplus u_{m_1-k_1+1}^1 \ominus u_{m_n-k_n+1}^n \oplus u_{m_n-k_n}^n \right) \right\rangle		\notag \\
& 
\otimes \left\vert \phi \left(v_1^1,v_2^1,\dots,v_{k_1}^1,v_1^2,v_2^2,\dots,v_{k_2}^2,\dots,
v_1^n,v_2^n,\dots,v_{k_n}^n \right)\right\rangle,
\end{align}
where
\begin{gather}
\label{second-extreme-case-conditions}
v_j^1 = u_{m_1-k_1+j}^1 \ominus u_{m_1-k_1+1}^1, \phantom{i} j =2,3,\dots,k_1,			
\notag \\
v_{j}^i = v_1^i \ominus u_{m_i-k_i+1}^i \oplus u_{m_i - k_i + j}^i, \phantom{i} i =2,3,\dots,n, \phantom{i} j =2,3,\dots,k_i.
\end{gather}
From Eqs. \ref{maximally-ES-contain-first-particle} and \ref{maximally-ES-without-first-particle},
one can derive the formulas of the entanglement swapping between a
d-level maximally entangled state and a d-level Bell state, which is proposed in Ref. \cite{KarimipourV652002}.
Assume that the measurement operator $\mathrm{\widehat{M}}$ is performed on
the first particle in each state, one can get
\begin{align}
\label{ES-cat-Bell-contain-first-particle}
&\left\vert\phi\left( u_1^1,u_2^1,\dots,u_m^1 \right)\right\rangle 
\otimes 
\left\vert\phi\left( u_1^2,u_2^2 \right)\right\rangle 	\notag \\
= & \frac {1}{d} \sum_{l_1,l_2=0}^{d-1} \zeta^{l_1 u_1^1 + l_2 u_1^2} 
\left\vert l_1, l_1 \oplus u_2^1, l_1 \oplus u_3^1,\dots,l_r \oplus u_{m}^1 \right\rangle_{1,2,\dots,m}
\left\vert l_2, l_2 \oplus u_2^2 \right\rangle_{\hat{1},\hat{2}}											\notag \\
\Rightarrow 
& \frac {1}{d} \sum_{l_1,l_2=0}^{d-1} \zeta^{l_1 u_1^1 + l_2 u_1^2}
\left\vert l_1,l_2 \right\rangle_{1,\hat{1}}
\left\vert l_1 \oplus u_2^1,l_1 \oplus u_3^1,\dots,l_1 \oplus u_m^1,l_2 \oplus u_{2}^2 \right\rangle_{2,\dots,m,\hat{2}} 	\notag \\
= & \frac {1}{d} \sum_{v_1, v_2=0}^{d-1} 
\zeta^{u_2^1 \left( u_1^1 + u_1^2 - v_1 \right) + u_1^2 v_2}
\left\vert\phi\left(v_1,v_2 \right)\right\rangle
\otimes 
\left\vert\phi\left( u_1^1 \oplus u_1^2 \ominus v_1,u_3^1 \ominus u_2^1,u_4^1 \ominus u_2^1,\dots,
u_m^1 \ominus u_2^1,v_2 \oplus u_2^2 \ominus u_2^1 \right)\right\rangle.
\end{align}
Assume that $\mathrm{\widehat{M}}$ is performed on the two particles
without involving the first particle in each state, one can get
\begin{align}
\label{ES-cat-Bell-without-first-particle}
&\left\vert \phi \left( u_1^1,u_2^1,\dots,u_m^1 \right) \right\rangle 
\otimes
\left\vert \phi \left( u_1^2,u_2^2 \right) \right\rangle 	\notag \\
= & \frac {1}{d} \sum_{l_1,l_2=0}^{d-1} \zeta^{l_1 u_1^1 + l_2 u_1^2} 
\left\vert l_1, l_1 \oplus u_2^1, l_1 \oplus u_3^1,\dots,l_1 \oplus u_m^1 \right\rangle_{1,2,\dots,s,\dots,m}
\left\vert l_2, l_2 \oplus u_2^2 \right\rangle_{\hat{1},\hat{2}}														\notag \\
\Rightarrow 
& \frac {1}{d} \sum_{l_1,l_2=0}^{d-1} \zeta^{l_1 u_1^1 + l_2 u_1^2}
\left\vert l_1, l_1 \oplus u_2^1,l_1 \oplus u_3^1,\dots,l_2 \oplus u_{2}^2,\dots,l_1 \oplus u_m^1 \right\rangle_{1,2,\dots,\hat{2},\dots,m}
\left\vert l_2,l_1 \oplus u_s^1 \right\rangle_{\hat{1},s}															\notag \\
= & \frac {1}{d} \sum_{v_1, v_2=0}^{d-1} 
\zeta^{\left( u_s^1 - v_2 \right) \left( u_1^2 - v_1 \right)}
\left\vert\phi\left( u_1^1 \oplus u_1^2 \ominus v_1,u_2^1,u_3^1,\dots,u_s^1 \oplus u_2^2 \ominus v_2,\dots,u_m^1 \right)\right\rangle
\otimes \left\vert\phi\left(v_1,v_2 \right)\right\rangle.
\end{align}
If $m=2$ in Eqs. \ref{ES-cat-Bell-contain-first-particle} and \ref{ES-cat-Bell-without-first-particle}, 
the entanglement swapping between d-level Bell states can be derived as follows
\cite{BoudaJ34202001,KarimipourV652002},
\begin{align}
& \left\vert\phi\left( u_1^1,u_2^1 \right)\right\rangle_{1,2} \otimes \left\vert\phi\left( u_1^2,u_2^2 \right)\right\rangle_{3,4}
\rightarrow
\frac {1}{d} \sum_{v_1, v_2=0}^{d-1}
\zeta^{\left( u_2^2 - v_2 \right) \left( u_1^2 - v_1 \right)}
\left\vert\phi\left(v_1,v_2 \right)\right\rangle_{1,4}
\otimes 
\left\vert\phi\left( u_1^1 \oplus u_1^2 \ominus v_1,u_2^1 \oplus u_2^2 \ominus v_2 \right)\right\rangle_{3,2},
\notag \\
& \left\vert\phi\left( u_1^1,u_2^1 \right)\right\rangle_{1,2} \otimes \left\vert\phi\left( u_1^2,u_2^2 \right)\right\rangle_{3,4}
\rightarrow
\frac {1}{d} \sum_{v_1, v_2=0}^{d-1}
\zeta^{ \left( u_2^1 + v_2 \right) u_1^2 - v_1 u_2^1}
\left\vert\phi\left( u_1^1 \oplus u_1^2 \ominus v_1,u_2^1 \oplus u_2^2 \oplus v_2 \right)\right\rangle_{1,4}
\otimes \left\vert\phi\left(v_1,v_2 \right)\right\rangle_{2,3},
\notag \\
& \left\vert\phi\left( u_1^1,u_2^1 \right)\right\rangle_{1,2} \otimes \left\vert\phi\left( u_1^2,u_2^2 \right)\right\rangle_{3,4}
\rightarrow
\frac {1}{d} \sum_{v_1, v_2=0}^{d-1}
\zeta^{u_1^2 v_2 - u_2^1 \left( u_1^1 + u_1^2 - v_1 \right)}
\left\vert\phi\left(v_1,v_2 \right)\right\rangle_{1,3}
\otimes 
\left\vert\phi\left( u_1^1 \oplus u_1^2 \ominus v_1,v_2 \oplus u_2^2 \ominus u_2^1 \right)\right\rangle_{2,4},
\notag \\
& \left\vert\phi\left( u_1^1,u_2^1 \right)\right\rangle_{1,2} \otimes \left\vert\phi\left( u_1^2,u_2^2 \right)\right\rangle_{3,4}
\rightarrow
\frac {1}{d} \sum_{v_1, v_2=0}^{d-1}
\zeta^{ \left( u_2^1 - u_2^2 - v_2 \right) \left( u_1^2 - v_1 \right)}
\left\vert \phi \left( u_1^1 \oplus u_1^2 \ominus v_1,u_2^1 \ominus u_2^2 \ominus v_2 \right)\right\rangle_{1,3}
\otimes \left\vert\phi\left(v_1,v_2 \right)\right\rangle_{4,2},
\end{align}
which are realized by performing $\mathrm{\widehat{M}}$ on the particles (1,4), (2,3), (1,3) and (4,2), respectively.

The final special case we would like to characterize is the entanglement swapping of 
the generalized GHZ states which have the form
$\left\vert \psi \right\rangle = \frac{1}{\sqrt d} \sum_{l=0}^{d-1} \left\vert l, l,\dots,l \right\rangle$.
Let us assume that there are $n$ generalized GHZ states containing 
$m_1,m_2,\dots,m_n$ particles each, and mark them by
$\left\vert \psi_1 \right\rangle$, $\left\vert \psi_2 \right\rangle,\dots,\left\vert \psi_n \right\rangle$ respectively.
Without losing generality, let us assume that the measurement operator $\mathrm{\widetilde{M}}$
is performed on the first $k_r$ particles in $\left\vert \psi_r \right\rangle$,
then the entanglement swapping can be expressed as
\begin{align}
\bigotimes_{r=1}^{n} \left\vert \psi_r \right\rangle													 	
= & \frac {1}{d^{n/2}} \sum_{l_1,l_2,\dots,l_n=0}^{d-1}
\bigotimes_{r=1}^{n} \left\vert l_r,l_r,\dots,l_r \right\rangle_{1,2,\dots,m_r}										\notag \\
\Rightarrow 
& \frac {1}{d^{n/2}} \sum_{l_1,l_2,\dots,l_n=0}^{d-1}
\bigotimes_{r=1}^{n} 
\left\vert l_r,l_r,\dots,l_r \right\rangle_{1,2,\dots,k_r}
\bigotimes_{r=1}^{n}
\left\vert l_r,l_r,\dots,l_r \right\rangle_{k_r+1,k_r+2,\dots,m_r} 									\notag \\
= & \frac {1}{d^{n/2}} \sum_{v_1^1,v_1^2,\dots,v_1^n=0}^{d-1}
\left\vert \phi \left(v_1^1,v_2^1,v_3^1,\dots,v_{k_1}^1,v_1^2,v_2^2,\dots,v_{k_2}^2,\dots,v_1^n,v_2^n,\dots,v_{k_n}^n \right)
\right\rangle																			\notag \\
& \otimes
\left\vert \phi \left( d - v_1^1,v_2^1,v_3^1,\dots,v_{k_1}^1,v_1^2,v_2^2,\dots,v_{k_2}^2,\dots,v_1^n,v_2^n,\dots,v_{k_n}^n \right)
\right\rangle,
\end{align}
where
\begin{gather}
v_j^1 = 0, \phantom{i} j =2,3,\dots,k_1,			
\notag \\
v_1^i = v_2^i = \dotsm = v_{k_i}^i, \phantom{i} i =2,3,\dots,n.
\end{gather}
It is easy to derive the the entanglement swapping formula for two bipartite generalized GHZ states,
\begin{align}
\bigotimes_{r=1}^{2} \left\vert \psi_r \right\rangle_{2r-1,2r}
\rightarrow
\sum_{u,v=0}^{d-1}	\left\vert \phi \left( u,v \right) \right\rangle_{1,3} \left\vert \phi \left( d - u,v \right) \right\rangle_{2,4},
\end{align}
which is realized by performing the measurement operator $\mathrm{\widehat{M}}$ on the first particle in each state.

In what follows, we will consider the entanglement swapping case proposed in
Ref. \cite{KarimipourV652002}, which is realized by performing $\mathrm{\widehat{M}}$
on one particle in a d-level multi-particle maximally entangled state and one particle in a d-level Bell state.
We employ mathematical induction, which will be frequently adopted in what follows,
to provide a proof for the entanglement swapping results.

Without losing generality, let us assume that there is a d-level $(n+1)$-particle maximally entangled state
and mark it by $\left\vert \phi \left( u_0^0,u_1^0,\dots,u_n^0 \right) \right\rangle$, 
and that there are $n$ d-level Bell states, denoted as
$\left\vert \phi \left( u_1^1,u_2^1 \right) \right\rangle$, $\left\vert \phi \left( u_1^2,u_2^2 \right) \right\rangle,\dots$, 
$\left\vert \phi \left( u_1^n,u_2^n \right) \right\rangle$ respectively.
Let us assume that $\mathrm{\widehat{M}}$ is performed on 
the particle with the label $u_r^0$ in $\left\vert \phi \left( u_0^0,u_1^0,\dots,u_n^0 \right) \right\rangle$ 
and the one with the label $u_2^r$ in $\left\vert \phi \left( u_1^r,u_2^r \right) \right\rangle$
for $r=1,2,\dots,n$, and that the measurement results are
$\left\vert \phi \left( v_1^1,v_2^1 \right) \right\rangle$, $\left\vert \phi \left( v_1^2,v_2^2 \right) \right\rangle$,
$\dots,\left\vert \phi \left( v_1^n,v_2^n \right) \right\rangle$, respectively,
then the remaining particles are projected onto
\begin{align}
\label{ES-a-cat-n-Bell}
\left\vert\phi\left( \oplus_{r=0}^n u_1^r \ominus \oplus_{r=1}^n v_1^r,
u_1^0 \oplus u_2^1 \ominus v_2^1,u_2^0 \oplus u_2^2 \ominus v_2^2,\dots,u_n^0 \oplus u_2^n \ominus v_2^n \right)\right\rangle,
\end{align}
The conclusion can be demonstrated by mathematical induction as follows,
\begin{proof}

From Eq. \ref{ES-cat-Bell-without-first-particle}, the entanglement swapping results after the first measurement
are given by
\begin{align}
\frac{1}{d} \sum_{v_1^1, v_2^1=0}^{d-1} 
\zeta^{\left( u_1^0 - v_2^1 \right) \left( u_1^1 - v_1^1 \right)}
\left\vert\phi\left( u_1^0 \oplus u_1^1 \ominus v_1^1,u_1^0 \oplus u_2^1 \ominus v_2^1,u_2^0,u_3^0,\dots,u_n^0 \right)\right\rangle
\otimes \left\vert\phi\left(v_1^1,v_2^1 \right)\right\rangle.
\end{align}
The entanglement swapping results after the second measurement are given by
\begin{align}
& \left\vert\phi\left( u_1^0 \oplus u_1^1 \ominus v_1^1,u_1^0 \oplus u_2^1 \ominus v_2^1,u_2^0,u_3^0,\dots,u_n^0 \right)\right\rangle 
\otimes
\left\vert \phi \left( u_1^2,u_2^2 \right) \right\rangle 	\notag \\
= & \frac {1}{d} \sum_{l_1,l_2=0}^{d-1} 
\zeta^{l_1 \left( u_1^0 + u_1^1 - v_1^1 \right) + l_2 u_1^2} 
\left\vert l_1,l_1 \oplus u_1^0 \oplus u_2^1 \ominus v_2^1, l_1 \oplus u_2^0,l_1 \oplus u_3^0,
\dots,l_1 \oplus u_n^0 \right\rangle
\otimes
\left\vert l_2, l_2 \oplus u_2^2 \right\rangle													\notag \\
\Rightarrow 
& \frac {1}{d} \sum_{l_1,l_2=0}^{d-1} 
\zeta^{l_1 \left( u_1^0 + u_1^1 - v_1^1 \right) + l_2 u_1^2} 
\left\vert l_1,l_1 \oplus u_1^0 \oplus u_2^1 \ominus v_2^1,l_2 \oplus u_2^2,l_1 \oplus u_3^0,l_1 \oplus u_4^0,
\dots,l_1 \oplus u_n^0 \right\rangle
\otimes
\left\vert l_2,l_1 \oplus u_2^0 \right\rangle														\notag \\
= & \frac {1}{d} \sum_{v_1^2,v_2^2=0}^{d-1} 
\zeta^{\left( u_2^0 - v_2^2 \right) \left( u_1^2 - v_1^2 \right)}
\left\vert\phi\left( \oplus_{r=0}^2 u_1^r \ominus \oplus_{r=1}^2 v_1^r,
u_1^0 \oplus u_2^1 \ominus v_2^1,u_2^0 \oplus u_2^2 \ominus v_2^2,u_3^0,u_4^0,\dots,u_n^0 \right)\right\rangle
\otimes \left\vert\phi\left(v_1^2,v_2^2 \right)\right\rangle.
\end{align}
It can be seen that the results meet Eq. \ref{ES-a-cat-n-Bell}.
Let us now suppose that the entanglement swapping results through the $(n-1)$-th measurement meet 
Eq. \ref{ES-a-cat-n-Bell}, such that the results after the final measurement are given by
\begin{align}
& \left\vert\phi\left( \oplus_{r=0}^{n-1} u_1^r \ominus \oplus_{r=1}^{n-1} v_1^r,u_1^0 \oplus u_2^1 \ominus v_2^1,
u_2^0 \oplus u_2^2 \ominus v_2^2,\dots,u_{n-1}^0 \oplus u_2^{n-1} \ominus v_2^{n-1},u_n^0 \right)\right\rangle 
\otimes
\left\vert \phi \left( u_1^n,u_2^n \right) \right\rangle 	\notag \\
= & \frac {1}{d} \sum_{l_1,l_2=0}^{d-1}
\zeta^{l_1 \left( \sum_{r=0}^{n-1} u_1^r - \sum_{r=1}^{n-1} v_1^r \right) + l_2 u_1^n} 
\left\vert l_1,l_1 \oplus u_1^0 \oplus u_2^1 \ominus v_2^1, l_1 \oplus u_2^0 \oplus u_2^2 \ominus v_2^2,
\dots,l_1 \oplus u_2^0 \oplus u_2^2 \ominus v_2^2,l_1 \oplus u_n^0 \right\rangle
\otimes
\left\vert l_2, l_2 \oplus u_2^n \right\rangle													\notag \\
\Rightarrow 
& \frac {1}{d} \sum_{l_1,l_2=0}^{d-1} 
\zeta^{l_1 \left( \sum_{r=0}^{n-1} u_1^r - \sum_{r=1}^{n-1} v_1^r \right) + l_2 u_1^n} 
\left\vert l_1,l_1 \oplus u_1^0 \oplus u_2^1 \ominus v_2^1, l_1 \oplus u_2^0 \oplus u_2^2 \ominus v_2^2,
\dots,l_1 \oplus u_2^0 \oplus u_2^2 \ominus v_2^2,l_2 \oplus u_2^n \right\rangle
\otimes
\left\vert l_2,l_1 \oplus u_n^0 \right\rangle														\notag \\
= & \frac {1}{d} \sum_{v_1^n,v_2^n=0}^{d-1} 
\zeta^{\left( u_n^0 - v_2^n \right) \left( u_1^n - v_1^n \right)}
\left\vert\phi\left( \oplus_{r=0}^n u_1^r \ominus \oplus_{r=1}^n v_1^r,
u_1^0 \oplus u_2^1 \ominus v_2^1,u_2^0 \oplus u_2^2 \ominus v_2^2,\dots,u_n^0 \oplus u_2^n \ominus v_2^n \right)\right\rangle
\otimes 
\left\vert\phi\left(v_1^n,v_2^n \right)\right\rangle.
\end{align}

\end{proof}

We would like to give a formula, which can directly prove Eq. \ref{ES-a-cat-n-Bell}
or derive the results shown in Eq. \ref{ES-a-cat-n-Bell}, rather than multiple calculations like the above proof.
Such a formula is given by
\begin{align}
\label{ES-a-cat-n-Bell-one-formula}
& \left\vert \phi \left( u_0^0,u_1^0,\dots,u_n^0 \right) \right\rangle
\bigotimes_{r=1}^n
\left\vert \phi \left( u_1^r,u_2^r \right) \right\rangle 	\notag \\
= & \frac {1}{d^{(n+1)/2}} \sum_{l_0,l_1,\dots,l_n=0}^{d-1} 
\zeta^{\sum_{r=0}^n l_r u_1^r} 
\left\vert l_0,l_0 \oplus u_1^0,l_0 \oplus u_2^0,\dots,l_1 \oplus u_n^0 \right\rangle
\bigotimes_{r=1}^n
\left\vert l_r, l_r \oplus u_2^r \right\rangle													\notag \\
\Rightarrow
& \frac {1}{d^{(n+1)/2}} \sum_{l_0,l_1,\dots,l_n=0}^{d-1} 
\zeta^{\sum_{r=0}^n l_r u_1^r} 
\left\vert l_0,l_1 \oplus u_2^1,l_2 \oplus u_2^2,\dots,l_n \oplus u_2^n \right\rangle
\bigotimes_{r=1}^n
\left\vert l_r, l_0 \oplus u_r^0 \right\rangle															\notag \\
= & \frac {1}{d^n} \sum_{v_1^1,v_2^1,v_1^2,v_2^2,\dots,v_1^n,v_2^n=0}^{d-1} 
\zeta^{\sum_{r=1}^n \left( u_r^0 - v_2^r \right) \left( u_1^r - v_1^r \right)}
\left\vert\phi\left( \oplus_{r=0}^n u_1^r \ominus \oplus_{r=1}^n v_1^r,
u_1^0 \oplus u_2^1 \ominus v_2^1,u_2^0 \oplus u_2^2 \ominus v_2^2,\dots,u_n^0 \oplus u_2^n \ominus v_2^n \right)\right\rangle
\notag \\
& \bigotimes_{r=1}^n
\left\vert\phi\left(v_1^r,v_2^r \right)\right\rangle,
\end{align}
where the coefficient can be obtained by multiplying the coefficients in the calculation results in stages, that is,
\begin{align}
\frac {1}{d^n} \sum_{v_1^1,v_2^1,v_1^2,v_2^2,\dots,v_1^n,v_2^n=0}^{d-1} 
\zeta^{\sum_{r=1}^n \left( u_r^0 - v_2^r \right) \left( u_1^r - v_1^r \right)}
\equiv
\prod_{r=1}^n \frac {1}{d} \sum_{v_1^r,v_2^r=0}^{d-1}
\zeta^{\left( u_r^0 - v_2^r \right) \left( u_1^r - v_1^r \right)}.
\end{align}

\subsubsection{Entanglement swapping of 2-level maximally entangled states}

\noindent
Below we study entanglement swapping for 2-level maximally entangled states.
Let us start by introducing the entanglement swapping between two Bell states shown in Eq. \ref{2-level-Bell-states},
which was proposed in Ref. \cite{ZukowskiM711993}.
Assuming that the measurement operator $\mathcal{\widehat{M}}$ is performed on the first particle in each state,
we can express the entanglement swapping between two Bell states as
\begin{align}
& \bigotimes_{r=1}^{2} \left\vert \tilde{\phi}\left( u_1^r,u_2^r \right) \right\rangle_{2r-1,2r}
\notag \\
= \quad & \frac {1}{2}
\left[ \left\vert 0 u_2^1 0 u_2^2 \right\rangle + (-1)^{u_1^2} \left\vert 0 u_2^1 1 \bar{u}_2^2 \right\rangle 
+ (-1)^{u_1^1} \left\vert 1 \bar{u}_2^1 0 u_2^2 \right\rangle
+ (-1)^{u_1^1 + u_1^2} \left\vert 1 \bar{u}_2^1 1 \bar{u}_2^2 \right\rangle \right]_{1234} 							\notag \\
\Rightarrow
\quad & \frac {1}{2}
\left[ \left\vert 0 0 u_2^1 u_2^2 \right\rangle + (-1)^{u_1^2} \left\vert 0 1 u_2^1 \bar{u}_2^2 \right\rangle 
+ (-1)^{u_1^1} \left\vert 1 0 \bar{u}_2^1 u_2^2 \right\rangle 
+ (-1)^{u_1^1 + u_1^2} \left\vert 1 1 \bar{u}_2^1 \bar{u}_2^2 \right\rangle \right]_{1324} 							\notag \\
= \quad & \frac {1}{4} \left[ 1 + (-1)^{u_1^1 + u_1^2} \right] 
\left( \left\vert \tilde{\phi}\left( 0,0 \right) \right\rangle \left\vert \tilde{\phi}\left( 0,u_1^2 \right) \right\rangle 
\pm \left\vert \tilde{\phi}\left( 1,0 \right) \right\rangle \left\vert \tilde{\phi}\left( 1,\bar{u}_1^2 \right) \right\rangle  \right)		\notag \\
 + &
\frac {1}{4} \left[ 1 - (-1)^{u_1^1 + u_1^2} \right] 
\left( \left\vert \tilde{\phi}\left( 1,0 \right) \right\rangle \left\vert \tilde{\phi}\left( 0,u_1^2 \right) \right\rangle 
\pm \left\vert \tilde{\phi}\left( 0,0 \right) \right\rangle \left\vert \tilde{\phi}\left( 1,\bar{u}_1^2 \right) \right\rangle  \right)  		\notag \\
 + &
\frac {1}{4} \left[ (-1)^{u_1^2} + (-1)^{u_1^1} \right]
\left( \left\vert \tilde{\phi}\left( 0,1 \right) \right\rangle \left\vert \tilde{\phi}\left( 0,u_2^2 \right) \right\rangle 
\pm \left\vert \tilde{\phi}\left( 1,1 \right) \right\rangle \left\vert \tilde{\phi}\left( 1,\bar{u}_2^2 \right) \right\rangle  \right)  		\notag \\
 + &
\frac {1}{4} \left[ (-1)^{u_1^2} - (-1)^{u_1^1} \right]
\left( \left\vert \tilde{\phi}\left( 1,1 \right) \right\rangle \left\vert \tilde{\phi}\left( 0,u_2^2 \right) \right\rangle 
\pm \left\vert \tilde{\phi}\left( 0,1 \right) \right\rangle \left\vert \tilde{\phi}\left( 1,\bar{u}_2^2 \right) \right\rangle  \right),
\label{2-level-Bell-ES-all}
\end{align}
where the subscripts 1,2 and 3,4 are used to mark the particles in the two Bell states, respectively.
Similarly, one can derive the entanglement swapping between two GHZ states (see Eq. \ref{2-level-GHZ-states}),
\begin{align}
\bigotimes_{r=1}^{2} \left\vert \tilde{\phi}\left( u_1^r,u_2^r,u_3^r \right) \right\rangle
\rightarrow 
& \frac {1}{4} \left[ 1 + (-1)^{u_1^1 + u_1^2} \right] 
\left( \left\vert \tilde{\phi}\left( 0,0 \right) \right\rangle \left\vert \tilde{\phi}\left( 0,u_3^1,u_2^2,u_3^2 \right) \right\rangle 
\pm \left\vert \tilde{\phi}\left( 1,0 \right) \right\rangle \left\vert \tilde{\phi}\left( 1,\bar{u}_3^1,\bar{u}_2^2,\bar{u}_3^2 \right) \right\rangle  
\right) \notag \\
 + &
\frac {1}{4} \left[ 1 - (-1)^{u_1^1 + u_1^2} \right] 
\left( \left\vert \tilde{\phi}\left( 1,0 \right) \right\rangle \left\vert \tilde{\phi}\left( 0,u_3^1,u_2^2,u_3^2 \right) \right\rangle 
\pm \left\vert \tilde{\phi}\left( 0,0 \right) \right\rangle \left\vert \tilde{\phi}\left( 1,\bar{u}_3^1,\bar{u}_2^2,\bar{u}_3^2 \right) \right\rangle  
\right) \notag \\
 + &
\frac {1}{4} \left[ (-1)^{u_1^2} + (-1)^{u_1^1} \right]
\left( \left\vert \tilde{\phi}\left( 0,1 \right) \right\rangle \left\vert \tilde{\phi}\left( 0,u_3^1,\bar{u}_2^2,\bar{u}_3^2 \right) \right\rangle 
\pm \left\vert \tilde{\phi}\left( 1,1 \right) \right\rangle \left\vert \tilde{\phi}\left( 1,\bar{u}_3^1,u_2^2,u_3^2 \right) \right\rangle  \right) \notag \\
 + &
\frac {1}{4} \left[ (-1)^{u_1^2} - (-1)^{u_1^1} \right]
\left( \left\vert \tilde{\phi}\left( 1,1 \right) \right\rangle \left\vert \tilde{\phi}\left( 0,\bar{u}_3^1,\bar{u}_2^2,\bar{u}_3^2 \right) \right\rangle 
\pm \left\vert \tilde{\phi}\left( 0,1 \right) \right\rangle \left\vert \tilde{\phi}\left( 1,\bar{u}_3^1,u_2^2,u_3^2 \right) \right\rangle  \right),
\end{align}
which is realized by performing $\mathcal{\widehat{M}}$ on the first particle in each GHZ state.

Let us then consider the entanglement swapping between CAT states (see Eq. \ref{CAT-state}).
Let us suppose that there are $n$ CAT states and mark them by
$\left\{ \left\vert \varphi \left( w_1^r,w_2^r,\dots,w_{m_r}^r \right) \right\rangle \right\}_{r=1}^{n}$, where $w_1^r \in \{0,1\}$.
Without losing generality, let us suppose that the first $k_r$ particles in 
$\left\vert \varphi \left( w_1^r,w_2^r,\dots,w_{m_r}^r \right) \right\rangle $ are selected,
and that the measurement operator $\mathbb{\widehat{M}}$ is performed on them.

Let us number these particles as $1,2,\dots,K$ and the remaining particles $1,2,\dots,R$,
where $K=\sum_{r=1}^n k_r$ and $R = \sum_{r=1}^n m_r - K$ in this paper,
such that we can arrive at
\begin{align}
\label{revised-cat-state-ES}
& 	\bigotimes_{r=1}^{n} \left\vert \varphi \left( w_1^r,w_2^r,\dots,w_{m_r}^r \right) \right\rangle	\notag\\
= &
	\bigotimes_{r=1}^{n} 
	\left\vert e_1^1,w_2^r,w_3^r,\dots,w_{m_r}^r \right\rangle
	+(-1)^{w_1^n}
	\bigotimes_{r=1}^{n-1} \left\vert e_1^1,w_2^r,w_3^r,\dots,w_{m_r}^r \right\rangle
	\left\vert e_2^1,\widehat{w}_2^n,\widehat{w}_3^n,\dots,\widehat{w}_{m_n}^n \right\rangle
	\notag	\\
&	+(-1)^{w_1^{n-1}}
	\bigotimes_{r=1}^{n-2} \left\vert e_1^1,w_2^r,w_3^r,\dots,w_{m_r}^r \right\rangle
	\left\vert e_2^1,\widehat{w}_2^{n-1},\widehat{w}_3^{n-1},\dots,\widehat{w}_{m_{n-1}}^{n-1} \right\rangle
	\left\vert e_1^1,w_2^n,w_3^n,\dots,w_{m_n}^n \right\rangle
	+ \dotsm
	\notag	\\
&	\dotsm
	+(-1)^{\sum_{r=1}^{n} w_1^r}
	\bigotimes_{r=1}^{n} \left\vert e_2^1,\widehat{w}_2^r,\widehat{w}_3^r,\dots,\widehat{w}_{m_r}^r \right\rangle
	\notag	\\
\Rightarrow &
	\bigotimes_{r=1}^{n} \left\vert e_1^1,w_2^r,w_3^r,\dots,w_{k_r}^r \right\rangle
	\bigotimes_{r=1}^{n} \left\vert w_{k_r+1}^r,w_{k_r+2}^r,\dots,w_{m_r}^r \right\rangle
	\notag	\\
&	+(-1)^{w_1^n}
	\bigotimes_{r=1}^{n-1} \left\vert e_1^1,w_2^r,w_3^r,\dots,w_{k_r}^r \right\rangle
	\left\vert e_2^1,\widehat{w}_2^n,\widehat{w}_3^n,\dots,\widehat{w}_{k_n}^n \right\rangle
	\bigotimes_{r=1}^{n-1} \left\vert w_{k_r+1}^r,w_{k_r+2}^r,\dots,w_{m_r}^r \right\rangle
	\left\vert \widehat{w}_{k_n+1}^n,\widehat{w}_{k_n+2}^n,\dots,\widehat{w}_{m_n}^n \right\rangle
	\notag	\\
&	+(-1)^{w_1^{n-1}}
	\bigotimes_{r=1}^{n-2} \left\vert e_1^1,w_2^r,w_3^r,\dots,w_{k_r}^r \right\rangle
	\left\vert e_2^1,\widehat{w}_2^{n-1},\widehat{w}_3^{n-1},\dots,\widehat{w}_{k_{n-1}}^{n-1} \right\rangle
	\left\vert e_1^1,w_2^n,w_3^n,\dots,w_{k_n}^n \right\rangle
	\notag	\\
&	\qquad \qquad \phantom{i}
	\bigotimes_{r=1}^{n-2} \left\vert w_{k_r+1}^r,w_{k_r+2}^r,\dots,w_{m_r}^r \right\rangle
	\left\vert \widehat{w}_{k_{n-1}+1}^{n-1},\widehat{w}_{k_{n-1}+2}^{n-1},\dots,\widehat{w}_{m_{n-1}}^{n-1} \right\rangle
	\left\vert w_{k_n+1}^n,w_{k_n+2}^n,\dots,w_{m_n}^n \right\rangle
	\notag	\\
&	+ \dots 
	+(-1)^{\sum_{r=1}^{n} w_1^r}
	\bigotimes_{r=1}^{n} \left\vert e_2^1,\widehat{w}_2^r,\widehat{w}_3^r,\dots,\widehat{w}_{k_r}^r \right\rangle
	\bigotimes_{r=1}^{n} \left\vert \widehat{w}_{k_r+1}^r,\widehat{w}_{k_r+2}^r,\dots,\widehat{w}_{m_r}^r \right\rangle
	\notag	\\
= &
	\sum_{j=1}^{2^{n-1}}
	\left[
	(-1)^{\omega_j}					
	\left\vert e_1^1 \right\rangle
	\bigotimes_{i=2}^K \left\vert \upomega_{i}^j \right\rangle
	\bigotimes_{i=1}^R \left\vert \upupsilon_{i}^j \right\rangle
	+
	(-1)^{ \omega_{2^n-j+1}}
	\left\vert e_2^1 \right\rangle
	\bigotimes_{i=2}^K \left\vert \widehat{\upomega}_{i}^{2^n-j+1} \right\rangle
	\bigotimes_{i=1}^R \left\vert \widehat{\upupsilon}_{i}^{2^n-j+1} \right\rangle
	\right]_{1,2,\dots,K,1,2,\dots,R}
	\notag\\
= &
\begin{dcases}
\quad
	\sum_{j}
	(-1)^{\omega_j}
	\left[ 
	\left\vert \varphi_j^+ \left(K \right) \right\rangle + \left\vert \varphi_j^- \left(K \right) \right\rangle
	\right]
	\left[
	\left\vert \varphi_j^+ \left(R \right) \right\rangle + \left\vert \varphi_j^- \left(R \right) \right\rangle
	\right]
\\
	+
	\sum_{j}
	(-1)^{\omega_{2^n-j+1}}
	\left[
	\left\vert \varphi_j^+ \left(K \right) \right\rangle - \left| \varphi_j^- \left(K \right) \right\rangle
	\right]
	\left[
	\left\vert \varphi_j^+ \left(R \right) \right\rangle - \left\vert \varphi_j^- \left(R \right) \right\rangle
	\right], 									 
	\phantom{i} \upupsilon_1^j= 0,
	\\
	\\
\quad	
	\sum_{j}
	(-1)^{\omega_j}	
	\left[
	\left\vert \varphi_j^+ \left(K \right) \right\rangle + \left\vert \varphi_j^- \left(K \right) \right\rangle
	\right]
	\left[
	\left\vert \varphi_j^+ \left(R \right) \right\rangle - \left\vert \varphi_j^- \left(R \right) \right\rangle
	\right]
\\
	+
	\sum_{j}
	(-1)^{\omega_{2^n-j+1}}
	\left[
	\left\vert \varphi_j^+ \left(K \right) \right\rangle - \left\vert \varphi_j^- \left(K \right) \right\rangle
	\right]
	\left[
	\left\vert \varphi_j^+ \left(R \right) \right\rangle + \left\vert \varphi_j^- \left(R \right) \right\rangle 
	\right],				
	\phantom{i} \upupsilon_1^j= 1,
\end{dcases}
	\notag\\
= & 
\begin{dcases}
	\quad
	\sum_{j}
\mu
	\left[
	\left\vert \varphi_j^+ \left( K \right) \right\rangle \left\vert \varphi_j^+ \left( R \right) \right\rangle
	+
	\left\vert \varphi_j^- \left( K \right) \right\rangle \left\vert \varphi_j^- \left( R \right) \right\rangle
	\right]
	+
	\sum_{j}
\nu
	\left[
	\left\vert \varphi_j^+ \left( K \right) \right\rangle \left\vert \varphi_j^- \left( R \right) \right\rangle 
	+
	\left\vert \varphi_j^- \left( K \right) \right\rangle \left\vert \varphi_j^+ \left( R \right) \right\rangle 
	\right],
	\\
	\quad
	\sum_{j}
\mu
	\left[
	\left\vert \varphi_j^+ \left( K \right) \right\rangle \left\vert \varphi_j^+ \left( R \right) \right\rangle
	-
	\left\vert \varphi_j^- \left( K \right) \right\rangle \left\vert \varphi_j^- \left( R \right) \right\rangle
	\right]
	-
	\sum_{j}
\nu
	\left[
	\left\vert \varphi_j^+ \left( K \right) \right\rangle \left\vert \varphi_j^- \left( R \right) \right\rangle 
	-
	\left\vert \varphi_j^- \left( K \right) \right\rangle \left\vert \varphi_j^+ \left( R \right) \right\rangle 
	\right],
\end{dcases}
\end{align}
where
\begin{gather}
\mu = (-1)^{\omega_j} + (-1)^{\omega_{2^n-j+1}},
\phantom{i}
\nu = (-1)^{\omega_j} - (-1)^{\omega_{2^n-j+1}},
\phantom{i}
\omega_j = \sum_{r=1}^n \delta \left({\upomega_{1+\sum_{i=1}^r k_i}^j,e_2^1}\right),
\phantom{i}
\omega_{2^n-j+1} = \sum_{r=1}^{n} w_1^r - \omega_j,	
\notag \\
\left\vert \varphi_j^{+} \left( K \right) \right\rangle
= \left\vert \varphi \left( 0,\upomega_2^j,\upomega_3^j,\dots,\upomega_{K}^j \right) \right\rangle
= \frac {1}{\sqrt{2}}
\left( \left\vert e_1^1,\upomega_2^j,\upomega_3^j,\dots,\upomega_{K}^j \right\rangle
+
\left\vert e_2^1,\widehat{\upomega}_2^j,\widehat{\upomega}_3^j,\dots,\widehat{\upomega}_{K}^j \right\rangle \right),
\notag \\
\left\vert \varphi_j^{-} \left( K \right) \right\rangle
= \left\vert \varphi \left( 1,\upomega_2^j,\upomega_3^j,\dots,\upomega_{K}^j \right) \right\rangle
= \frac {1}{\sqrt{2}}
\left( \left\vert e_1^1,\upomega_2^j,\upomega_3^j,\dots,\upomega_{K}^j \right\rangle
-
\left\vert e_2^1,\widehat{\upomega}_2^j,\widehat{\upomega}_3^j,\dots,\widehat{\upomega}_{K}^j \right\rangle \right),
\notag \\
\left\vert \varphi_j^{+} \left( R \right) \right\rangle
= \left\vert \varphi \left( 0,\upomega_2^j,\upomega_3^j,\dots,\upomega_{R}^j \right) \right\rangle
= \frac {1}{\sqrt{2}}
\left( \left\vert e_1^1,\upomega_2^j,\upomega_3^j,\dots,\upomega_{R}^j \right\rangle
+
\left\vert e_2^1,\widehat{\upomega}_2^j,\widehat{\upomega}_3^j,\dots,\widehat{\upomega}_{R}^j \right\rangle \right),
\notag \\
\left\vert \varphi_j^{-} \left( R \right) \right\rangle
= \left\vert \varphi \left( 1,\upomega_2^j,\upomega_3^j,\dots,\upomega_{R}^j \right) \right\rangle
= \frac {1}{\sqrt{2}}
\left( \left\vert e_1^1,\upomega_2^j,\upomega_3^j,\dots,\upomega_{R}^j \right\rangle
-
\left\vert e_2^1,\widehat{\upomega}_2^j,\widehat{\upomega}_3^j,\dots,\widehat{\upomega}_{R}^j \right\rangle \right).
\notag
\end{gather}
Note that some inessential coefficients are ignored in Eq. \ref{revised-cat-state-ES}.
For CAT states, if one sets $\left\{ \left\vert e_1^j \right\rangle,\left\vert e_2^j \right\rangle \right\} = 
\left\{ \left\vert 0 \right\rangle,\left\vert 1 \right\rangle \right\}$ 
or $\left\{ \left\vert + \right\rangle,\left\vert - \right\rangle \right\}$,
where $\left\vert \pm \right\rangle = 1/\sqrt{2} \left( \left\vert 0 \right\rangle \pm \left\vert 1 \right\rangle \right)$,
then the entanglement swapping considered above is enlightening for the entanglement swapping 
of graph states \cite{RaussendorfR6822003,HeinM06020962006}. 
Let us provide a simple case for the entanglement swapping of graph states:
the entanglement swapping between two four-particle cluster states, which has the form \cite{RaussendorfR6822003}
\begin{gather}
\left\vert \mathcal{C} \right\rangle = \frac{1}{2}
\left(
\left\vert +,0,+,0 \right\rangle + \left\vert +,0,-,1 \right\rangle + \left\vert -,1,-,0 \right\rangle + \left\vert -,1,+,1 \right\rangle
\right).
\label{cluster-states}
\end{gather}
Let us assume that the measurement operator $\mathbb{\widetilde{M}}$
is performed on the first two particles in each state, then the entanglement swapping can be expressed as
\begin{align}
\label{cluster-states-ES}
& \left\vert \mathcal{C} \right\rangle_{\tilde{1},\tilde{2},\tilde{3},\tilde{4}} 
\otimes
\left\vert \mathcal{C} \right\rangle_{\hat{1},\hat{2},\hat{3},\hat{4}}						\notag \\
\rightarrow
& \frac{1}{4} 
\left( 
   \left\vert \varsigma_1^+ \right\rangle \left\vert \varsigma_1^+ \right\rangle
+ \left\vert \varsigma_1^- \right\rangle \left\vert \varsigma_1^+ \right\rangle
+ \left\vert \varsigma_1^+ \right\rangle \left\vert \varsigma_2^+ \right\rangle
+ \left\vert \varsigma_1^- \right\rangle \left\vert \varsigma_2^+ \right\rangle
+ \left\vert \varsigma_2^+ \right\rangle \left\vert \varsigma_3^+ \right\rangle
+ \left\vert \varsigma_2^- \right\rangle \left\vert \varsigma_3^+ \right\rangle
+ \left\vert \varsigma_2^+ \right\rangle \left\vert \varsigma_4^+ \right\rangle
+ \left\vert \varsigma_2^- \right\rangle \left\vert \varsigma_4^+ \right\rangle
\right.
\notag \\
& \left. 
+ \left\vert \varsigma_2^+ \right\rangle \left\vert \varsigma_5^+ \right\rangle
- \left\vert \varsigma_2^- \right\rangle \left\vert \varsigma_5^+ \right\rangle
+ \left\vert \varsigma_2^+ \right\rangle \left\vert \varsigma_6^+ \right\rangle
- \left\vert \varsigma_2^- \right\rangle \left\vert \varsigma_6^+ \right\rangle
+ \left\vert \varsigma_1^+ \right\rangle \left\vert \varsigma_7^+ \right\rangle
- \left\vert \varsigma_1^- \right\rangle \left\vert \varsigma_7^+ \right\rangle
+ \left\vert \varsigma_1^+ \right\rangle \left\vert \varsigma_8^+ \right\rangle
- \left\vert \varsigma_1^- \right\rangle \left\vert \varsigma_8^+ \right\rangle
\right)_{\tilde{1},\tilde{2},\hat{1},\hat{2},\tilde{3},\tilde{4},\hat{3},\hat{4}},
\end{align}
where the subscripts $\tilde{1},\tilde{2},\tilde{3},\tilde{4}$ and $\hat{1},\hat{2},\hat{3},\hat{4}$ are used to
indicate the particles in the two states, respectively, and
\begin{align}
& \left\vert \varsigma_1^{+} \right\rangle
= \left\vert \varphi \left( 0,0,+,0 \right) \right\rangle 
= \frac {1}{\sqrt{2}} \left( \left\vert + 0 + 0 \right\rangle + \left\vert - 1 - 1 \right\rangle \right),
\phantom{i}
\left\vert \varsigma_1^{-} \right\rangle
= \left\vert \varphi \left( 1,0,+,0 \right) \right\rangle 
= \frac {1}{\sqrt{2}} \left( \left\vert +0+0 \right\rangle - \left\vert - 1 - 1 \right\rangle \right),
\notag \\
& \left\vert \varsigma_2^{+} \right\rangle
= \left\vert \varphi \left( 0,0,-,1 \right) \right\rangle 
= \frac {1}{\sqrt{2}} \left( \left\vert +0-1 \right\rangle + \left\vert -1+0 \right\rangle \right),
\phantom{i}
\left\vert \varsigma_2^{-} \right\rangle
= \left\vert \varphi \left( 1,0,+,0 \right) \right\rangle 
= \frac {1}{\sqrt{2}} \left( \left\vert + 0 + 0 \right\rangle - \left\vert - 1 - 1 \right\rangle \right),
\notag \\
& \left\vert \varsigma_3^{+} \right\rangle
= \left\vert \varphi \left( 1,0,-,0 \right) \right\rangle 
= \frac {1}{\sqrt{2}} \left( \left\vert + 0 - 0 \right\rangle + \left\vert - 1 + 1 \right\rangle \right),
\phantom{i}
\left\vert \varsigma_3^{-} \right\rangle
= \left\vert \varphi \left( 1,0,+,0 \right) \right\rangle 
= \frac {1}{\sqrt{2}} \left( \left\vert + 0 + 0 \right\rangle - \left\vert - 1 - 1 \right\rangle \right),
\notag \\
& \left\vert \varsigma_4^{+} \right\rangle
= \left\vert \varphi \left( 1,0,+,1 \right) \right\rangle
= \frac {1}{\sqrt{2}} \left( \left\vert + 0 + 1 \right\rangle + \left\vert - 1 - 0 \right\rangle \right),
\phantom{i}
\left\vert \varsigma_4^{-} \right\rangle
= \left\vert \varphi \left( 1,0,+,0 \right) \right\rangle
= \frac {1}{\sqrt{2}} \left( \left\vert + 0 + 1 \right\rangle - \left\vert - 1 - 0 \right\rangle \right),
\notag \\
& \left\vert \varsigma_5^{+} \right\rangle
= \left\vert \varphi \left( 1,0,+,1 \right) \right\rangle 
= \frac {1}{\sqrt{2}} \left( \left\vert + 0 + 1 \right\rangle + \left\vert - 1 - 0 \right\rangle \right),
\phantom{i}
\left\vert \varsigma_5^{-} \right\rangle
= \left\vert \varphi \left( 1,1,-,1 \right) \right\rangle 
= \frac {1}{\sqrt{2}} \left( \left\vert + 1 - 1 \right\rangle - \left\vert - 0 + 0 \right\rangle \right),
\notag \\
& \left\vert \varsigma_6^{+} \right\rangle
= \left\vert \varphi \left( 1,1,+,0 \right) \right\rangle 
= \frac {1}{\sqrt{2}} \left( \left\vert + 1 + 0 \right\rangle + \left\vert - 0 - 1 \right\rangle \right),
\phantom{i}
\left\vert \varsigma_6^{-} \right\rangle
= \left\vert \varphi \left( 1,1,-,1 \right) \right\rangle 
= \frac {1}{\sqrt{2}} \left( \left\vert + 1 - 1 \right\rangle - \left\vert - 0 + 0 \right\rangle \right),
\notag \\
& \left\vert \varsigma_7^{+} \right\rangle
= \left\vert \varphi \left( 1,1,+,0 \right) \right\rangle 
= \frac {1}{\sqrt{2}} \left( \left\vert + 1 + 0 \right\rangle + \left\vert - 0 - 1 \right\rangle \right),
\phantom{i}
\left\vert \varsigma_7^{-} \right\rangle
= \left\vert \varphi \left( 1,1,+,1 \right) \right\rangle 
= \frac {1}{\sqrt{2}} \left( \left\vert + 1 + 1 \right\rangle - \left\vert - 0 - 0 \right\rangle \right),
\notag \\
& \left\vert \varsigma_8^{+} \right\rangle
= \left\vert \varphi \left( 1,1,+,0 \right) \right\rangle
= \frac {1}{\sqrt{2}} \left( \left\vert + 1 + 0 \right\rangle + \left\vert - 0 - 1 \right\rangle \right),
\phantom{i}
\left\vert \varsigma_8^{-} \right\rangle
= \left\vert \varphi \left( 1,1,-,0 \right) \right\rangle
= \frac {1}{\sqrt{2}} \left( \left\vert + 1 - 0 \right\rangle - \left\vert - 0 + 1 \right\rangle \right).
\end{align}
If each subsystem is limited on the basis $\left\{ \left\vert 0 \right\rangle,\left\vert 1 \right\rangle \right\}$,
one can get the result of the entanglement swapping between cat states, which was proposed in Ref. \cite{BoseS5721998}.
Furthermore, for the entanglement swapping of any number of the Bell states and GHZ states
shown in Eqs. \ref{2-level-maximally-entangled-states} and \ref{2-level-Bell-states},
one can derive the formulas by setting the first particle in the state
$1/\sqrt{2} \left( \left\vert 0 \right\rangle \pm \left\vert 1 \right\rangle \right)$,
which was discussed in Ref. \cite{JiZX5852022}.

\subsection{Entanglement swapping between generalized pure states and maximally entangled states}

\noindent
Below we will characterize entanglement swapping between generalized pure states and maximally entangled states.
For simplicity and not losing generality, we consider entanglement swapping for any number of bipartite systems.
Let us suppose that there are $n_1$ generalized pure states (see Eq. \ref{generalized-pure-states})
and $n_2$ bipartite maximally entangled states, denoted as
$\left\{ \left\vert \mathscr{P}_{pure}^r \right\rangle \right\}_{r=1}^{n_1}$ and 
$\left\{ \left\vert \phi \left( u_1^r,u_2^r \right) \right\rangle \right\}_{r=1}^{n_2}$ respectively,
and that the first particle in each of $\left\{ \left\vert \mathscr{P}_{pure}^r \right\rangle \right\}_{r=1}^{n_1}$ and
$\left\{ \left\vert \phi \left( u_1^r,u_2^r \right) \right\rangle \right\}_{r=1}^{t}$,
and the second particle in each of $\left\{ \left\vert \phi \left( u_1^r,u_2^r \right) \right\rangle \right\}_{r=t+1}^{n_2}$, are selected,
then the entanglement swapping realized by performing $\mathrm{\widetilde{M}}$ on the selected particles can be 
expressed as
\begin{align}
\label{generalized-pure-and-maximally-entangled-states-ES}
&	\bigotimes_{r=1}^{n_1} \left\vert \mathscr{P}_{pure}^r \right\rangle
	\bigotimes_{r=1}^{n_2} \left\vert \phi \left( u_1^r,u_2^r \right) \right\rangle							\notag \\		
= & 
\frac{1}{d^{n_2/2}}
\sum_{l_1^1,l_2^1,l_1^2,l_2^2,\dots,l_1^{n_1},l_2^{n_1}=0}^{d-1}
\sum_{l_1,l_2,\dots,l_n = 0}^{d-1}
\prod_{r=1}^{n} \lambda_{l_1^r,l_2^r} \zeta^{\sum_{r=1}^{n_2} l_r^{\prime} u_1^r}
\bigotimes_{r=1}^{n} \left\vert l_1^r,l_2^r \right\rangle 
\bigotimes_{r=1}^{n} \left\vert l_r,l_r \oplus u_2^r \right\rangle							\notag \\
\Rightarrow & 
\frac{1}{d^{n_2/2}}
\sum_{l_1^1,l_2^1,l_1^2,l_2^2,\dots,l_1^{n_1},l_2^{n_1}=0}^{d-1}
\sum_{l_1,l_2,\dots,l_n = 0}^{d-1}
\prod_{r=1}^{n} \lambda_{l_1^r,l_2^r} \zeta^{\sum_{r=1}^{n_2} l_r^{\prime} u_1^r}
	\bigotimes_{r=1}^{n_1} \left\vert l_1^r \right\rangle
	\bigotimes_{r=1}^{t} \left\vert l_r \right\rangle
	\bigotimes_{r=t+1}^{n_2} \left\vert l_r \oplus u_2^r \right\rangle
	\bigotimes_{r=1}^{n_1} \left\vert l_2^r \right\rangle
	\bigotimes_{r=1}^{t} \left\vert l_r \oplus u_2^r \right\rangle
	\bigotimes_{r=t+1}^{n_2} \left\vert l_r \right\rangle									\notag \\
= &
	\frac{\mathcal{P}}{\sqrt{d^{n_2+1}}}
	\sum_{v_1^1,v_2^1,\dotsm,v_{n_1}^1,v_1^2,v_2^2,\dotsm,v_{n_2}^2=0}^{d-1}
	\sum_{l_1^1,l_2^1,l_2^2,\dots,l_2^{n_1}=0}^{d-1}
	\lambda_{l_1^1,l_2^1} \prod_{r=2}^{n_1} \lambda_{l_1^1 \oplus v_r^1,l_2^r}
	\zeta^{\sum_{r=1}^{n_2} \left( l_1^1 + v_r^2 \right) u_1^r - \sum_{r=t+1}^{n_2} u_2^r u_1^r - l_1^1 v_1^1} 		\notag \\
& 	\times
\left\vert \phi \left( v_1^1,v_2^1,\dotsm,v_{n_1}^1,v_1^2,v_2^2,\dotsm,v_{n_2}^2 \right) \right\rangle
\otimes 
\left\vert l_2^1,l_2^2,\dots,l_2^{n_1},
l_1^1 \oplus v_1^2 \oplus u_2^1,l_1^1 \oplus v_2^2 \oplus u_2^2,\dots,l_1^1 \oplus v_t^2 \oplus u_2^t,				
\right.	
\notag \\
& \left.
l_1^1 \oplus v_{t+1}^2 \ominus u_2^{t+1}, l_1^1 \oplus v_{t+2}^2 \ominus u_2^{t+2},\dots,l_1^1 \oplus v_{n_2}^2 \ominus u_2^{n_2}
\right\rangle,
\end{align}
where
\begin{gather}
\mathcal{P} = \frac{1}{\sqrt{
	\sum_{v_2^1,\dotsm,v_{n_1}^1=0}^{d-1}
	\sum_{l_1^1,l_2^1,l_2^2,\dots,l_2^{n_1}=0}^{d-1}
\left\vert \lambda_{l_1^1,l_2^1} \prod_{r=2}^{n_1} \lambda_{l_1^1 \oplus v_r^1,l_2^r} \right\vert^2}}.		\notag
\end{gather}

In particular, if $\mathrm{\widetilde{M}}$ acts on the first particle in all states,
the entanglement swapping result is given by
\begin{align}
&
	\frac{\mathcal{P}}{\sqrt{d^{n_2+1}}}
	\sum_{v_1^1,v_2^1,\dotsm,v_{n_1}^1,v_1^2,v_2^2,\dotsm,v_{n_2}^2=0}^{d-1}
	\sum_{l_1^1,l_2^1,l_2^2,\dots,l_2^{n_1}=0}^{d-1}
	\lambda_{l_1^1,l_2^1} \prod_{r=2}^{n_1} \lambda_{l_1^1 \oplus v_r^1,l_2^r}
	\zeta^{\sum_{r=1}^{n_2} \left( l_1^1 + v_r^2 \right) u_1^r - l_1^1 v_1^1} 								\notag \\
& 	\times
\left\vert \phi \left( v_1^1,v_2^1,\dotsm,v_{n_1}^1,v_1^2,v_2^2,\dotsm,v_{n_2}^2 \right) \right\rangle
\otimes
\left\vert
l_2^1,l_2^2,\dots,l_2^{n_1},
l_1^1 \oplus v_1^2 \oplus u_2^1,l_1^1 \oplus v_2^2 \oplus u_2^2,\dots,l_1^1 \oplus v_{n_2}^2 \oplus u_2^{n_2}
\right\rangle,
\end{align}
where $\mathcal{P}$ is the same as that in Eq. \ref{generalized-pure-and-maximally-entangled-states-ES}.
If $\mathrm{\widetilde{M}}$ acts on the second particle in all states, the entanglement swapping result is given by
\begin{align}
&
	\frac{\mathcal{P}}{\sqrt{d^{n_2+1}}}
	\sum_{v_1^1,v_2^1,\dotsm,v_{n_1}^1,v_1^2,v_2^2,\dotsm,v_{n_2}^2=0}^{d-1}
	\sum_{l_2^1,l_1^1,l_1^2,\dots,l_1^{n_1}=0}^{d-1}
	\lambda_{l_1^1,l_2^1} \prod_{r=2}^{n_1} \lambda_{l_1^r,l_2^1 \oplus v_r^1}
	\zeta^{\sum_{r=1}^{n_2} \left( l_2^1 + v_r^2 - u_2^r \right) u_1^r - l_2^1 v_1^1} 								\notag \\
& 	\times
\left\vert
l_1^1,l_1^2,\dots,l_1^{n_1},
l_2^1 \oplus v_1^1 \ominus u_2^1,l_2^1 \oplus v_2^2 \ominus u_2^2,\dots,l_2^1 \oplus v_{n_2}^2 \ominus u_2^{n_2}
\right\rangle
\otimes
\left\vert \phi \left( v_1^1,v_2^1,\dotsm,v_{n_1}^1,v_1^2,v_2^2,\dotsm,v_{n_2}^2 \right) \right\rangle,
\end{align}
where $\mathcal{P}$ is also the same as that in Eq. \ref{generalized-pure-and-maximally-entangled-states-ES}.
By the way, as mentioned earlier, selecting either the first particle or the second particle in each generalized pure state is equivalent.

As a special case of the entanglement swapping shown in Eq. \ref{generalized-pure-and-maximally-entangled-states-ES},
the entanglement swapping between general pure states and bipartite maximally entangled states can be characterized as
\begin{align}
\label{general-pure-and-maximally-entangled-states-ES}
\bigotimes_{r=1}^{n_1} \left\vert \mathscr{P}_{\uppercase\expandafter{\romannumeral1}}^r \right\rangle
	\bigotimes_{r=1}^{n_2} \left\vert \phi \left( u_1^r,u_2^r \right) \right\rangle
\rightarrow &
	\frac{\mathcal{P}}{\sqrt{d^{n_2+1}}}
	\sum_{v_1^1,v_2^1,\dots,v_{n_1}^1,v_1^2,v_2^2,\dots,v_{n_2}^2,l_1=0}^{d-1}
	\lambda_{l_1} \prod_{r=2}^{n_1} \lambda_{l_1 \oplus v_r^1}
	\zeta^{\sum_{r=1}^{n_2} \left( l_1 + v_r^2 \right) u_1^r - \sum_{r=t+1}^{n_2} u_2^r u_1^r -l_1 v_1^1} 				
\notag \\
&
\times
\left\vert \phi \left( v_1^1,v_2^1,\dotsm,v_{n_1}^1,v_1^2,v_2^2,\dotsm,v_{n_2}^2 \right) \right\rangle
\otimes
\left\vert
l_1,l_1 \oplus v_2^1,l_1 \oplus v_3^1,\dots,l_1 \oplus v_{n_1}^1,
\right.
\notag \\
&
l_1 \oplus v_1^2 \oplus u_2^1,l_1 \oplus v_2^2 \oplus u_2^2,\dots,l_1 \oplus v_t^2 \oplus u_2^t,
\notag \\
&
\left.
l_1 \oplus v_{t+1}^2 \ominus u_2^{t+1},l_1 \oplus v_{t+2}^2 \ominus u_2^{t+2},\dots,l_1 \oplus v_{n_2}^2 \ominus u_2^{n_2}
\right\rangle,
\end{align}
where
\begin{gather}
\mathcal{P} = \frac{1}{\sqrt{
	\sum_{v_2^1,\dotsm,v_{n_1}^1,l_1=0}^{d-1}
\left\vert \lambda_{l_1} \prod_{r=2}^{n_1} \lambda_{l_1 \oplus v_r^1} \right\vert^2}}.		\notag
\end{gather}
Further, as a simple special case of Eq. \ref{general-pure-and-maximally-entangled-states-ES},
the entanglement swapping between a 2-level general pure state and a 2-level maximally entangled state, is given by
\begin{align}
& \left\vert \mathscr{P}_{\uppercase\expandafter{\romannumeral2}}^r \right\rangle_{1,2} 
\otimes \left\vert \tilde{\phi} \left( u_1,u_2 \right) \right\rangle_{3,4} 									\notag \\
\rightarrow
& 
\left\vert \tilde{\phi} \left( 0,0 \right)\right\rangle_{1,3} \otimes \frac{\mathcal{P}}{2}
\left(
\lambda_1 \left\vert 0 u_2 \right\rangle + \lambda_2 (-1)^{u_1} \left\vert 1 \bar{u}_2 \right\rangle
\right)_{2,4}
+
\left\vert \tilde{\phi} \left( 0,1 \right)\right\rangle_{1,3} \otimes \frac{\mathcal{P}}{2}
\left(
\lambda_1 \left\vert 0 u_2 \right\rangle - \lambda_2 (-1)^{u_1} \left\vert 1 \bar{u}_2 \right\rangle
\right)_{2,4}
\notag \\
+
& \left\vert \tilde{\phi} \left( 1,0 \right)\right\rangle_{1,3} \otimes \frac{\mathcal{P}}{2}
\left(
\lambda_1 (-1)^{u_1} \left\vert 0 u_2 \right\rangle + \lambda_2 \left\vert 1 \bar{u}_2 \right\rangle
\right)_{2,4}
+
\left\vert \tilde{\phi} \left( 1,1 \right)\right\rangle_{1,3} \otimes \frac{\mathcal{P}}{2}
\left(
\lambda_1 (-1)^{u_1} \left\vert 0 u_2 \right\rangle - \lambda_2 \left\vert 1 \bar{u}_2 \right\rangle
\right)_{2,4},
\end{align}
where
\begin{align}
\mathcal{P} = 
\frac{1}{\sqrt{2 \left\vert \lambda_1 \right\vert^2 + 2 \left\vert \lambda_2 \right\vert^2}}.			\notag
\end{align}


\section{Mixed state entanglement swapping}

\noindent
In this section, we study entanglement swapping of mixed states.
We will first consider the entanglement swapping for $X$ states in d-level systems,
which actually include the case of 2-level systems studied in Refs. \cite{RoaL8962014}.
Then, we will the entanglement swapping cases for mixed maximally entangled states,
and formulate the entanglement swapping of mixed d-level Bell state, as the special cases.

\subsection{Entanglement swapping of $X$ states}

\noindent
The main characteristic of an $X$ state is that non-zero elements only appear on the diagonals of its density matrix,
including the major diagonal and minor one. An $X$ state having the following form,
\begin{gather}
\rho_{X_m} = \sum_{l_1,l_2,\dots,l_m=0}^{d-1}
\left(
\lambda_{a,a}
\left\vert l_1,l_2,\dots,l_m \right\rangle \left\langle l_1,l_2,\dots,l_m \right\vert +
\lambda_{a,\bar{a}}
\left\vert l_1,l_2,\dots,l_m \right\rangle \left\langle d-1- l_1,d-1-l_2,\dots,d-1-l_m \right\vert
\right),																			\\
\sum_{a,\bar{a}} \left( \lambda_{a,a} + \lambda_{a,\bar{a}} \right) = 1,								\notag
\end{gather}
where $a$ and $\bar{a}$ are the decimal representations of $l_1 l_2 \cdots l_m$ and 
$\left( d-1- l_1 \right) \left( d-1-l_2 \right) \cdots \left( d-1-l_m \right)$, respectively.
For example, suppose that $d=3$ and $l_1 l_2 \cdots l_m = 210$, then $a = 21$ and $\bar{a} = 5$.
As the general pure states considered before, below we would only like to consider 
the entanglement swapping cases for the bipartite states
\begin{gather}
\rho_{X_2} = \sum_{l_1,l_2=0}^{d-1}
\left(
\lambda_{a,a}
\left\vert l_1,l_2 \right\rangle \left\langle l_1,l_2 \right\vert +
\lambda_{a,\bar{a}}
\left\vert l_1,l_2 \right\rangle \left\langle d-1- l_1,d-1-l_2 \right\vert
\right).
\end{gather}
In the form of matrix, $\rho_{X_2}$ can be expressed as:
\begin{gather}
\begin{bmatrix}
\lambda_{0,0} & 0 & 0 & \cdots & 0 & 0 & \lambda_{0,d^2-1}				\\
0 & \lambda_{1,1} & 0 & \cdots & 0 & \lambda_{1,d^2-2} & 0				\\
0 & 0 & \lambda_{2,2}  & \cdots & \lambda_{2,d^2-3} & 0 & 0				\\
\hdotsfor{7}														\\
0 & 0 & \lambda_{d^2-3,2}  & \cdots & \lambda_{d^2-3,d^2-3} & 0 & 0			\\
0 & \lambda_{d^2-2,1} & 0 & \cdots & 0 & \lambda_{d^2-2,d^2-2} & 0			\\
\lambda_{d^2-1,0} & 0 & 0 & \cdots & 0 & 0 & \lambda_{d^2-1,d^2-1}
\end{bmatrix}.
\end{gather}
From $\rho_{X}$, one can get a special case described in Ref. \cite{RoaL8962014}, i.e. the 
2-level $X$ states, which is given by
\begin{align}
\label{2-level-X-states}
\hat{\rho}_{X_2} 
= & 
\lambda_{0,0}
\left\vert 0,0 \right\rangle \left\langle 0,0 \right\vert +
\lambda_{0,3}
\left\vert 0,0 \right\rangle \left\langle 1,1 \right\vert +
\lambda_{1,1}
\left\vert 0,1 \right\rangle \left\langle 0,1 \right\vert +
\lambda_{1,2}
\left\vert 0,1 \right\rangle \left\langle 1,0 \right\vert
\notag \\
&
+ \lambda_{2,2}
\left\vert 1,0 \right\rangle \left\langle 1,0 \right\vert +
\lambda_{2,1}
\left\vert 1,0 \right\rangle \left\langle 0,1 \right\vert +
\lambda_{3,3}
\left\vert 1,1 \right\rangle \left\langle 1,1 \right\vert +
\lambda_{3,0}
\left\vert 1,1 \right\rangle \left\langle 0,0 \right\vert
\notag \\ 
= & 
\begin{bmatrix}
\lambda_{0,0} & 0 & 0 & \lambda_{0,3}						\\
0 & \lambda_{1,1} & \lambda_{1,2} & 0						\\
0 & \lambda_{2,1} & \lambda_{2,2} & 0						\\
\lambda_{3,0} & 0 & 0 & \lambda_{3,3}				
\end{bmatrix}.
\end{align}

Now, let us start to formulate the entanglement swapping for any number of $X$ states in the state $\rho_{X_2}$.
Suppose that there are $n$ $X$ states, $\rho_{X_2}^1,\rho_{X_2}^2,\dots,\rho_{X_2}^n$.
For clarity, let us mark them by
\begin{gather}
\rho_{X_2}^r = \sum_{l_1^r,l_2^r=0}^{d-1}
\left(
\lambda_{\left[ l_1^r,l_2^r \right],\left[ l_1^r,l_2^r \right]}
\left\vert l_1^r,l_2^r \right\rangle \left\langle l_1^r,l_2^r \right\vert +
\lambda_{\left[ l_1^r,l_2^r \right],\left[ d-1- l_1^r,d-1-l_2^r \right]}
\left\vert l_1^r,l_2^r \right\rangle \left\langle d-1- l_1^r,d-1-l_2^r \right\vert
\right),
\end{gather}
where $r=1,2,\dots,n$. Let us suppose that the measurement operator
$\mathrm{\widetilde{M}}$ is performed on the second particle in each $X$ states.
then the entanglement swapping formula is given by
\begin{align}
\label{X-states-ES}
& \bigotimes_{r=1}^{n} \rho_{X_2}^r
\notag \\
= & \sum_{l_1^1,l_2^1,l_1^2,l_2^2,\dots,l_1^n,l_2^n = 0}^{d-1} 
\prod_{r=1}^n \lambda_{\left[ l_1^r,l_2^r \right],\left[ l_1^r,l_2^r \right]}
\bigotimes_{r=1}^{n} \left\vert l_1^r,l_2^r \right\rangle \left\langle l_1^r,l_2^r \right\vert
\tag{a} \\
& +
\sum_{l_1^1,l_2^1,l_1^2,l_2^2,\dots,l_1^n,l_2^n = 0}^{d-1} 
\prod_{r=1}^n \lambda_{\left[ l_1^r,l_2^r \right],\left[ d-1-l_1^r,d-1-l_2^r \right]}
\bigotimes_{r=1}^{n} \left\vert l_1^r,l_2^r \right\rangle \left\langle d-1-l_1^r,d-1-l_2^r \right\vert
\tag{b} \\
& +
\sum_{l_1^{k_1},l_2^{k_1},l_1^{k_2},l_2^{k_2},\dots,l_1^{k_n},l_2^{k_n} = 0}^{d-1}
\lambda_{\left[ l_1^{k_1},l_2^{k_1} \right],\left[ l_1^{k_1},l_2^{k_1} \right]}
\prod_{\begin{subarray}{c}
	n_i \in \left\{ 2,3,\dots,n \right\} \\
	i=1,2,\dots,t
	\end{subarray}}
\lambda_{\left[ l_1^{k_{n_i}},l_2^{k_{n_i}} \right],\left[ l_1^{k_{n_i}},l_2^{k_{n_i}} \right]}
\prod_{j \in \left\{ 1,2,\dots,n \right\} \setminus \left\{ n_1,n_2,\dots,n_t \right\}}
\lambda_{\left[ l_1^{k_j},l_2^{k_j} \right],\left[ d-1-l_1^{k_j},d-1-l_2^{k_j} \right]}
\notag \\
& \times
\left\vert l_1^{k_1},l_2^{k_1},l_1^{k_2},l_2^{k_2},\dots,
l_1^{k_{n_1-1}},l_2^{k_{n_1-1}},l_1^{k_{n_1}},l_2^{k_{n_1}},l_1^{k_{n_1+1}},l_2^{k_{n_1+1}},\dots,
l_1^{k_{n_2-1}},l_2^{k_{n_2-1}},l_1^{k_{n_2}},l_2^{k_{n_2}},l_1^{k_{n_2+1}},l_2^{k_{n_2+1}},\dots
\right.
\notag \\
& \left.
l_1^{k_{n_t-1}},l_2^{k_{n_t-1}},l_1^{k_{n_t}},l_2^{k_{n_t}},l_1^{k_{n_t+1}},l_2^{k_{n_t+1}},\dots,l_1^{k_n},l_2^{k_n} \right\rangle
\left\langle l_1^{k_1},l_2^{k_1},d-1-l_1^{k_2},d-1-l_2^{k_2},\dots,
\right.
\notag \\
& \left.
d-1-l_1^{k_{n_1-1}},d-1-l_2^{k_{n_1-1}},l_1^{k_{n_1}},l_2^{k_{n_1}},d-1-l_1^{k_{n_1+1}},d-1-l_2^{k_{n_1+1}},\dots
\right.
\notag \\
& \left.
d-1-l_1^{k_{n_2-1}},d-1-l_2^{k_{n_2-1}},l_1^{k_{n_2}},l_2^{k_{n_2}},d-1-l_1^{k_{n_2+1}},d-1-l_2^{k_{n_2+1}},\dots
\right.
\notag \\
& \left.
d-1-l_1^{k_{n_t-1}},d-1-l_2^{k_{n_t-1}},l_1^{k_{n_t}},l_2^{k_{n_t}},d-1-l_1^{k_{n_t+1}},d-1-l_2^{k_{n_t+1}},\dots,
d-1-l_1^{k_n},d-1-l_2^{k_n} \right\vert
\tag{c} \\
& +
\sum_{l_1^{k_1},l_2^{k_1},l_1^{k_2},l_2^{k_2},\dots,l_1^{k_n},l_2^{k_n} = 0}^{d-1}
\prod_{\begin{subarray}{c}
	n_i \in \left\{ 2,3,\dots,n \right\} \\
	i=1,2,\dots,t
	\end{subarray}}
\lambda_{\left[ l_1^{k_{n_i}},l_2^{k_{n_i}} \right],\left[ l_1^{k_{n_i}},l_2^{k_{n_i}} \right]}
\prod_{j \in \left\{ 1,2,\dots,n \right\} \setminus \left\{ n_1,n_2,\dots,n_t \right\}}
\lambda_{\left[ l_1^{k_j},l_2^{k_j} \right],\left[ d-1-l_1^{k_j},d-1-l_2^{k_j} \right]}
\notag \\
& \times
\left\vert l_1^{k_1},l_2^{k_1},l_1^{k_2},l_2^{k_2},\dots,
l_1^{k_{n_1-1}},l_2^{k_{n_1-1}},l_1^{k_{n_1}},l_2^{k_{n_1}},l_1^{k_{n_1+1}},l_2^{k_{n_1+1}},\dots,
l_1^{k_{n_2-1}},l_2^{k_{n_2-1}},l_1^{k_{n_2}},l_2^{k_{n_2}},l_1^{k_{n_2+1}},l_2^{k_{n_2+1}},\dots,
\right.
\notag \\
& \left.
l_1^{k_{n_t-1}},l_2^{k_{n_t-1}},l_1^{k_{n_t}},l_2^{k_{n_t}},l_1^{k_{n_t+1}},l_2^{k_{n_t+1}},\dots,l_1^{k_n},l_2^{k_n} \right\rangle
\left\langle d-1-l_1^{k_1},d-1-l_2^{k_1},d-1-l_1^{k_2},d-1-l_2^{k_2},\dots,
\right.
\notag \\
& \left.
d-1-l_1^{k_{n_1-1}},d-1-l_2^{k_{n_1-1}},l_1^{k_{n_1}},l_2^{k_{n_1}},d-1-l_1^{k_{n_1+1}},d-1-l_2^{k_{n_1+1}},\dots,
\right.
\notag \\
& \left.
d-1-l_1^{k_{n_2-1}},d-1-l_2^{k_{n_2-1}},l_1^{k_{n_2}},l_2^{k_{n_2}},d-1-l_1^{k_{n_2+1}},d-1-l_2^{k_{n_2+1}},\dots,
\right.
\notag \\
& \left.
d-1-l_1^{k_{n_t-1}},d-1-l_2^{k_{n_t-1}},l_1^{k_{n_t}},l_2^{k_{n_t}},d-1-l_1^{k_{n_t+1}},d-1-l_2^{k_{n_t+1}},\dots,
d-1-l_1^{k_n},d-1-l_2^{k_n} \right\vert
\tag{d} \\
\Rightarrow
& \sum_{l_1^1,l_2^1,l_1^2,l_2^2,\dots,l_1^n,l_2^n = 0}^{d-1} 
\prod_{r=1}^n \lambda_{\left[ l_1^r,l_2^r \right],\left[ l_1^r,l_2^r \right]}
\bigotimes_{r=1}^{n} \left\vert l_1^r \right\rangle \left\langle l_1^r \right\vert
\bigotimes_{r=1}^{n} \left\vert l_2^r \right\rangle \left\langle l_2^r \right\vert
\notag \\
& +
\sum_{l_1^1,l_2^1,l_1^2,l_2^2,\dots,l_1^n,l_2^n = 0}^{d-1} 
\prod_{r=1}^n \lambda_{\left[ l_1^r,l_2^r \right],\left[ d-1-l_1^r,d-1-l_2^r \right]}
\bigotimes_{r=1}^{n} \left\vert l_1^r \right\rangle \left\langle d-1-l_1^r \right\vert
\bigotimes_{r=1}^{n} \left\vert l_2^r \right\rangle \left\langle d-1-l_2^r \right\vert
\notag \\
& +
\sum_{l_1^{k_1},l_2^{k_1},l_1^{k_2},l_2^{k_2},\dots,l_1^{k_n},l_2^{k_n} = 0}^{d-1}
\lambda_{\left[ l_1^{k_1},l_2^{k_1} \right],\left[ l_1^{k_1},l_2^{k_1} \right]}
\prod_{\begin{subarray}{c}
	n_i \in \left\{ 2,3,\dots,n \right\} \\
	i=1,2,\dots,t
	\end{subarray}}
\lambda_{\left[ l_1^{k_{n_i}},l_2^{k_{n_i}} \right],\left[ l_1^{k_{n_i}},l_2^{k_{n_i}} \right]}
\prod_{j \in \left\{ 1,2,\dots,n \right\} \setminus \left\{ n_1,n_2,\dots,n_t \right\}}
\lambda_{\left[ l_1^{k_j},l_2^{k_j} \right],\left[ d-1-l_1^{k_j},d-1-l_2^{k_j} \right]}
\notag \\
& \times
\left\vert l_1^{k_1},l_1^{k_2},\dots,l_1^{k_{n_1-1}},l_1^{k_{n_1}},l_1^{k_{n_1+1}},\dots,
l_1^{k_{n_2-1}},l_1^{k_{n_2}},l_1^{k_{n_2+1}},\dots,
l_1^{k_{n_t-1}},l_1^{k_{n_t}},l_1^{k_{n_t+1}},\dots,l_1^{k_n} \right\rangle
\left\langle l_1^{k_1},d-1-l_1^{k_2},\dots,
\right.
\notag \\
& \left.
d-1-l_1^{k_{n_1-1}},l_1^{k_{n_1}},d-1-l_1^{k_{n_1+1}},\dots,
d-1-l_1^{k_{n_2-1}},l_1^{k_{n_2}},d-1-l_1^{k_{n_2+1}},\dots,
d-1-l_1^{k_{n_t-1}},l_1^{k_{n_t}},d-1-l_1^{k_{n_t+1}},\dots,
\right.
\notag \\
& \left.
d-1-l_1^{k_n} \right\vert
\otimes
\left\vert l_2^{k_1},l_2^{k_2},\dots,l_2^{k_{n_1-1}},l_2^{k_{n_1}},l_2^{k_{n_1+1}},\dots,
l_2^{k_{n_2-1}},l_2^{k_{n_2}},l_2^{k_{n_2+1}},\dots,
l_2^{k_{n_t-1}},l_2^{k_{n_t}},l_2^{k_{n_t+1}},\dots,l_2^{k_n} \right\rangle
\left\langle l_2^{k_1},d-1-l_2^{k_2},\dots,
\right.
\notag \\
& \left.
d-1-l_2^{k_{n_1-1}},l_2^{k_{n_1}},d-1-l_2^{k_{n_1+1}},\dots,
d-1-l_2^{k_{n_2-1}},l_2^{k_{n_2}},d-1-l_2^{k_{n_2+1}},\dots,
d-1-l_2^{k_{n_t-1}},l_2^{k_{n_t}},d-1-l_2^{k_{n_t+1}},\dots,
\right.
\notag \\
& \left.
d-1-l_2^{k_n} \right\vert
\notag \\
& +
\sum_{l_1^{k_1},l_2^{k_1},l_1^{k_2},l_2^{k_2},\dots,l_1^{k_n},l_2^{k_n} = 0}^{d-1}
\prod_{\begin{subarray}{c}
	n_i \in \left\{ 2,3,\dots,n \right\} \\
	i=1,2,\dots,t
	\end{subarray}}
\lambda_{\left[ l_1^{k_{n_i}},l_2^{k_{n_i}} \right],\left[ l_1^{k_{n_i}},l_2^{k_{n_i}} \right]}
\prod_{j \in \left\{ 1,2,\dots,n \right\} \setminus \left\{ n_1,n_2,\dots,n_t \right\}}
\lambda_{\left[ l_1^{k_j},l_2^{k_j} \right],\left[ d-1-l_1^{k_j},d-1-l_2^{k_j} \right]}
\notag \\
& \times
\left\vert l_1^{k_1},l_1^{k_2},\dots,l_1^{k_{n_1-1}},l_1^{k_{n_1}},l_1^{k_{n_1+1}},\dots,
l_1^{k_{n_2-1}},l_1^{k_{n_2}},l_1^{k_{n_2+1}},\dots,
l_1^{k_{n_t-1}},l_1^{k_{n_t}},l_1^{k_{n_t+1}},\dots,l_1^{k_n} \right\rangle
\left\langle d-1-l_1^{k_1},d-1-l_1^{k_2},\dots,
\right.
\notag \\
& \left.
d-1-l_1^{k_{n_1-1}},l_1^{k_{n_1}},d-1-l_1^{k_{n_1+1}},\dots,
d-1-l_1^{k_{n_2-1}},l_1^{k_{n_2}},d-1-l_1^{k_{n_2+1}},\dots,
d-1-l_1^{k_{n_t-1}},l_1^{k_{n_t}},d-1-l_1^{k_{n_t+1}},\dots,
\right.
\notag \\
& \left.
d-1-l_1^{k_n} \right\vert
\otimes
\left\vert l_2^{k_1},l_2^{k_2},\dots,l_2^{k_{n_1-1}},l_2^{k_{n_1}},l_2^{k_{n_1+1}},\dots,
l_2^{k_{n_2-1}},l_2^{k_{n_2}},l_2^{k_{n_2+1}},\dots,
l_2^{k_{n_t-1}},l_2^{k_{n_t}},l_2^{k_{n_t+1}},\dots,l_2^{k_n} \right\rangle
\notag \\
&
\left\langle d-1-l_2^{k_1},d-1-l_2^{k_2},\dots,
d-1-l_2^{k_{n_1-1}},l_2^{k_{n_1}},d-1-l_2^{k_{n_1+1}},\dots,
d-1-l_2^{k_{n_2-1}},l_2^{k_{n_2}},d-1-l_2^{k_{n_2+1}},\dots,
\right.
\notag \\
& \left.
d-1-l_2^{k_{n_t-1}},l_2^{k_{n_t}},d-1-l_2^{k_{n_t+1}},\dots,d-1-l_2^{k_n} \right\vert
\notag \\
= & 
\begin{dcases}
\quad 
\frac{\mathcal{P}_1}{d} \sum_{l_1^1,l_1^2,\dots,l_1^n,l_2^1 = 0}^{d-1}
\left[
\sum_{v_1^1,v_2^1,\dots,v_n^1 = 0}^{d-1}
\lambda_{\left[ l_1^1,l_2^1 \right],\left[ l_1^1,l_2^1 \right]}
\prod_{r=2}^n \lambda_{\left[ l_1^r,v_r^1 \oplus l_2^1 \right],\left[ l_1^r,v_r^1 \oplus l_2^1 \right]}
\bigotimes_{r=1}^{n} \left\vert l_1^r \right\rangle \left\langle l_1^r \right\vert
\otimes \left\vert \phi \left( v_1^1,v_2^1,\dots,v_n^1 \right) \right\rangle 
\left\langle \phi \left( v_1^1,v_2^1,\dots,v_n^1 \right) \right\vert
\right.
\\
+
\sum_{v_1^2 = 0}^{d-1}
\sum_{v_2^2,v_3^2,\dots,v_n^2 = \frac{d}{2},0}
\lambda_{\left[ l_1^1,l_2^1 \right],\left[ d-1-l_1^1,d-1- l_2^1 \right]}
\prod_{r=2}^n \lambda_{\left[ l_1^r,l_2^1 \oplus v_r^2 \right],\left[ d-1-l_1^r,d-1- \left( l_2^1 \oplus v_r^2 \right) \right]}
\bigotimes_{r=1}^{n} \left\vert l_1^r \right\rangle \left\langle d-1-l_1^r \right\vert
\\
\left.
\quad
\otimes \left\vert \phi \left( v_1^2,v_2^2,\dots,v_n^2 \right) \right\rangle 
\left\langle \phi \left( v_1^2,v_2^2,\dots,v_n^2 \right) \right\vert
\right],
\quad \textrm{if} \phantom{i} d \phantom{i} \textrm{is an even number},
\\
\\
\quad
\frac{\mathcal{P}_2}{d}
\sum_{l_1^1,l_1^2,\dots,l_1^n,l_2^1 = 0}^{d-1}
\left[
\sum_{v_1^1,v_2^1,\dots,v_n^1 = 0}^{d-1}
\lambda_{\left[ l_1^1,l_2^1 \right],\left[ l_1^1,l_2^1 \right]}
\prod_{r=2}^n \lambda_{\left[ l_1^r,v_r^1 \oplus l_2^1 \right],\left[ l_1^r,v_r^1 \oplus l_2^1 \right]}
\bigotimes_{r=1}^{n} \left\vert l_1^r \right\rangle \left\langle l_1^r \right\vert
\otimes \left\vert \phi \left( v_1^1,v_2^1,\dots,v_n^1 \right) \right\rangle 
\left\langle \phi \left( v_1^1,v_2^1,\dots,v_n^1 \right) \right\vert
\right.
\\
\left.
+
\sum_{v_1^2 = 0}^{d-1}
\prod_{r=1}^n \lambda_{\left[ l_1^r,l_2^1 \right],\left[ d-1-l_1^r,d-1- l_2^1 \right]}
\bigotimes_{r=1}^{n} \left\vert l_1^r \right\rangle \left\langle d-1-l_1^r \right\vert
\otimes \left\vert \phi \left( v_1^2,0,0,\dots,0 \right) \right\rangle 
\left\langle \phi \left( v_1^2,0,0,\dots,0 \right) \right\vert
\right]
\\
+
\frac{\mathcal{P}_2}{d}
\sum_{l_1^{k_1},l_1^{k_2},\dots,l_1^{k_n},l_2^{k_1} = 0}^{d-1}
\left[
\sum_{\begin{subarray}{c}
	v_1^3,v_{n_1}^3,v_{n_2}^3,\dots,v_{n_t}^3 = 0,1,\dots,d-1 \\
	n_i \in \left\{ 2,3,\dots,n \right\}, i=1,2,\dots,t
	\end{subarray}}
\lambda_{\left[ l_1^{k_1},l_2^{k_1} \right],\left[ l_1^{k_1},l_2^{k_1} \right]}
\prod_{i=1}^t \lambda_{\left[ l_1^{k_{n_i}},v_{n_i}^3 \oplus l_2^{k_1} \right],\left[ l_1^{k_{n_i}},v_{n_i}^3 \oplus l_2^{k_1} \right]}
\right.
\\
\quad \times
\prod_{j \in \left\{ 1,2,\dots,n \right\} \setminus \left\{ n_1,n_2,\dots,n_t \right\}}
\lambda_{\left[ l_1^{k_j},\frac{d-1}{2} \right],\left[ d-1-l_1^{k_j},\frac{d-1}{2} \right]}
\left\vert l_1^{k_1},l_1^{k_2},\dots,l_1^{k_{n_1-1}},l_1^{k_{n_1}},l_1^{k_{n_1+1}},\dots,
l_1^{k_{n_2-1}},l_1^{k_{n_2}},l_1^{k_{n_2+1}},\dots,
l_1^{k_{n_t-1}},l_1^{k_{n_t}},l_1^{k_{n_t+1}},
\right.
\\
\quad
\left.
\dots,l_1^{k_n} \right\rangle
\left\langle l_1^{k_1},d-1-l_1^{k_2},\dots,
d-1-l_1^{k_{n_1-1}},l_1^{k_{n_1}},d-1-l_1^{k_{n_1+1}},\dots,
d-1-l_1^{k_{n_2-1}},l_1^{k_{n_2}},d-1-l_1^{k_{n_2+1}},\dots,
\right.
\\
\quad
\left.
d-1-l_1^{k_{n_t-1}},l_1^{k_{n_t}},d-1-l_1^{k_{n_t+1}},\dots,d-1-l_1^{k_n} \right\vert
\\
\quad
\otimes 
\left\vert \phi \left( v_1^3,\frac{d-1}{2} \ominus l_2^{k_1},\frac{d-1}{2} \ominus l_2^{k_1},\dots,\frac{d-1}{2} \ominus l_2^{k_1},
v_{n_1}^3,\frac{d-1}{2} \ominus l_2^{k_1},\dots,\frac{d-1}{2} \ominus l_2^{k_1},v_{n_2}^3,\frac{d-1}{2} \ominus l_2^{k_1},\dots,
\right.
\right.
\\
\quad
\left.
\left.
\frac{d-1}{2} \ominus l_2^{k_1},v_{n_t}^3,\frac{d-1}{2} \ominus l_2^{k_1},\dots,\frac{d-1}{2} \ominus l_2^{k_1} \right) \right\rangle 
\left\langle \phi \left( v_1^3,\frac{d-1}{2} \ominus l_2^{k_1},\frac{d-1}{2} \ominus l_2^{k_1},\dots,
\right.
\right.
\\
\quad
\frac{d-1}{2} \ominus l_2^{k_1},
v_{n_1}^3,\frac{d-1}{2} \ominus l_2^{k_1},\dots,\frac{d-1}{2} \ominus l_2^{k_1},v_{n_2}^3,\frac{d-1}{2} \ominus l_2^{k_1},\dots,
\\
\quad
\left.
\left.
\frac{d-1}{2} \ominus l_2^{k_1},v_{n_t}^3,\frac{d-1}{2} \ominus l_2^{k_1},\dots,\frac{d-1}{2} \ominus l_2^{k_1} \right) \right\vert
\\
+
\sum_{\begin{subarray}{c}
	v_{n_1}^4,v_{n_2}^4,\dots,v_{n_t}^4 = 0,1,\dots,d-1 \\
	n_i \in \left\{ 2,3,\dots,n \right\}, i=1,2,\dots,t
	\end{subarray}}
\prod_{i=1}^t \lambda_{\left[ l_1^{k_{n_i}},v_{n_i}^4 \oplus \frac{d-1}{2} \right],\left[ l_1^{k_{n_i}},v_{n_i}^4 \oplus \frac{d-1}{2} \right]}
\prod_{j \in \left\{ 1,2,\dots,n \right\} \setminus \left\{ n_1,n_2,\dots,n_t \right\}}
\lambda_{\left[ l_1^{k_j},l_2^{k_1} \right],\left[ d-1-l_1^{k_j},d-1-l_2^{k_1} \right]}
\zeta^{ - \left( 2 l_1^{k_1} + 1 \right) v_1^4}
\\
\quad
\left\vert l_1^{k_1},l_1^{k_2},\dots,l_1^{k_{n_1-1}},l_1^{k_{n_1}},l_1^{k_{n_1+1}},\dots,
l_1^{k_{n_2-1}},l_1^{k_{n_2}},l_1^{k_{n_2+1}},\dots,
l_1^{k_{n_t-1}},l_1^{k_{n_t}},l_1^{k_{n_t+1}},\dots,l_1^{k_n} \right\rangle
\left\langle d-1-l_1^{k_1},d-1-l_1^{k_2},\dots,
\right.
\\
\quad
\left.
d-1-l_1^{k_{n_1-1}},l_1^{k_{n_1}},d-1-l_1^{k_{n_1+1}},\dots,
d-1-l_1^{k_{n_2-1}},l_1^{k_{n_2}},d-1-l_1^{k_{n_2+1}},\dots,
\right.
\\
\quad
\left.
d-1-l_1^{k_{n_t-1}},l_1^{k_{n_t}},d-1-l_1^{k_{n_t+1}},\dots,d-1-l_1^{k_n} \right\vert
\\
\quad
\otimes
\left\vert \phi \left( v_1^3,0,0,\dots,0,v_{n_1}^3,0,\dots,0,v_{n_2}^3,0,\dots,0,v_{n_t}^3,0,\dots,0 \right) \right\rangle 
\left\langle \phi \left( v_1^3,0,0,\dots,0,v_{n_1}^3,0,\dots,0,v_{n_2}^3,0,\dots,
\right.
\right.
\\
\quad
\left.
\left.
\left.
0,v_{n_t}^3,0,\dots,0 \right) \right\vert
\right],
\quad \textrm{if} \phantom{i} d \phantom{i} \textrm{is an odd number},
\end{dcases}
\end{align}
where
\begin{align}
& \mathcal{P}_1 = \frac{d}{\mathcal{P}_1'}, \quad \mathcal{P}_2 = \frac{d}{\mathcal{P}_2'},
\notag \\
& \mathcal{P}_1' = 
\sum_{l_1^1,l_1^2,\dots,l_1^n,l_2^1 = 0}^{d-1}
\left[
\sum_{v_1^1,v_2^1,\dots,v_n^1 = 0}^{d-1}
\lambda_{\left[ l_1^1,l_2^1 \right],\left[ l_1^1,l_2^1 \right]}
\prod_{r=2}^n \lambda_{\left[ l_1^r,v_r^1 \oplus l_2^1 \right],\left[ l_1^r,v_r^1 \oplus l_2^1 \right]}
+
\sum_{v_1^2 = 0}^{d-1}
\sum_{v_2^2,v_3^2,\dots,v_n^2 = \frac{d}{2},0}
\lambda_{\left[ l_1^1,l_2^1 \right],\left[ d-1-l_1^1,d-1- l_2^1 \right]}
\right.
\notag 
\\
& \qquad \quad \times
\left.
\prod_{r=2}^n \lambda_{\left[ l_1^r,l_2^1 \oplus v_r^2 \right],\left[ d-1-l_1^r,d-1- \left( l_2^1 \oplus v_r^2 \right) \right]}
\right],
\notag 
\\
& \mathcal{P}_2' = \sum_{l_1^1,l_1^2,\dots,l_1^n,l_2^1 = 0}^{d-1}
\left[
\sum_{v_1^1,v_2^1,\dots,v_n^1 = 0}^{d-1}
\lambda_{\left[ l_1^1,l_2^1 \right],\left[ l_1^1,l_2^1 \right]}
\prod_{r=2}^n \lambda_{\left[ l_1^r,v_r^1 \oplus l_2^1 \right],\left[ l_1^r,v_r^1 \oplus l_2^1 \right]}
+
\sum_{v_1^2 = 0}^{d-1}
\prod_{r=1}^n \lambda_{\left[ l_1^r,l_2^1 \right],\left[ d-1-l_1^r,d-1- l_2^1 \right]}
\right]
\notag 
\\
& \qquad
+ 
\sum_{l_1^{k_1},l_1^{k_2},\dots,l_1^{k_n},l_2^{k_1} = 0}^{d-1}
\left[
\sum_{\begin{subarray}{c}
	v_1^3,v_{n_1}^3,v_{n_2}^3,\dots,v_{n_t}^3 = 0,1,\dots,d-1 \\
	n_i \in \left\{ 2,3,\dots,n \right\}, i=1,2,\dots,t
	\end{subarray}}
\lambda_{\left[ l_1^{k_1},l_2^{k_1} \right],\left[ l_1^{k_1},l_2^{k_1} \right]}
\prod_{i=1}^t \lambda_{\left[ l_1^{k_{n_i}},v_{n_i}^3 \oplus l_2^{k_1} \right],\left[ l_1^{k_{n_i}},v_{n_i}^3 \oplus l_2^{k_1} \right]}
\right. 
\notag \\
& \qquad \quad \times
\prod_{j \in \left\{ 1,2,\dots,n \right\} \setminus \left\{ n_1,n_2,\dots,n_t \right\}}
\lambda_{\left[ l_1^{k_j},\frac{d-1}{2} \right],\left[ d-1-l_1^{k_j},\frac{d-1}{2} \right]}
\notag \\
& \qquad +
\left.
\sum_{\begin{subarray}{c}
	v_{n_1}^4,v_{n_2}^4,\dots,v_{n_t}^4 = 0,1,\dots,d-1 \\
	n_i \in \left\{ 2,3,\dots,n \right\}, i=1,2,\dots,t
	\end{subarray}}
\prod_{i=1}^t \lambda_{\left[ l_1^{k_{n_i}},v_{n_i}^4 \oplus \frac{d-1}{2} \right],\left[ l_1^{k_{n_i}},v_{n_i}^4 \oplus \frac{d-1}{2} \right]}
\prod_{j \in \left\{ 1,2,\dots,n \right\} \setminus \left\{ n_1,n_2,\dots,n_t \right\}}
\lambda_{\left[ l_1^{k_j},l_2^{k_1} \right],\left[ d-1-l_1^{k_j},d-1-l_2^{k_1} \right]}
\right].
\notag
\end{align}
Note that in the above formula, Eqs. (a) and (b) represent the first and last terms of the expansion of 
$\bigotimes_{r=1}^{n} \rho_{X_2}^r$ respectively, while (c) and (d) the remainder.
In addition, the following parameter substitution for the orders of particles are made for the remaining items:
\begin{gather}
\begin{matrix}
1 & 2 & 3 & \dots &  n \\
\downarrow & \downarrow & \downarrow & \dots &  \downarrow \\
k_1 & k_2 & k_3 & \dots &  k_n \\
\end{matrix}.
\end{gather}

As a special case of the entanglement swapping shown above, let us give the entanglement swapping formulas for two $X$ sates.
When $d$ is an even number, we have
\begin{align}
\label{two-X-states-ES-first-case}
& \bigotimes_{r=1}^{2} \rho_{X_2}^r
\notag \\
\rightarrow &
\frac{\mathcal{P}_1}{d} 
\sum_{l_1^1,l_1^2,l_2^1 = 0}^{d-1}
\left[
\sum_{v_1^1,v_2^1 = 0}^{d-1}
\lambda_{\left[ l_1^1,l_2^1 \right],\left[ l_1^1,l_2^1 \right]}
\lambda_{\left[ l_1^2,v_2^1 \oplus l_2^1 \right],\left[ l_1^2,v_2^1 \oplus l_2^1 \right]}
\bigotimes_{r=1}^{2} \left\vert l_1^r \right\rangle \left\langle l_1^r \right\vert
\otimes \left\vert \phi \left( v_1^1,v_2^1 \right) \right\rangle 
\left\langle \phi \left( v_1^1,v_2^1 \right) \right\vert
\right.
\notag \\
& \left.
+
\sum_{v_1^2 = 0}^{d-1}
\sum_{v_2^2 = 0,\frac{d}{2}}
\lambda_{\left[ l_1^1,l_2^1 \right],\left[ d-1-l_1^1,d-1- l_2^1 \right]}
\lambda_{\left[ l_1^2,l_2^1 \oplus v_2^2 \right],\left[ d-1-l_1^2,d-1- \left( l_2^1 \oplus v_2^2 \right) \right]}
\bigotimes_{r=1}^{2} \left\vert l_1^r \right\rangle \left\langle d-1-l_1^r \right\vert
\left\vert \phi \left( v_1^2,v_2^2 \right) \right\rangle 
\left\langle \phi \left( v_1^2,v_2^2 \right) \right\vert
\right],
\end{align}
and when $d$ is an odd number, we have
\begin{align}
\label{two-X-states-ES-second-case}
& \bigotimes_{r=1}^{2} \rho_{X_2}^r
\notag \\
\rightarrow & 
\frac{\mathcal{P}_2}{d}
\sum_{l_1^1,l_1^2,l_2^1 = 0}^{d-1}
\left[
\sum_{v_1^1,v_2^1 = 0}^{d-1}
\lambda_{\left[ l_1^1,l_2^1 \right],\left[ l_1^1,l_2^1 \right]}
\lambda_{\left[ l_1^2,v_2^1 \oplus l_2^1 \right],\left[ l_1^2,v_2^1 \oplus l_2^1 \right]}
\bigotimes_{r=1}^{2} \left\vert l_1^r \right\rangle \left\langle l_1^r \right\vert
\otimes \left\vert \phi \left( v_1^1,v_2^1 \right) \right\rangle 
\left\langle \phi \left( v_1^1,v_2^1 \right) \right\vert
\right.
\notag \\
& \left.
+
\sum_{v_1^2 = 0}^{d-1}
\prod_{r=1}^2 \lambda_{\left[ l_1^r,l_2^1 \right],\left[ d-1-l_1^r,d-1- l_2^1 \right]}
\bigotimes_{r=1}^{2} \left\vert l_1^r \right\rangle \left\langle d-1-l_1^r \right\vert
\otimes \left\vert \phi \left( v_1^2,0 \right) \right\rangle 
\left\langle \phi \left( v_1^2,0 \right) \right\vert
\right]
\notag \\
& +
\frac{\mathcal{P}_2}{d}
\sum_{l_1^1,l_1^2 = 0}^{d-1}
\left[
\sum_{v_1^3 = 0}^{d-1}
\lambda_{\left[ l_1^1,l_2^1 \right],\left[ l_1^1,l_2^1 \right]}
\lambda_{\left[ l_1^2,\frac{d-1}{2} \right],\left[ d-1-l_1^2,\frac{d-1}{2} \right]}
\left\vert l_1^1,l_1^2 \right\rangle \left\langle l_1^1,d-1-l_1^2 \right\vert
\otimes
\left\vert \phi \left( v_1^3,\frac{d-1}{2} \ominus l_2^1 \right) \right\rangle
\left\langle \phi \left( v_1^3,\frac{d-1}{2} \ominus l_2^1 \right) \right\vert
\right.
\notag \\
& \left.
+
\sum_{v_2^4 = 0}^{d-1}
\lambda_{\left[ l_1^1,l_2^1 \right],\left[ d-1-l_1^1,d-1-l_2^1 \right]}
\lambda_{\left[ l_1^2,v_2^4 \oplus \frac{d-1}{2} \right],\left[ l_1^2,v_2^4 \oplus \frac{d-1}{2} \right]}
\zeta^{ - \left( 2 l_1^1 + 1 \right) v_1^4}
\left\vert l_1^1,l_1^2 \right\rangle \left\langle d-1-l_1^1,l_1^2 \right\vert
\otimes
\left\vert \phi \left( v_1^3,v_2^3 \right) \right\rangle \left\langle \phi \left( v_1^3,v_2^3 \right) \right\vert
\right],
\end{align}
where
\begin{align}
& \mathcal{P}_1 = \frac{d}{\mathcal{P}_1'}, \quad \mathcal{P}_2 = \frac{d}{\mathcal{P}_2'},
\notag \\
& \mathcal{P}_1' = 
\sum_{l_1^1,l_1^2,l_2^1 = 0}^{d-1}
\left[
\sum_{v_1^1,v_2^1 = 0}^{d-1}
\lambda_{\left[ l_1^1,l_2^1 \right],\left[ l_1^1,l_2^1 \right]}
\lambda_{\left[ l_1^r,v_2^1 \oplus l_2^1 \right],\left[ l_1^r,v_2^1 \oplus l_2^1 \right]}
+
\sum_{v_1^2 = 0}^{d-1}
\sum_{v_2^2 = 0,\frac{d}{2}}
\lambda_{\left[ l_1^1,l_2^1 \right],\left[ d-1-l_1^1,d-1- l_2^1 \right]}
\lambda_{\left[ l_1^2,l_2^1 \oplus v_2^2 \right],\left[ d-1-l_1^2,d-1- \left( l_2^1 \oplus v_2^2 \right) \right]}
\right],
\notag 
\\
& \mathcal{P}_2' = \sum_{l_1^1,l_1^2,l_2^1 = 0}^{d-1}
\left[
\sum_{v_1^1,v_2^1 = 0}^{d-1}
\lambda_{\left[ l_1^1,l_2^1 \right],\left[ l_1^1,l_2^1 \right]}
\lambda_{\left[ l_1^2,v_2^1 \oplus l_2^1 \right],\left[ l_1^r,v_2^1 \oplus l_2^1 \right]}
+
\sum_{v_1^2 = 0}^{d-1}
\prod_{r=1}^2 \lambda_{\left[ l_1^r,l_2^1 \right],\left[ d-1-l_1^r,d-1- l_2^1 \right]}
\right]
\notag 
\\
& \qquad
+ 
\sum_{l_1^1,l_1^2 = 0}^{d-1}
\left[
\sum_{v_1^3 = 0}^{d-1}
\lambda_{\left[ l_1^1,l_2^1 \right],\left[ l_1^1,l_2^1 \right]}
\lambda_{\left[ l_1^2,\frac{d-1}{2} \right],\left[ d-1-l_1^2,\frac{d-1}{2} \right]}
+
\sum_{v_2^4 = 0}^{d-1}
\lambda_{\left[ l_1^1,l_2^1 \right],\left[ d-1-l_1^1,d-1-l_2^1 \right]}
\lambda_{\left[ l_1^2,v_2^4 \oplus \frac{d-1}{2} \right],\left[ l_1^2,v_{n_i}^4 \oplus \frac{d-1}{2} \right]}
\right].
\notag
\end{align}

Further, one can get the entanglement swapping between two 2-level $X$ states (see Eq. \ref{2-level-X-states}),
which was presented in Ref. \cite{RoaL8962014}. We can describe the entanglement swapping as
\begin{align}
\label{two-2-level-X-states-ES}
& \bigotimes_{r=1}^{2} \hat{\rho}_{X_2}
\notag \\
\rightarrow \quad
& \frac{\mathcal{P}}{2}
\left[
\left( \lambda_{00}\lambda_{00} + \lambda_{11}\lambda_{11}  \right) \left\vert 00 \right\rangle \left\langle 00 \right\vert
+
\left( \lambda_{03}\lambda_{03} + \lambda_{12}\lambda_{12}  \right) \left\vert 00 \right\rangle \left\langle 11 \right\vert
\right.
\notag \\
& \phantom{[} + 
\left( \lambda_{00}\lambda_{22} + \lambda_{11}\lambda_{33}  \right) \left\vert 01 \right\rangle \left\langle 01 \right\vert
+
\left( \lambda_{03}\lambda_{21} + \lambda_{12}\lambda_{30}  \right) \left\vert 01 \right\rangle \left\langle 10 \right\vert
\notag \\
& \phantom{[} + 
\left( \lambda_{21}\lambda_{03} + \lambda_{30}\lambda_{12}  \right) \left\vert 10 \right\rangle \left\langle 01 \right\vert
+
\left( \lambda_{22}\lambda_{00} + \lambda_{33}\lambda_{11}  \right) \left\vert 10 \right\rangle \left\langle 10 \right\vert
\notag \\
& \phantom{[} + 
\left.
\left( \lambda_{21}\lambda_{21} + \lambda_{30}\lambda_{30}  \right) \left\vert 11 \right\rangle \left\langle 00 \right\vert
+
\left( \lambda_{22}\lambda_{22} + \lambda_{33}\lambda_{33}  \right) \left\vert 11 \right\rangle \left\langle 11 \right\vert
\right] \otimes \left\vert \tilde{\phi}\left( 0,0 \right) \right\rangle \left\langle \tilde{\phi}\left( 0,0 \right) \right\vert
\notag \\
+
& \frac{\mathcal{P}}{2}
\left[
\left( \lambda_{00}\lambda_{00} + \lambda_{11}\lambda_{11}  \right) \left\vert 00 \right\rangle \left\langle 00 \right\vert
-
\left( \lambda_{03}\lambda_{03} + \lambda_{12}\lambda_{12}  \right) \left\vert 00 \right\rangle \left\langle 11 \right\vert
\right.
\notag \\
& \phantom{[} + 
\left( \lambda_{00}\lambda_{22} + \lambda_{11}\lambda_{33}  \right) \left\vert 01 \right\rangle \left\langle 01 \right\vert
-
\left( \lambda_{03}\lambda_{21} + \lambda_{12}\lambda_{30}  \right) \left\vert 01 \right\rangle \left\langle 10 \right\vert
\notag \\
& \phantom{[} 
- 
\left( \lambda_{21}\lambda_{03} + \lambda_{30}\lambda_{12}  \right) \left\vert 10 \right\rangle \left\langle 01 \right\vert
+
\left( \lambda_{22}\lambda_{00} + \lambda_{33}\lambda_{11}  \right) \left\vert 10 \right\rangle \left\langle 10 \right\vert
\notag \\
& \phantom{[} 
\left.
-
\left( \lambda_{21}\lambda_{21} + \lambda_{30}\lambda_{30}  \right) \left\vert 11 \right\rangle \left\langle 00 \right\vert
+
\left( \lambda_{22}\lambda_{22} + \lambda_{33}\lambda_{33}  \right) \left\vert 11 \right\rangle \left\langle 11 \right\vert
\right] \otimes \left\vert \tilde{\phi}\left( 1,0 \right) \right\rangle \left\langle \tilde{\phi}\left( 1,0 \right) \right\vert
\notag \\
+
& \frac{\mathcal{P}}{2}
\left[
\left( \lambda_{00}\lambda_{11} + \lambda_{11}\lambda_{00}  \right) \left\vert 00 \right\rangle \left\langle 00 \right\vert
+
\left( \lambda_{03}\lambda_{12} + \lambda_{12}\lambda_{03}  \right) \left\vert 00 \right\rangle \left\langle 11 \right\vert
\right.
\notag \\
& \phantom{[} + 
\left( \lambda_{00}\lambda_{33} + \lambda_{11}\lambda_{22}  \right) \left\vert 01 \right\rangle \left\langle 01 \right\vert
+
\left( \lambda_{03}\lambda_{30} + \lambda_{12}\lambda_{21}  \right) \left\vert 01 \right\rangle \left\langle 10 \right\vert
\notag \\
& \phantom{[} + 
\left( \lambda_{21}\lambda_{12} + \lambda_{30}\lambda_{03}  \right) \left\vert 10 \right\rangle \left\langle 01 \right\vert
+
\left( \lambda_{22}\lambda_{11} + \lambda_{33}\lambda_{00}  \right) \left\vert 10 \right\rangle \left\langle 10 \right\vert
\notag \\
& \phantom{[} + 
\left.
\left( \lambda_{21}\lambda_{30} + \lambda_{30}\lambda_{21}  \right) \left\vert 11 \right\rangle \left\langle 00 \right\vert
+
\left( \lambda_{22}\lambda_{33} + \lambda_{33}\lambda_{22}  \right) \left\vert 11 \right\rangle \left\langle 11 \right\vert
\right] \otimes \left\vert \tilde{\phi}\left( 0,1 \right) \right\rangle \left\langle \tilde{\phi}\left( 0,1 \right) \right\vert
\notag \\
+
& \frac{\mathcal{P}}{2}
\left[
\left( \lambda_{00}\lambda_{11} + \lambda_{11}\lambda_{00}  \right) \left\vert 00 \right\rangle \left\langle 00 \right\vert
-
\left( \lambda_{03}\lambda_{12} + \lambda_{12}\lambda_{03}  \right) \left\vert 00 \right\rangle \left\langle 11 \right\vert
\right.
\notag \\
& \phantom{[} + 
\left( \lambda_{00}\lambda_{33} + \lambda_{11}\lambda_{22}  \right) \left\vert 01 \right\rangle \left\langle 01 \right\vert
-
\left( \lambda_{03}\lambda_{30} + \lambda_{12}\lambda_{21}  \right) \left\vert 01 \right\rangle \left\langle 10 \right\vert
\notag \\
& \phantom{[} 
-
\left( \lambda_{21}\lambda_{12} + \lambda_{30}\lambda_{03}  \right) \left\vert 10 \right\rangle \left\langle 01 \right\vert
+
\left( \lambda_{22}\lambda_{11} + \lambda_{33}\lambda_{00}  \right) \left\vert 10 \right\rangle \left\langle 10 \right\vert
\notag \\
& \phantom{[} 
\left.
-
\left( \lambda_{21}\lambda_{30} + \lambda_{30}\lambda_{21}  \right) \left\vert 11 \right\rangle \left\langle 00 \right\vert
+
\left( \lambda_{22}\lambda_{33} + \lambda_{33}\lambda_{22}  \right) \left\vert 11 \right\rangle \left\langle 11 \right\vert
\right] \otimes \left\vert \tilde{\phi}\left( 1,1 \right) \right\rangle \left\langle \tilde{\phi}\left( 1,1 \right) \right\vert,
\end{align}
where
\begin{align}
& \mathcal{P} = \frac{1}{\mathcal{P}'},
\notag \\
& \mathcal{P}' = 
\lambda_{00}\lambda_{00} + \lambda_{11}\lambda_{11} + \lambda_{03}\lambda_{03} + \lambda_{12}\lambda_{12}
+
\lambda_{00}\lambda_{22} + \lambda_{11}\lambda_{33} + \lambda_{03}\lambda_{21} + \lambda_{12}\lambda_{30}
\notag \\
& \quad +
\lambda_{21}\lambda_{03} + \lambda_{30}\lambda_{12} + \lambda_{22}\lambda_{00} + \lambda_{33}\lambda_{11}
+
\lambda_{21}\lambda_{21} + \lambda_{30}\lambda_{30} + \lambda_{22}\lambda_{22} + \lambda_{33}\lambda_{33}
\notag \\
& \quad +
\lambda_{00}\lambda_{11} + \lambda_{11}\lambda_{00} + \lambda_{03}\lambda_{12} + \lambda_{12}\lambda_{03}
+
\lambda_{00}\lambda_{33} + \lambda_{11}\lambda_{22} + \lambda_{03}\lambda_{30} + \lambda_{12}\lambda_{21}
\notag \\
& \quad +
\lambda_{21}\lambda_{12} + \lambda_{30}\lambda_{03} + \lambda_{22}\lambda_{11} + \lambda_{33}\lambda_{00}
+
\lambda_{21}\lambda_{30} + \lambda_{30}\lambda_{21} + \lambda_{22}\lambda_{33} + \lambda_{33}\lambda_{22}.
\notag
\end{align}

\subsection{Entanglement swapping of mixed maximally entangled states}

\noindent
In what follows we consider the entanglement swapping of mixed the maximally entangled states, 
where the state has the form
\begin{gather}
\rho \left( u_1, u_2, \dots, u_m \right) = \sum_{u_1,u_2,\dots,u_m=0}^{d-1} \lambda_{u_1 u_2 \cdots u_m}
\left\vert\phi\left( u_1,u_2,\dots,u_m \right)\right\rangle \left\langle \phi\left( u_1,u_2,\dots,u_m \right) \right\vert,		\\
\phantom{i}
\sum_{u_1 u_2 \cdots u_m = 0}^{d-1} \lambda_{u_1 u_2 \cdots u_m} = 1.									\notag
\end{gather}
Let us assume that there are $n$ such mixed states containing
$m_1,m_2,\dots,m_n$ particles each, and denote them as
$\rho \left( u_1^1,u_2^1,\dots,u_{m_1}^1 \right),
\rho \left( u_1^2,u_2^2,\dots,u_{m_2}^2 \right),
\dots,
\rho \left( u_1^n,u_2^n,\dots,u_{m_n}^n \right)$ in turn.
As before, we will first consider the general entanglement swapping case realized by performing
$\mathrm{\widetilde{M}}$ on the first $k_r$ particles in $\rho \left( u_1^r,u_2^r,\dots,u_{m_r}^r \right)$ for $r = 1,2,\dots,t$
and the last $k_r$ particles in $\rho \left( u_1^r,u_2^r,\dots,u_{m_r}^r \right)$ for $r = t+1,t+2,\dots,n$.
We will then consider two extreme cases: measuring the first $k_r$ particles in all mixed states
and measuring the last $k_r$ particles.
Finally, we will provide entanglement swapping formulas for mixed d-level Bell states, 
which are special cases of the entanglement swapping mentioned above.

The first case, entanglement swapping between $n$ mixed the maximally entangled states, can be expressed as
\begin{align}
\label{mixed-maximally-states-ES}
& \bigotimes_{r=1}^{n} \rho \left( u_1^r,u_2^r,\dots,u_{m_r}^r \right) 
\notag \\
= & \frac {1}{d^n}  \sum_{u_1^1,u_2^1,\dots,u_{m_1}^1=0}^{d-1} \sum_{u_1^2,u_2^2,\dots,u_{m_1}^2=0}^{d-1} \cdots
\sum_{u_1^n,u_2^n,\dots,u_{m_n}^n=0}^{d-1} 
\sum_{l_1,l_2,\dots,l_n=0}^{d-1} 
\sum_{l_1',l_2',\dots,l_n'=0}^{d-1}
\prod_{r=1}^n \lambda_{u_1^r u_2^r \dots u_{m_r}^r}
\zeta^{\sum_{r=1}^{n} \left( l_r - l_r' \right) u_1^r}
\notag \\
& \bigotimes_{r=1}^{n} \left\vert l_r, l_r \oplus u_2^r,l_r \oplus u_3^r,\dots,l_r \oplus u_{m_r}^r \right\rangle
\left\langle l_r', l_r' \oplus u_2^r,l_r' \oplus u_3^r,\dots,l_r' \oplus u_{m_r}^r \right\vert
\notag \\
\Rightarrow 
& \frac{1}{d^n} \sum_{u_1^1,u_2^1,\dots,u_{m_1}^1=0}^{d-1} \sum_{u_1^2,u_2^2,\dots,u_{m_1}^2=0}^{d-1} \cdots
\sum_{u_1^n,u_2^n,\dots,u_{m_n}^n=0}^{d-1} 
\sum_{l_1,l_2,\dots,l_n=0}^{d-1} \sum_{l_1',l_2',\dots,l_n'=0}^{d-1}
\prod_{r=1}^n \lambda_{u_1^r u_2^r \dots u_{m_r}^r}
\zeta^{\sum_{r=1}^{n} \left( l_r - l_r' \right) u_1^r}
\notag \\
& \bigotimes_{r=1}^{t}
\left\vert l_r, l_r \oplus u_2^r,l_r \oplus u_3^r,\dots,l_r \oplus u_{k_r}^r \right\rangle 
\left\langle l_r', l_r' \oplus u_2^r,l_r' \oplus u_3^r,\dots,l_r' \oplus u_{k_r}^r \right\vert
\notag \\
&	
\bigotimes_{r=t+1}^n \left\vert l_r \oplus u_{m_r-k_r+1}^r, l_r \oplus u_{m_r-k_r+2}^r,\dots,l_r \oplus u_{m_r}^r \right\rangle
\left\langle l_r' \oplus u_{m_r-k_r+1}^r, l_r' \oplus u_{m_r-k_r+2}^r,\dots,l_r' \oplus u_{m_r}^r \right\vert
\notag \\
&
\bigotimes_{r=1}^t
\left\vert l_r \oplus u_{k_r+1}^r,l_r \oplus u_{k_r+2}^r,\dots,l_r \oplus u_{m_r}^r \right\rangle
\left\langle l_r' \oplus u_{k_r+1}^r,l_r' \oplus u_{k_r+2}^r,\dots,l_r' \oplus u_{m_r}^r \right\vert
\notag \\
&
\bigotimes_{r=t+1}^n
\left\vert l_r,l_r \oplus u_2^r,l_r \oplus u_3^r,\dots,l_r \oplus u_{m_r-k_r}^r \right\rangle
\left\langle l_r',l_r' \oplus u_2^r,l_r' \oplus u_3^r,\dots,l_r' \oplus u_{m_r-k_r}^r \right\vert							\notag \\
= & \frac {\mathcal{P}}{d^{n+1}} 
\sum_{u_1^1,u_2^1,\dots,u_{m_1}^1=0}^{d-1} 
\sum_{u_1^2,u_2^2,\dots,u_{m_1}^2=0}^{d-1} \cdots
\sum_{u_1^n,u_2^n,\dots,u_{m_n}^n=0}^{d-1} 
\sum_{v_1^1,v_1^2,\dots,v_1^n=0}^{d-1}
\sum_{l,l'=0}^{d-1} 
\zeta^{\left( \sum_{r=1}^{n} u_1^r - v_1^1 \right) \left( l - l' \right)}
\prod_{r=1}^n \lambda_{u_1^r u_2^r \dots u_{m_r}^r}
\notag \\
& \times 
\left\vert \phi \left(v_1^1,v_2^1,\dots,v_{k_1}^1,v_1^2,v_2^2,\dots,v_{k_2}^2,\dots,v_1^n,v_2^n,\dots,v_{k_n}^n \right)\right\rangle
\left\langle \phi \left(v_1^1,v_2^1,\dots,v_{k_1}^1,v_1^2,v_2^2,\dots,v_{k_2}^2,\dots,v_1^n,v_2^n,\dots,v_{k_n}^n \right)\right\vert 
\notag \\
& \otimes
\left\vert l \oplus u_{k_1+1}^1,l \oplus u_{k_1+2}^1,\dots,l \oplus u_{m_1}^1,
l \oplus v_1^2 \oplus u_{k_2+1}^2,l \oplus v_1^2 \oplus u_{k_2+2}^2,\dots,l \oplus v_1^2 \oplus u_{m_2}^2,
\right.
\notag \\
&
l \oplus v_1^3 \oplus u_{k_3+1}^3,l \oplus v_1^3 \oplus u_{k_3+2}^3,\dots,l \oplus v_1^3 \oplus u_{m_3}^3,\dots,
l \oplus v_1^t \oplus u_{k_t+1}^t,l \oplus v_1^t \oplus u_{k_t+2}^t,\dots,l \oplus v_1^t \oplus u_{m_t}^t,
\notag \\
& \dots,
l \oplus v_1^{t+1} \ominus u_{m_{t+1} - k_{t+1} + 1}^{t+1},
l \oplus v_1^{t+1} \ominus u_{m_{t+1} - k_{t+1} + 1}^{t+1} \oplus u_2^{t+1},
l \oplus v_1^{t+1} \ominus u_{m_{t+1} - k_{t+1} + 1}^{t+1} \oplus u_3^{t+1},
\notag \\
& \dots,
l \oplus v_1^{t+1} \ominus u_{m_{t+1} - k_{t+1} + 1}^{t+1} \oplus u_{m_{t+1} - k_{t+1}}^{t+1},
l \oplus v_1^{t+2} \ominus u_{m_{t+2} - k_{t+2} + 1}^{t+2},
l \oplus v_1^{t+2} \ominus u_{m_{t+2} - k_{t+2} + 1}^{t+2} \oplus u_2^{t+2},
\notag \\
&
l \oplus v_1^{t+2} \ominus u_{m_{t+2} - k_{t+2} + 1}^{t+2} \oplus u_3^{t+2},
\dots,
l \oplus v_1^{t+2} \ominus u_{m_{t+2} - k_{t+2} + 1}^{t+2} \oplus u_{m_{t+2} - k_{t+2}}^{t+2},
\dots,
l \oplus v_1^n \ominus u_{m_n - k_n + 1}^n,
\notag \\
& \left.
l \oplus v_1^n \ominus u_{m_n - k_n + 1}^n \oplus u_2^n,
l \oplus v_1^n \ominus u_{m_n - k_n + 1}^n \oplus u_3^n,
\dots,
l \oplus v_1^n \ominus u_{m_n - k_n + 1}^n \oplus u_{m_n - k_n}^n
\right\rangle
\notag \\
&
\times \left\langle 
l' \oplus u_{k_1+1}^1,l' \oplus u_{k_1+2}^1,\dots,l' \oplus u_{m_1}^1,
l' \oplus v_1^2 \oplus u_{k_2+1}^2,l' \oplus v_1^2 \oplus u_{k_2+2}^2,\dots,l' \oplus v_1^2 \oplus u_{m_2}^2,
\right.
\notag \\
&
l' \oplus v_1^3 \oplus u_{k_3+1}^3,l' \oplus v_1^3 \oplus u_{k_3+2}^3,\dots,l' \oplus v_1^3 \oplus u_{m_3}^3,\dots
l' \oplus v_1^t \oplus u_{k_t+1}^t,l' \oplus v_1^t \oplus u_{k_t+2}^t,\dots,l' \oplus v_1^t \oplus u_{m_t}^t,
\notag \\
& \dots,
l' \oplus v_1^{t+1} \ominus u_{m_{t+1} - k_{t+1} + 1}^{t+1},
l' \oplus v_1^{t+1} \ominus u_{m_{t+1} - k_{t+1} + 1}^{t+1} \oplus u_2^{t+1},
l' \oplus v_1^{t+1} \ominus u_{m_{t+1} - k_{t+1} + 1}^{t+1} \oplus u_3^{t+1},
\notag \\
& \dots,
l' \oplus v_1^{t+1} \ominus u_{m_{t+1} - k_{t+1} + 1}^{t+1} \oplus u_{m_{t+1} - k_{t+1}}^{t+1},
l' \oplus v_1^{t+2} \ominus u_{m_{t+2} - k_{t+2} + 1}^{t+2},
l' \oplus v_1^{t+2} \ominus u_{m_{t+2} - k_{t+2} + 1}^{t+2} \oplus u_2^{t+2},
\notag \\
&
l' \oplus v_1^{t+2} \ominus u_{m_{t+2} - k_{t+2} + 1}^{t+2} \oplus u_3^{t+2},
\dots,
l' \oplus v_1^{t+2} \ominus u_{m_{t+2} - k_{t+2} + 1}^{t+2} \oplus u_{m_{t+2} - k_{t+2}}^{t+2},
\dots,
l' \oplus v_1^n \ominus u_{m_n - k_n + 1}^n,
\notag \\
& \left.
l' \oplus v_1^n \ominus u_{m_n - k_n + 1}^n \oplus u_2^n,
l' \oplus v_1^n \ominus u_{m_n - k_n + 1}^n \oplus u_3^n,
\dots,
l' \oplus v_1^n \ominus u_{m_n - k_n + 1}^n \oplus u_{m_n - k_n}^n
\right\vert,
\end{align}
where $v_j^i$ satisfy Eq. \ref{general-maximally-ES-conditions} for $i =1,2,\dots,n$ and $j =2,3,\dots,n$, and
\begin{gather}
\mathcal{P} = 
\frac{1}{d
\sum_{u_1^1,u_2^1,\dots,u_{m_1}^1=0}^{d-1} 
\sum_{u_1^2,u_2^2,\dots,u_{m_1}^2=0}^{d-1} 
\cdots
\sum_{u_1^n,u_2^n,\dots,u_{m_n}^n=0}^{d-1}
\prod_{r=1}^n \lambda_{u_1^r u_2^r \dots u_{m_r}^r}}.
\label{general-case-probability}
\end{gather}

Next, let us characterize the two extreme cases mentioned above, the first case can be expressed as
\begin{align}
\label{mixed-maximally-states-ES-first-extreme-case}
& \bigotimes_{r=1}^{n} \rho \left( u_1^r,u_2^r,\dots,u_{m_r}^r \right) 
\notag \\
= & \frac{1}{d^n} \sum_{u_1^1,u_2^1,\dots,u_{m_1}^1=0}^{d-1} \sum_{u_1^2,u_2^2,\dots,u_{m_1}^2=0}^{d-1} \cdots
\sum_{u_1^n,u_2^n,\dots,u_{m_n}^n=0}^{d-1} 
\sum_{l_1,l_2,\dots,l_n=0}^{d-1} 
\sum_{l_1',l_2',\dots,l_n'=0}^{d-1}
\prod_{r=1}^n \lambda_{u_1^r u_2^r \dots u_{m_r}^r}
\zeta^{\sum_{r=1}^{n} \left( l_r - l_r' \right) u_1^r}
\notag \\
& \bigotimes_{r=1}^{n} \left\vert l_r, l_r \oplus u_2^r,l_r \oplus u_3^r,\dots,l_r \oplus u_{m_r}^r \right\rangle
\left\langle l_r', l_r' \oplus u_2^r,l_r' \oplus u_3^r,\dots,l_r' \oplus u_{m_r}^r \right\vert
\notag \\
\Rightarrow 
& \frac{1}{d^n} \sum_{u_1^1,u_2^1,\dots,u_{m_1}^1=0}^{d-1} \sum_{u_1^2,u_2^2,\dots,u_{m_1}^2=0}^{d-1} \cdots
\sum_{u_1^n,u_2^n,\dots,u_{m_n}^n=0}^{d-1} 
\sum_{l_1,l_2,\dots,l_n=0}^{d-1} \sum_{l_1',l_2',\dots,l_n'=0}^{d-1}
\prod_{r=1}^n \lambda_{u_1^r u_2^r \dots u_{m_r}^r}
\zeta^{\sum_{r=1}^{n} \left( l_r - l_r' \right) u_1^r}
\notag \\
& \bigotimes_{r=1}^n
\left\vert l_r, l_r \oplus u_2^r,l_r \oplus u_3^r,\dots,l_r \oplus u_{k_r}^r \right\rangle
\left\langle l_r', l_r' \oplus u_2^r,l_r' \oplus u_3^r,\dots,l_r' \oplus u_{k_r}^r \right\vert	
\notag \\
& \bigotimes_{r=1}^n
\left\vert l_r \oplus u_{k_r+1}^r,l_r \oplus u_{k_r+2}^r,\dots,l_r \oplus u_{m_r}^r \right\rangle
\left\langle l_r' \oplus u_{k_r+1}^r,l_r' \oplus u_{k_r+2}^r,\dots,l_r' \oplus u_{m_r}^r \right\vert						\notag \\
= & \frac {\mathcal{P}}{d^{n+1}} 
\sum_{u_1^1,u_2^1,\dots,u_{m_1}^1=0}^{d-1} 
\sum_{u_1^2,u_2^2,\dots,u_{m_1}^2=0}^{d-1} \cdots
\sum_{u_1^n,u_2^n,\dots,u_{m_n}^n=0}^{d-1} 
\sum_{v_1^1,v_1^2,\dots,v_1^n=0}^{d-1}
\sum_{l,l'=0}^{d-1} 
\zeta^{\left( \sum_{r=1}^{n} u_1^r - v_1^1 \right) \left( l - l' \right)}
\prod_{r=1}^n \lambda_{u_1^r u_2^r \dots u_{m_r}^r}
\notag \\
& \times 
\left\vert 
\phi \left(v_1^1,v_2^1,\dots,v_{k_1}^1,v_1^2,v_2^2,\dots,v_{k_2}^2,\dots,v_1^n,v_2^n,\dots,v_{k_n}^n \right)\right\rangle
\left\langle \phi \left(v_1^1,v_2^1,\dots,v_{k_1}^1,
v_1^2,v_2^2,\dots,v_{k_2}^2,\dots,v_1^n,v_2^n,\dots,v_{k_n}^n \right)\right\vert 
\notag \\
& \otimes
\left\vert l \oplus u_{k_1+1}^1,l \oplus u_{k_1+2}^1,\dots,l \oplus u_{m_1}^1,
l \oplus v_1^2 \oplus u_{k_2+1}^2,l \oplus v_1^2 \oplus u_{k_2+2}^2,\dots,l \oplus v_1^2 \oplus u_{m_2}^2,
\right.
\notag \\
& \left. 
l \oplus v_1^3 \oplus u_{k_3+1}^3,l \oplus v_1^3 \oplus u_{k_3+2}^3,\dots,l \oplus v_1^3 \oplus u_{m_3}^3,\dots,
l \oplus v_1^n \oplus u_{k_n+1}^n,l \oplus v_1^n \oplus u_{k_n+2}^n,\dots,l \oplus v_1^n \oplus u_{m_n}^n
\right\rangle
\notag \\
&
\times \left\langle
l' \oplus u_{k_1+1}^1,l' \oplus u_{k_1+2}^1,\dots,l' \oplus u_{m_1}^1,
l' \oplus v_1^2 \oplus u_{k_2+1}^2,l' \oplus v_1^2 \oplus u_{k_2+2}^2,\dots,l' \oplus v_1^2 \oplus u_{m_2}^2,
\right.
\notag \\
& \left. 
l' \oplus v_1^3 \oplus u_{k_3+1}^3,l' \oplus v_1^3 \oplus u_{k_3+2}^3,\dots,l' \oplus v_1^3 \oplus u_{m_3}^3,\dots,
l' \oplus v_1^n \oplus u_{k_n+1}^n,l' \oplus v_1^n \oplus u_{k_n+2}^n,\dots,l' \oplus v_1^n \oplus u_{m_n}^n
\right\vert,
\end{align}
where $v_j^i$ satisfy Eq. \ref{first-extreme-case-conditions} for $i =1,2,\dots,n$ and $j =2,3,\dots,n$, and
$\mathcal{P}$ satisfy Eq. \ref{general-case-probability}.
The second extreme case can be expressed as
\begin{align}
\label{mixed-maximally-states-ES-second-extreme-case}
& \bigotimes_{r=1}^{n} \rho \left( u_1^r,u_2^r,\dots,u_{m_r}^r \right) 
\notag \\
= & \frac{1}{d^n} \sum_{u_1^1,u_2^1,\dots,u_{m_1}^1=0}^{d-1} \sum_{u_1^2,u_2^2,\dots,u_{m_1}^2=0}^{d-1} \cdots
\sum_{u_1^n,u_2^n,\dots,u_{m_n}^n=0}^{d-1}
\sum_{l_1,l_2,\dots,l_n=0}^{d-1}
\sum_{l_1',l_2',\dots,l_n'=0}^{d-1}
\prod_{r=1}^n \lambda_{u_1^r u_2^r \dots u_{m_r}^r}
\zeta^{\sum_{r=1}^{n} \left( l_r - l_r' \right) u_1^r}
\notag \\
& \bigotimes_{r=1}^{n} \left\vert l_r, l_r \oplus u_2^r,l_r \oplus u_3^r,\dots,l_r \oplus u_{m_r}^r \right\rangle
\left\langle l_r', l_r' \oplus u_2^r,l_r' \oplus u_3^r,\dots,l_r' \oplus u_{m_r}^r \right\vert
\notag \\
\Rightarrow 
& \frac{1}{d^n}
\sum_{u_1^1,u_2^1,\dots,u_{m_1}^1=0}^{d-1} 
\sum_{u_1^2,u_2^2,\dots,u_{m_1}^2=0}^{d-1} \cdots
\sum_{u_1^n,u_2^n,\dots,u_{m_n}^n=0}^{d-1} 
\sum_{l_1,l_2,\dots,l_n=0}^{d-1} \sum_{l_1',l_2',\dots,l_n'=0}^{d-1}
\prod_{r=1}^n \lambda_{u_1^r u_2^r \dots u_{m_r}^r}
\zeta^{\sum_{r=1}^{n} \left( l_r - l_r' \right) u_1^r}
\notag \\
& \bigotimes_{r=1}^n
\left\vert l_r, l_r \oplus u_2^r,l_r \oplus u_3^r,\dots,l_r \oplus u_{m_r - k_r}^r \right\rangle
\left\langle l_r', l_r' \oplus u_2^r,l_r' \oplus u_3^r,\dots,l_r' \oplus u_{m_r - k_r}^r \right\vert
\notag \\
& \bigotimes_{r=1}^n
\left\vert l_r \oplus u_{m_r - k_r + 1}^r,l_r \oplus u_{m_r - k_r + 2}^r,\dots,l_r \oplus u_{m_r}^r \right\rangle
\left\langle l_r' \oplus u_{m_r - k_r + 1}^r,l_r' \oplus u_{m_r - k_r + 2}^r,\dots,l_r' \oplus u_{m_r}^r \right\vert 		\notag \\
= & \frac {\mathcal{P}}{d^{n+1}}
\sum_{u_1^1,u_2^1,\dots,u_{m_1}^1=0}^{d-1} 
\sum_{u_1^2,u_2^2,\dots,u_{m_1}^2=0}^{d-1} \cdots
\sum_{u_1^n,u_2^n,\dots,u_{m_n}^n=0}^{d-1} 
\sum_{v_1^1,v_1^2,\dots,v_1^n=0}^{d-1}
\sum_{l,l'=0}^{d-1} 
\zeta^{\left( \sum_{r=1}^{n} u_1^r - v_1^1 \right) \left( l - l' \right)}
\prod_{r=1}^n \lambda_{u_1^r u_2^r \dots u_{m_r}^r}
\notag \\
& \otimes
\left\vert 
l,l \oplus u_2^1,l \oplus u_3^1,\dots,l \oplus u_{m_1 - k_1}^1,
l \oplus v_1^2 \oplus u_{m_1 + k_1 + 1}^1 \ominus u_{m_2 - k_2 +1}^2,
l \oplus v_1^2 \oplus u_{m_1 + k_1 + 1}^1 \ominus u_{m_2 - k_2 +1}^2 \oplus u_2^2,
\right.
\notag \\
&
l \oplus v_1^2 \oplus u_{m_1 + k_1 + 1}^1 \ominus u_{m_2 - k_2 +1}^2 \oplus u_3^2,
\dots,
l \oplus v_1^2 \oplus u_{m_1 + k_1 + 1}^1 \ominus u_{m_2 - k_2 +1}^2 \oplus u_{m_2-k_2}^2,
\dots,
l \oplus v_1^n \oplus u_{m_1 + k_1 + 1}^1 \ominus u_{m_n - k_n +1}^n,
\notag \\
& \left.
l \oplus v_1^n \oplus u_{m_1 + k_1 + 1}^1 \ominus u_{m_n - k_n +1}^n \oplus u_2^n,
l \oplus v_1^n \oplus u_{m_1 + k_1 + 1}^1 \ominus u_{m_n - k_n +1}^n \oplus u_3^n,
\dots,
l \oplus v_1^n \oplus u_{m_1 + k_1 + 1}^1 \ominus u_{m_n - k_n +1}^n \oplus u_{m_n-k_n}^n
\right\rangle
\notag \\
\times &
\left\langle
l',l' \oplus u_2^1,l' \oplus u_3^1,\dots,l' \oplus u_{m_1 - k_1}^1,
l' \oplus v_1^2 \oplus u_{m_1 + k_1 + 1}^1 \ominus u_{m_2 - k_2 +1}^2,
l' \oplus v_1^2 \oplus u_{m_1 + k_1 + 1}^1 \ominus u_{m_2 - k_2 +1}^2 \oplus u_2^2,
\right.
\notag \\
&
l' \oplus v_1^2 \oplus u_{m_1 + k_1 + 1}^1 \ominus u_{m_2 - k_2 +1}^2 \oplus u_3^2,
\dots,
l' \oplus v_1^2 \oplus u_{m_1 + k_1 + 1}^1 \ominus u_{m_2 - k_2 +1}^2 \oplus u_{m_2-k_2}^2,
\dots,
l' \oplus v_1^n \oplus u_{m_1 + k_1 + 1}^1 \ominus u_{m_n - k_n +1}^n,
\notag \\
& \left.
l' \oplus v_1^n \oplus u_{m_1 + k_1 + 1}^1 \ominus u_{m_n - k_n +1}^n \oplus u_2^n,
l' \oplus v_1^n \oplus u_{m_1 + k_1 + 1}^1 \ominus u_{m_n - k_n +1}^n \oplus u_3^n,
\dots,
l' \oplus v_1^n \oplus u_{m_1 + k_1 + 1}^1 \ominus u_{m_n - k_n +1}^n \oplus u_{m_n-k_n}^n
\right\vert
\notag \\
\otimes & 
\left\vert \phi \left(v_1^1,v_2^1,\dots,v_{k_1}^1,v_1^2,v_2^2,\dots,v_{k_2}^2,\dots,v_1^n,v_2^n,\dots,v_{k_n}^n \right)\right\rangle
\left\langle \phi \left(v_1^1,v_2^1,\dots,v_{k_1}^1,v_1^2,v_2^2,\dots,v_{k_2}^2,\dots,v_1^n,v_2^n,\dots,v_{k_n}^n \right)\right\vert,
\end{align}
where $v_j^i$ satisfy Eq. \ref{first-extreme-case-conditions} for $i =1,2,\dots,n$ and $j =2,3,\dots,n$, 
and $\mathcal{P}$ also satisfies Eq. \ref{general-case-probability}.

Finally, the special cases mentioned above, the entanglement swapping of the mixed d-level Bell states,
can be described as
\begin{align}
& \bigotimes_{r=1}^{n} \rho \left( u_1^r,u_2^r \right)
\notag \\
\rightarrow & \frac {\mathcal{P}}{d^{n+1}}
\sum_{u_1^1,u_2^1,u_1^2,u_2^2,\dots,u_1^n,u_2^n=0}^{d-1}
\sum_{v_1,v_2,\dots,v_n=0}^{d-1}
\sum_{l,l'=0}^{d-1}
\zeta^{\left( \sum_{r=1}^{n} u_1^r - v_1 \right) \left( l - l' \right)}
\prod_{r=1}^n \lambda_{u_1^r u_2^r \dots u_{m_r}^r}
\left\vert \phi \left( v_1,v_2,\dots,v_n \right) \right\rangle
\left\langle \phi \left( v_1,v_2,\dots,v_n \right)\right\vert
\notag \\
& \otimes 
\left\vert l \oplus u_2^1,l \oplus v_2 \oplus u_2^2,l \oplus v_3 \oplus u_2^3,\dots,l \oplus v_t \oplus u_2^t,
l \oplus v_{t+1} \ominus u_2^{t+1},l \oplus v_{t+2} \ominus u_2^{t+2},\dots,l \oplus v_n \ominus u_2^n
\right\rangle
\notag \\
& \times \left\langle l' \oplus u_2^1,l' \oplus v_2 \oplus u_2^2,l' \oplus v_3 \oplus u_2^3,\dots,l' \oplus v_t \oplus u_2^t,
l' \oplus v_{t+1} \ominus u_2^{t+1},l' \oplus v_{t+2} \ominus u_2^{t+2},\dots,l' \oplus v_n \ominus u_2^n \right\vert,
\end{align}
\begin{align}
& \bigotimes_{r=1}^{n} \rho \left( u_1^r,u_2^r \right)
\notag \\
\rightarrow & \frac {\mathcal{P}}{d^{n+1}}
\sum_{u_1^1,u_2^1,u_1^2,u_2^2,\dots,u_1^n,u_2^n=0}^{d-1}
\sum_{v_1,v_2,\dots,v_n=0}^{d-1}
\sum_{l,l'=0}^{d-1}
\zeta^{\left( \sum_{r=1}^{n} u_1^r - v_1 \right) \left( l - l' \right)}
\prod_{r=1}^n \lambda_{u_1^r u_2^r \dots u_{m_r}^r}
\left\vert \phi \left( v_1,v_2,\dots,v_n \right) \right\rangle
\left\langle \phi \left( v_1,v_2,\dots,v_n \right)\right\vert
\notag \\
&
\otimes 
\left\vert l \oplus u_2^1,l \oplus v_2 \oplus u_2^2,l \oplus v_3 \oplus u_2^3,\dots,l \oplus v_n \oplus u_2^n \right\rangle
\left\langle l' \oplus u_2^1,l' \oplus v_2 \oplus u_2^2,l' \oplus v_3 \oplus u_2^3,\dots,l' \oplus v_n \oplus u_2^n \right\vert,
\end{align}
\begin{align}
& \bigotimes_{r=1}^{n} \rho \left( u_1^r,u_2^r \right)
\notag \\
\rightarrow & \frac {\mathcal{P}}{d^{n+1}}
\sum_{u_1^1,u_2^1,u_1^2,u_2^2,\dots,u_1^n,u_2^n=0}^{d-1}
\sum_{v_1,v_2,\dots,v_n=0}^{d-1}
\sum_{l,l'=0}^{d-1}
\zeta^{\left( \sum_{r=1}^{n} u_1^r - v_1 \right) \left( l - l' \right)}
\prod_{r=1}^n \lambda_{u_1^r u_2^r \dots u_{m_r}^r}
\left\vert \phi \left( v_1,v_2,\dots,v_n \right) \right\rangle
\left\langle \phi \left( v_1,v_2,\dots,v_n \right)\right\vert
\notag \\
& \otimes 
\left\vert l,l \oplus v_2 \oplus u_2^1 \ominus u_2^2,l \oplus v_3 \oplus u_2^1 \ominus u_3^2,\dots,
l \oplus v_n \oplus u_2^1 \ominus u_2^n \right\rangle
\left\langle l',l' \oplus v_2 \oplus u_2^1 \ominus u_2^2,l' \oplus v_3 \oplus u_2^1 \ominus u_3^2,
\dots
\right.
\notag \\
& \left.
\dots,l' \oplus v_n \oplus u_2^1 \ominus u_2^n \right\vert,
\end{align}
where all the $\mathcal{P}$s in the above three formulas are the same and satisfy Eq. \ref{general-case-probability}.

\section{Entanglement swapping chains}

\noindent
Entanglement swapping chain is a key support for achieving remote quantum communication and building quantum networks.
It is an application of the entanglement swapping between two entangled states,
which is realized by by performing quantum measurements at all the intermediate locations of 
more than two entangled states \cite{JiZX5852022,HardyL6252000}.
Suppose that there are $n$ entangled states, $\left\{ \rho_{2r-1,2r} \right\}_{r=1}^n$,
where the subscripts $(2r-1,2r)$ indicate two particles in each states,
such that a entanglement swapping chain can be realized by
performing the measurement operator $\mathrm{\widehat{M}}$ on the particles $(2,3),(4,5),\dots,(2n-2,2n-1)$ in turn.
Below we will formulate the entanglement swapping chains for pure states and mixed states, respectively.
For the former, as before, we will first consider generalized pure states,
and then characterize some special cases, including the entanglement swapping chain for maximally entangled states. 
For the latter, we will only consider mixed maximally entangled states.

\subsection{Entanglement swapping chains for pure states}

Let us begin with the formulation of the entanglement swapping chains for generalized pure states.
As the entanglement swapping shown in Eq. \ref{generalized-pure-states-ES},
let us assume that there are $n$ generalized pure states,
$\left\{ \left\vert \mathscr{P}_{pure}^r \right\rangle \right\}_{r=1}^n$,
then performing $\mathrm{\widehat{M}}$ on the particles $(2,3),(4,5),\dots,(2n-2,2n-1)$ in turn and 
marking the measurement results as
$\left\vert \phi \left( \hat{v}_1^1,\hat{v}_2^1 \right) \right\rangle, \left\vert \phi \left( \hat{v}_1^2,\hat{v}_2^2 \right) \right\rangle,
\dots, \left\vert \phi \left( \hat{v}_1^{n-1},\hat{v}_2^{n-1} \right) \right\rangle$ respectively,
will eventually make the particles 1 and $2n$ collapse onto
\begin{gather}
\label{generalized-pure-states-ES-chain}
\frac{\mathcal{P}}{\sqrt{d^{n-1}}}
\sum_{l_1^1,l_2^1,l_2^2,\dots,l_2^n=0}^{d-1}
\lambda_{l_1^1,l_2^1} \prod_{r=2}^n  \lambda_{l_2^{r-1} \ominus \hat{v}_2^{r-1},l_2^r}
\zeta^{-\sum_{r=1}^{n-1} \left( l_2^r - \hat{v}_2^r \right) \hat{v}_1^r} 
\left\vert l_1^1,l_2^n \right\rangle_{1,2n},
\\
\mathcal{P} =
\sqrt{\frac{d^{n-1}}{
\sum_{l_1^1,l_2^1,l_2^2,\dots,l_2^n=0}^{d-1}
\left\vert \lambda_{l_1^1,l_2^1} \prod_{r=2}^n  \lambda_{l_2^{r-1} \ominus \hat{v}_2^{r-1},l_2^r} \right\vert^2}}. 	\notag
\end{gather}
This result can be proven through mathematical induction as follows:
\begin{proof}
The entanglement swapping result after the first measurement
which is performed on the particles (2,3), is given by
\begin{gather}
	\bigotimes_{r=1}^{2} \left\vert \mathscr{P}_{pure}^r \right\rangle_{2r-1,2r}
\rightarrow
	\frac{\mathcal{P}}{\sqrt d}
	\sum_{l_1^1,l_2^1,l_2^2=0}^{d-1}
	\sum_{v_1^1,v_2^1=0}^{d-1}
	\lambda_{l_1^1,l_2^1} \lambda_{l_2^1 \ominus v_2^1,l_2^2} 
	\zeta^{-\left( l_2^1-v_2^1 \right) v_1^1}
	\left\vert l_1^1,l_2^2 \right\rangle_{1,4}
	\left\vert \phi \left( v_1^1,v_2^1 \right) \right\rangle_{3,2},
\\
\mathcal{P} = \frac{1}{\sqrt{
	\sum_{l_1^1,l_2^1,l_2^2,v_2^1=0}^{d-1}
	\left\vert \lambda_{l_1^1,l_2^1} \lambda_{l_2^1 \ominus v_2^1,l_2^2}  \right\vert^2
	}}. 				\notag
\end{gather}
After the second measurement performed on the particles (4,5), one can arrive at
\begin{align}
&	\frac{\mathcal{P}}{\sqrt d}
	\sum_{l_1^1,l_2^1,l_2^2=0}^{d-1}
	\lambda_{l_1^1,l_2^1} \lambda_{l_2^1 \ominus \hat{v}_2^1,l_2^2} 
	\zeta^{-\left( l_2^1 - \hat{v}_2^1 \right) \hat{v}_1^1}
	\left\vert l_1^1,l_2^2 \right\rangle_{1,4}
\otimes
	\left\vert \mathscr{P}_{pure}^3 \right\rangle_{5,6}														\notag \\
= &	\frac{\mathcal{P}}{\sqrt d}
	\sum_{l_1^1,l_2^1,l_2^2=0}^{d-1}
	\lambda_{l_1^1,l_2^1} \lambda_{l_2^1 \ominus \hat{v}_2^1,l_2^2} \lambda_{l_1^3,l_2^3}
	\zeta^{-\left( l_2^1 - \hat{v}_2^1 \right) \hat{v}_1^1}
	\left\vert l_1^1,l_2^2 \right\rangle_{1,4}
	\left\vert l_1^3, l_2^3 \right\rangle_{5,6} 			\notag \\
\Rightarrow
& \frac{\mathcal{P}}{\sqrt d}
	\sum_{l_1^1,l_2^1,l_2^2=0}^{d-1}
	\lambda_{l_1^1,l_2^1} \lambda_{l_2^1 \ominus \hat{v}_2^1,l_2^2} \lambda_{l_1^3,l_2^3}
	\zeta^{-\left( l_2^1 - \hat{v}_2^1 \right) \hat{v}_1^1}
	\left\vert l_1^1,l_2^3 \right\rangle_{1,6}
	\left\vert l_1^3, l_2^2 \right\rangle_{5,4} 													\notag \\
= & \frac{\widehat{\mathcal{P}}}{d}
	\sum_{l_1^1,l_2^1,l_2^2,l_2^3=0}^{d-1}
	\sum_{v_1^2,v_2^2=0}^{d-1}
	\lambda_{l_1^1,l_2^1} \lambda_{l_2^1 \ominus \hat{v}_2^1,l_2^2} \lambda_{l_2^2 \ominus v_2^2,l_2^3}
	\zeta^{-\left( l_2^1 - \hat{v}_2^1 \right) \hat{v}_1^1 - \left( l_2^2 - v_2^2 \right) v_1^2}
	\left\vert l_1^1,l_2^3 \right\rangle_{1,6}
	\left\vert \phi \left( v_1^2,v_2^2 \right) \right\rangle_{5,4},									
\end{align}
where
\begin{gather}
\mathcal{P} = 
\sqrt{\frac{d}{
\sum_{l_1^1,l_2^1,l_2^2=0}^{d-1}
\left\vert \lambda_{l_1^1,l_2^1} \lambda_{l_2^1 \ominus \hat{v}_2^1,l_2^2} \right\vert^2}},
\quad
\widehat{\mathcal{P}} = 
\sqrt{\frac{d}{
\sum_{l_1^1,l_2^1,l_2^2,l_2^3,v_2^2=0}^{d-1}
\left\vert \lambda_{l_1^1,l_2^1} \lambda_{l_2^1 \ominus \hat{v}_2^1,l_2^2} \lambda_{l_2^2 \ominus v_2^2,l_2^3} \right\vert^2}}. 	\notag
\end{gather}
It can be seen that in these two entanglement swapping cases, both the states that the unmeasured particles collapse onto
satisfy Eq. \ref{generalized-pure-states-ES-chain}.
Assume that the entanglement swapping result after the $(n-2)$-th measurement is consistent
with Eq. \ref{generalized-pure-states-ES-chain}, then the entanglement swapping result
after the final measurement performed on the particles (2n-2,2n-1) is given by
\begin{align}
&	\frac{\mathcal{P}}{\sqrt{d^{n-2}}}
	\sum_{l_1^1,l_2^1,l_2^2,\dots,l_2^{n-1}=0}^{d-1}
	\lambda_{l_1^1,l_2^1} \prod_{r=1}^{n-2}  \lambda_{l_2^{r} \ominus \hat{v}_2^r,l_2^{r+1}}
	\zeta^{-\sum_{r=1}^{n-2} \left( l_2^r - \hat{v}_2^r \right) v_1^r} 
	\left\vert l_1^1,l_2^{n-1} \right\rangle_{1,2n-2}
\otimes
	\left\vert \mathscr{P}_{pure}^n \right\rangle_{2n-1,2n}											\notag \\
\rightarrow
& \frac{\widehat{\mathcal{P}}}{\sqrt{d^{n-1}}}
	\sum_{l_1^1,l_2^1,l_2^2,\dots,l_2^n=0}^{d-1}
	\sum_{v_1^{n-1},v_2^{n-1}=0}^{d-1}
	\lambda_{l_1^1,l_2^1} \lambda_{l_2^{n-1} \ominus v_2^{n-1},l_2^n}
	\prod_{r=1}^{n-2}  \lambda_{l_2^r \ominus \hat{v}_2^r,l_2^{r+1}}
	\zeta^{-\sum_{r=1}^{n-1} \left( l_2^r - \hat{v}_2^r \right) \hat{v}_1^r} 
	\left\vert l_1^1,l_2^n \right\rangle_{1,2n}
	\left\vert \phi \left( v_1^{n-1},v_2^{n-1} \right) \right\rangle_{2n-1,2n-2},
\\ & 
\begin{centering}
\mathcal{P} =
\sqrt{\frac{d^{n-2}}{
\sum_{l_1^1,l_2^1,l_2^2,\dots,l_2^{n-1}=0}^{d-1}
\left\vert \lambda_{l_1^1,l_2^1} \prod_{r=2}^{n-1}  \lambda_{l_2^{r-1} \ominus \hat{v}_2^{r-1},l_2^r} \right\vert^2}},
\quad
\widehat{\mathcal{P}}= 
\sqrt{\frac{d^{n-2}}{
\sum_{l_2^1,l_2^2,\dots,l_2^n} \sum_{l_1^1,v_2^{n-1}=0}^{d-1}
\left\vert 
\lambda_{l_1^1,l_2^1} \lambda_{l_2^{n-1} \ominus v_2^{n-1},l_2^n}
\prod_{r=1}^{n-2}  \lambda_{l_2^r \ominus \hat{v}_2^r,l_2^{r+1}} 
\right\vert^2}}.	\notag
\end{centering}
\end{align}
from which one can get the state that is consistent with Eq. \ref{generalized-pure-states-ES-chain}. QED.

\end{proof}

We would like to employ mathematical induction, as above, to provide a proof for the formulas of
the entanglement swapping chains for general pure states, which was proposed in Ref. \cite{HardyL6252000}.
Suppose that measurement results are
$\left\vert \phi \left( \hat{v}_1^1,\hat{v}_2^1 \right) \right\rangle, \left\vert \phi \left( \hat{v}_1^2,\hat{v}_2^2 \right) \right\rangle,
\dots, \left\vert \phi \left( \hat{v}_1^{n-1},\hat{v}_2^{n-1} \right) \right\rangle$ respectively,
then the particles (1,2n) collapse onto
\begin{gather}
\label{general-pure-states-ES-chain}
\frac{\mathcal{P}}{\sqrt{d^{n-1}}}
\sum_{l_1=0}^{d-1}
\lambda_{l_1} \prod_{r=1}^{n-1} \lambda_{l_1 \ominus \sum_{i=1}^r \hat{v}_2^i}
\zeta^{- \sum_{r=1}^{n-1} \left( l_1 - \sum_{i=1}^r \hat{v}_2^i \right) \hat{v}_1^r} 
\left\vert l_1, l_1 \ominus \sum_{r=1}^{n-1} \hat{v}_2^r \right\rangle,	\\
\mathcal{P} =
\sqrt{\frac{d^{n-1}}{\sum_{l_1=0}^{d-1}
\left\vert \lambda_{l_1} \prod_{r=1}^{n-1} \lambda_{l_1 \ominus \sum_{i=1}^r \hat{v}_2^i} \right\vert^2}}. 	\notag
\end{gather}

\begin{proof}

Via the first measurement, the entanglement swapping result is
\begin{gather}
\frac{\mathcal{P}}{\sqrt{d}} \sum_{l_1,v_1^1,v_2^1=0}^{d-1}
\lambda_{l_1} \lambda_{l_1 \ominus v_2^1} \zeta^{-v_1^1 \left( l_1 - v_2^1 \right)}
\left\vert l_1,l_1 \ominus v_2^1 \right\rangle_{1,4} 
\left\vert \phi \left( v_1^1,v_2^1 \right) \right\rangle_{3,2},	\\
\mathcal{P} =
\frac{1}{\sqrt{\sum_{l_1,v_2^1=0}^{d-1} \left\vert \lambda_{l_1} \lambda_{l_1 \ominus v_2^1}^2 \right\vert^2}}. 	\notag
\end{gather}
After the second measurement, the entanglement swapping result is
\begin{gather}
\frac{\mathcal{P}}{d} \sum_{l_1,v_1^2,v_2^2=0}^{d-1}
\lambda_{l_1} \lambda_{l_1 \ominus \hat{v}_2^1} \lambda_{l_1 \ominus \hat{v}_2^1 \ominus v_2^2} 
\zeta^{-\hat{v}_1^1 \left( l_1 - \hat{v}_2^1 \right) - v_1^2 \left( l_1 - \hat{v}_2^1 - v_2^2\right)}
\left\vert l_1,l_1 \ominus \hat{v}_2^1 \ominus v_2^2 \right\rangle_{1,6} 
\left\vert \phi \left( v_1^2,v_2^2 \right) \right\rangle_{5,4},	\\
\mathcal{P} =
\sqrt{\frac{d}{\sum_{l_1,v_2^2=0}^{d-1}
\left\vert \lambda_{l_1} \lambda_{l_1 \ominus \hat{v}_2^1} \lambda_{l_1 \ominus \hat{v}_2^1 \ominus v_2^2}  \right\vert^2}}. 	\notag
\end{gather}
It can be seen that the entanglement swapping results after the first two measurements
are consistent with Eq. \ref{general-pure-states-ES-chain}.
Let us now assume that the entanglement swapping result after the $(n-2)$-th measurement satisfies
Eq. \ref{general-pure-states-ES-chain}, then the entanglement swapping via the final measurement is given by
\begin{align}
&	\frac{\mathcal{P}}{\sqrt{d^{n-2}}} 
\sum_{l_1=0}^{d-1}
\lambda_{l_1} \prod_{r=1}^{n-2} \lambda_{l_1 \ominus \sum_{i=1}^r \hat{v}_2^i}
\zeta^{- \sum_{r=1}^{n-2} \left( l_1 - \sum_{i=1}^r \hat{v}_2^i \right) \hat{v}_1^r}
\left\vert l_1, l_1 \ominus \sum_{r=1}^{n-2} \hat{v}_2^r \right\rangle_{1,2n-2}
\otimes
\left\vert \mathscr{P}_{\uppercase\expandafter{\romannumeral1}} \right\rangle_{2n-1,2n}
\notag \\
= & \frac{\mathcal{P}}{\sqrt{d^{n-2}}} 
\sum_{l_1=0}^{d-1}
\lambda_{l_1} \lambda_{l_n} \prod_{r=1}^{n-2} \lambda_{l_1 \ominus \sum_{i=1}^r \hat{v}_2^i}
\zeta^{- \sum_{r=1}^{n-2} \left( l_1 - \sum_{i=1}^r \hat{v}_2^i \right) \hat{v}_1^r}
\left\vert l_1, l_1 \ominus \sum_{r=1}^{n-2} \hat{v}_2^r \right\rangle_{1,2n-2}
\left\vert l_n, l_n \right\rangle_{2n-1,2n} 			\notag \\
\Rightarrow 
& \frac{\mathcal{P}}{\sqrt{d^{n-2}}} 
\sum_{l_1=0}^{d-1}
\lambda_{l_1} \lambda_{l_n} \prod_{r=1}^{n-2} \lambda_{l_1 \ominus \sum_{i=1}^r \hat{v}_2^i}
\zeta^{- \sum_{r=1}^{n-2} \left( l - \sum_{i=1}^r \hat{v}_2^i \right) \hat{v}_1^r}
\left\vert l_1, l_n \right\rangle_{1,2n}
\left\vert l_n, l_1 \ominus \sum_{r=1}^{n-2} \hat{v}_2^r \right\rangle_{2n-1,2n-2} 			
\notag \\
= & \frac{\widehat{\mathcal{P}}}{\sqrt{d^{n-1}}}
\sum_{l_1,v_1^{n-1},v_2^{n-1}=0}^{d-1}
\lambda_{l_1} \lambda_{l_1 \ominus \sum_{r=1}^{n-2} \hat{v}_2^r \ominus v_2^{n-1}} 
\prod_{r=1}^{n-2} \lambda_{l_1 \ominus \sum_{i=1}^r \hat{v}_2^i}
\zeta^{- \sum_{r=1}^{n-2} \left( l_1 - \sum_{i=1}^r \hat{v}_2^i \right) \hat{v}_1^r 
- \left( l_1 - \sum_{i=1}^{n-2} \hat{v}_2^i - v_2^{n-1}\right) v_1^{n-1}}
\notag \\
& \times
\left\vert l_1, l_1 \ominus \sum_{r=1}^{n-2} \hat{v}_2^r \ominus v_2^{n-1} \right\rangle_{1,2n}
\left\vert \phi \left( v_1^{n-1},v_2^{n-1} \right) \right\rangle_{2n-2,2n-1},											
\end{align}
where
\begin{gather}
\mathcal{P} =
\sqrt{\frac{d^{n-2}}{\sum_{l_1=0}^{d-1} 
\left\vert \lambda_{l_1} \prod_{r=1}^{n-2} \lambda_{l_1 \ominus \sum_{i=1}^r \hat{v}_2^i} \right\vert^2}},
\phantom{i}
\widehat{\mathcal{P}} =
\sqrt{\frac{d^{n-2}}{\sum_{l_1,v_2^{n-1}=0}^{d-1}
\left\vert \lambda_{l_1} \lambda_{l_1 \ominus \sum_{r=1}^{n-2} \hat{v}_2^r \ominus v_2^{n-1}} 
\prod_{r=1}^{n-2} \lambda_{l_1 \ominus \sum_{i=1}^r \hat{v}_2^i} \right\vert^2}}. 
\notag
\end{gather}
It can be seen that the state that the particles (1,2n) collapse onto
is consistent with the state shown in Eq. \ref{general-pure-states-ES-chain}. QED.

\end{proof}

Let us now formulate entanglement swapping chains for maximally entangled states.
Let us assume that there are $n$ d-level $(2m)$-particle maximally entangled states,
and mark them by
$\left\{ \left\vert \phi\left(u_1^r,u_2^r,\dots,u_{2m}^r \right)\right\rangle_{2r-1,2r} \right\}_{r=1}^n$
where the subscripts $1,3,5,\dots,2n-1$ and $2,4,6,\dots,2n$ are used to mark the first $m$ particles and
the last $m$ particles respectively. As above, assume that the measurement operator
$\mathrm{\widetilde{M}}$ is performed on the particles $(2,3),(4,5),\dots,(2n-2,2n-1)$ in turn,
such that the particles (1,2n) collapse onto
\begin{align}
\label{maximally-entangled-states-ES-chain}
&	\frac {1}{d^{n-1}} \sum_{v_1^1,v_1^2,\dots,v_1^{n-1}=0}^{d-1} 
\sum_{v_{m+1}^1,v_{m+1}^2,\dots,v_{m+1}^{n-1}=0}^{d-1}
\zeta^{\sum_{r=1}^{n-1} \left( u_1^{r+1} - v_1^{r} \right) \sum_{k=1}^{r} \left( u_{m+1}^k - v_{m+1}^k \right)}
\left\vert\phi
\left(
\oplus_{r=1}^n u_1^r \ominus \oplus_{r=1}^{n-1} v_1^r,u_2^1,u_3^1,\dots,u_{m}^1,
\right.\right.			\notag \\
&
\left.\left.
\oplus_{r=1}^{n-1} \left( u_{m+1}^r \ominus v_{m+1}^r \right) \oplus u_{m+1}^n,
\oplus_{r=1}^{n-1}\left( u_{m+1}^r \ominus v_{m+1}^r \right) \oplus u_{m+2}^n,
\dots,\oplus_{r=1}^{n-1} \left( u_{m+1}^r \ominus v_{m+1}^r \right) \oplus u_{2m}^n
\right)
\right\rangle_{1,2n},
\end{align}
and the measurement results are
\begin{gather}
\left\{ 
\left\vert \phi\left( v_1^r,v_2^r,\dots,v_{2m}^r \right)\right\rangle_{2r+1,2r} 
\right\}_{r=1}^{n-1},
\notag \\
v_i^r = u_i^{r+1}, \phantom{i} i =2,3,\dots,m,
\notag \\
v_j^r = v_{m+1}^r \ominus u_{m+1}^r \oplus u_j^r, \phantom{i} j = m+2,m+3,\dots,2m.
\label{maximally-entangled-states-ES-chain-conditions}
\end{gather}

\begin{proof}

The entanglement swapping result after the first measurement is given by
\begin{align}
&\left\vert\phi\left( u_1^1,u_2^1,\dots,u_{2m}^1 \right)\right\rangle_{1,2}
\otimes
\left\vert\phi\left( u_1^2,u_2^2,\dots,u_{2m}^2 \right)\right\rangle_{3,4} 
\notag \\
\rightarrow 
& \frac {1}{d} \sum_{v_1^1, v_{m+1}^1=0}^{d-1} 
\zeta^{\left( u_{m+1}^1 - v_{m+1}^1 \right) \left( u_1^2 - v_1^1 \right)}
\left\vert\phi \left( u_1^1 \oplus u_1^2 \ominus v_1^1,u_2^1,u_3^1,\dots,u_m^1,u_{m+1}^1 \oplus u_{m+1}^2 \ominus v_{m+1}^1,
u_{m+1}^1 \oplus u_{m+2}^2 \ominus v_{m+1}^1,\dots
\right.\right.													\notag \\
& \left.\left.
\dots, u_{m+1}^1 \oplus u_{2m}^2 \ominus v_{m+1}^1 \right) \right\rangle_{1,4}
\otimes 
\left\vert\phi\left( v_1^1,v_2^1,\dots,v_m^1,v_{m+1}^1,v_{m+2}^1,\dots,v_{2m}^1 \right)\right\rangle_{3,2},
\end{align}
where
\begin{gather}
v_i^1 = u_i^2, \phantom{i} i =2,3,\dots,m,
\notag \\
v_j^1 = v_{m+1}^1 \ominus u_{m+1}^1 \oplus u_j^1, \phantom{i} j = m+2,m+3,\dots,2m.
\notag
\end{gather}
The entanglement swapping result after the second measurement is given by
\begin{align}
& \frac {1}{d} \sum_{v_1^1, v_{m+1}^1=0}^{d-1} 
\zeta^{\left( u_{m+1}^1 - v_{m+1}^1 \right) \left( u_1^2 - v_1^1 \right)}
\left\vert\phi \left( u_1^1 \oplus u_1^2 \ominus v_1^1,u_2^1,u_3^1,\dots,u_m^1,u_{m+1}^1 \oplus u_{m+1}^2 \ominus v_{m+1}^1,
u_{m+1}^1 \oplus u_{m+2}^2 \ominus v_{m+1}^1,\dots
\right.\right.													\notag \\
& \left.\left.
\dots, u_{m+1}^1 \oplus u_{2m}^2 \ominus v_{m+1}^1 \right) \right\rangle_{1,4}
\otimes 
\left\vert\phi\left( u_1^3,u_2^3,\dots,u_{2m}^3 \right)\right\rangle_{5,6}
\notag \\
\rightarrow &
\frac {1}{d^2} \sum_{v_1^1,v_1^2,v_{m+1}^1,v_{m+1}^2 =0}^{d-1}
\zeta^{\sum_{r=1}^2 \left( u_1^{r+1} - v_1^{r} \right) \sum_{k=1}^{r} \left( u_{m+1}^k - v_{m+1}^k \right)}
\left\vert\phi \left( \oplus_{r=1}^2 u_1^r \ominus v_1^1,u_2^1,u_3^1,\dots,u_m^1,
u_{m+1}^1 \oplus u_{m+1}^2 \ominus v_{m+1}^1,
u_{m+1}^1 \oplus u_{m+2}^2 \ominus v_{m+1}^1,\dots, 
\right.\right.													\notag \\
& \left.\left.
u_{m+1}^1 \oplus u_{2m}^2 \ominus v_{m+1}^1 \right) \right\rangle_{1,6}
\otimes 
\left\vert\phi\left( v_1^1,v_2^1,\dots,v_{2m}^1 \right)\right\rangle_{5,4},
\end{align}
where
\begin{gather}
v_i^2 = u_i^3, \phantom{i} i =2,3,\dots,m,
\notag \\
v_j^2 = v_{m+1}^2 \ominus u_{m+1}^2 \oplus u_j^2, \phantom{i} j = m+2,m+3,\dots,2m.
\notag
\end{gather}
It can be seen that the above results satisfy Eq. \ref{maximally-entangled-states-ES-chain}.
Let us suppose that the entanglement swapping results after the $(n-1)$-th measurement satisfy 
Eq. \ref{maximally-entangled-states-ES-chain}, such that the result after the final measurement is given by
\begin{align}
&	\frac {1}{d^{n-2}} 
\sum_{v_1^1,v_1^2,\dots,v_1^{n-2}=0}^{d-1} 
\sum_{v_{m+1}^1,v_{m+1}^2,\dots,v_{m+1}^{n-2}=0}^{d-1}
\zeta^{\sum_{r=1}^{n-2} \left( u_1^{r+1} - v_1^{r} \right) \sum_{k=1}^{r} \left( u_{m+1}^k - v_{m+1}^k \right)}
\left\vert\phi
\left(
\oplus_{r=1}^{n-1} u_1^r \ominus \oplus_{r=1}^{n-2} v_1^r,u_2^1,u_3^1,\dots,u_m^1,
\right.\right.																						\notag \\
&
\left.\left.
\oplus_{r=1}^{n-2} \left( u_{m+1}^r \ominus v_{m+1}^r \right) \oplus u_{m+1}^{n-1},
\oplus_{r=1}^{n-2}\left( u_{m+1}^r \ominus v_{m+1}^r \right) \oplus u_{m+2}^{n-1},
\dots,\oplus_{r=1}^{n-2} \left( u_{m+1}^r \ominus v_{m+1}^r \right) \oplus u_{2m}^{n-1}
\right)
\right\rangle_{1,2n-2}
\notag \\
& \otimes
\left\vert \phi\left(u_1^n,u_2^n,\dots,u_{2m}^n \right)\right\rangle_{2n-1,2n} 										\notag \\
= & \frac {1}{d^{n-1}}
\sum_{v_1^1,v_1^2,\dots,v_1^{n-2}=0}^{d-1}
\sum_{v_{m+1}^1,v_{m+1}^2,\dots,v_{m+1}^{n-2}=0}^{d-1}
\sum_{l_1,l_2=0}^{d-1}
\zeta^{\sum_{r=1}^{n-2} \left( u_1^{r+1} - v_1^{r} \right) \sum_{k=1}^{r} \left( u_{m+1}^k - v_{m+1}^k \right)
+ l_1 \left( \sum_{r=1}^{n-1} u_1^r - \sum_{r=1}^{n-2} v_1^r \right) + l_2 u_1^n}
\notag \\
& \times
\left\vert l_1, l_1 \oplus u_2^1, l_1 \oplus u_3^1,\dots,l_r \oplus u_{m}^1,
l_1 \oplus \oplus_{r=1}^{n-2} \left( u_{m+1}^r \ominus v_{m+1}^r \right) \oplus u_{m+1}^{n-1},
l_1 \oplus \oplus_{r=1}^{n-2}\left( u_{m+1}^r \ominus v_{m+1}^r \right) \oplus u_{m+2}^{n-1},\dots,
\right.																							
\notag \\
&
\left.
l_1 \oplus	 \oplus_{r=1}^{n-2} \left( u_{m+1}^r \ominus v_{m+1}^r \right) \oplus u_{2m}^{n-1}
\right\rangle
\otimes
\left\vert l_2, l_2 \oplus u_2^n,l_2 \oplus u_3^n,\dots,l_2 \oplus u_{2m}^n \right\rangle							
\notag \\
\Rightarrow
& \frac {1}{d^{n-1}}
\sum_{v_1^1,v_1^2,\dots,v_1^{n-2}=0}^{d-1}
\sum_{v_{m+1}^1,v_{m+1}^2,\dots,v_{m+1}^{n-2}=0}^{d-1}
\sum_{l_1,l_2=0}^{d-1}
\zeta^{\sum_{r=1}^{n-2} \left( u_1^{r+1} - v_1^{r} \right) \sum_{k=1}^{r} \left( u_{m+1}^k - v_{m+1}^k \right)
+ l_1 \left( \sum_{r=1}^{n-1} u_1^r - \sum_{r=1}^{n-2} v_1^r \right) + l_2 u_1^n}
\notag \\
& \times
\left\vert l_1,l_1 \oplus u_2^1,l_1 \oplus u_3^1,\dots,l_1 \oplus u_m^1,l_2 \oplus u_{m+1}^n,l_2 \oplus u_{m+2}^n,
\dots,l_2 \oplus u_{2m}^n \right\rangle_{1,2n}															
\notag \\
& \otimes
\left\vert l_2,l_2 \oplus u_2^n,l_2 \oplus u_3^n,\dots,l_2 \oplus u_m^n, 
l_1 \oplus \oplus_{r=1}^{n-2} \left( u_{m+1}^r \ominus v_{m+1}^r \right) \oplus u_{m+1}^{n-1},
l_1 \oplus \oplus_{r=1}^{n-2}\left( u_{m+1}^r \ominus v_{m+1}^r \right) \oplus u_{m+2}^{n-1},\dots,				
\right.																							
\notag \\
&
\left.
\dots,l_1 \oplus	 \oplus_{r=1}^{n-2} \left( u_{m+1}^r \ominus v_{m+1}^r \right) \oplus u_{2m}^{n-1}
\right\rangle_{2n-1,2n-2}	 				\notag \\
= & \frac {1}{d^{n-1}} 
\sum_{v_1^1,v_1^2,\dots,v_1^{n-1}=0}^{d-1}
\sum_{v_{m+1}^1,v_{m+1}^2,\dots,v_{m+1}^{n-1}=0}^{d-1}
\zeta^{\sum_{r=1}^{n-1} \left( u_1^{r+1} - v_1^{r} \right) \sum_{k=1}^{r} \left( u_{m+1}^k - v_{m+1}^k \right)}
\left\vert\phi
\left(
\oplus_{r=1}^n u_1^r \ominus \oplus_{r=1}^{n-1} v_1^r,u_2^1,u_3^1,\dots,u_{m}^1,
\right.\right.			\notag \\
&
\left.\left.
\oplus_{r=1}^{n-1} \left( u_{m+1}^r \ominus v_{m+1}^r \right) \oplus u_{m+1}^n,
\oplus_{r=1}^{n-1}\left( u_{m+1}^r \ominus v_{m+1}^r \right) \oplus u_{m+2}^n,
\dots,\oplus_{r=1}^{n-1} \left( u_{m+1}^r \ominus v_{m+1}^r \right) \oplus u_{2m}^n
\right)
\right\rangle_{1,2n}
\notag \\
&
\otimes 
\left\vert \phi\left(v_1^{n-1},v_2^{n-1},\dots,v_{2m}^{n-1} \right)\right\rangle,
\end{align}
where 
\begin{gather}
v_i^{n-1} = u_i^n, \phantom{i} i =2,3,\dots,m,
\notag \\
v_j^{n-1} = v_{m+1}^{n-1} \ominus u_{m+1}^{n-1} \oplus u_j^{n-1}, \phantom{i} j = m+2,m+3,\dots,2m.
\notag
\end{gather}
It can be seen that the joint state of the particles (1,2n) is consistent with Eq. \ref{maximally-entangled-states-ES-chain}. QED.

\end{proof}

From Eq. \ref{maximally-entangled-states-ES-chain}, we can derive the formula for the special case:
the entanglement swapping chain for d-level Bell states. Suppose that the measurement results are
$\left\{ \left\vert \phi\left( v_1^r,v_2^r \right)\right\rangle_{2r+1,2r} \right\}_{r=1}^{n-1}$,
then the particles (1,2n) collapse onto
\begin{align}
\label{d-level-Bell-states-ES-chain}
\frac {1}{d^{n-1}} 
\sum_{v_1^1,v_1^2,\dots,v_1^{n-1}=0}^{d-1}
\sum_{v_{m+1}^1,v_{m+1}^2,\dots,v_{m+1}^{n-1}=0}^{d-1}
\zeta^{\sum_{r=1}^{n-1} \left( u_1^{r+1} - v_1^{r} \right) \sum_{k=1}^{r} \left( u_2^k - v_2^k \right)}
\left\vert\phi
\left(
\oplus_{r=1}^n u_1^r \ominus \oplus_{r=1}^{n-1} v_1^r,
\oplus_{r=1}^n u_2^r \ominus \oplus_{r=1}^{n-1} v_2^r
\right)
\right\rangle_{1,2n}.
\end{align}

We have characterized entanglement swapping chains, and provided proofs for corresponding results 
through mathematical induction. Obviously, it requires at least three steps for deriving the final entanglement swapping results.
Below we would like to give a formula by which one can directly get the results of entanglement swapping chains.
Let us first give the formula of the entanglement swapping chain for generalized pure states,
\begin{align}
&	\bigotimes_{r=1}^{n} \left\vert \mathscr{P}_{pure}^r \right\rangle_{2r-1,2r}
\notag \\
= & 	\sum_{l_1^1,l_2^1,l_1^2,l_2^2,\dots,l_1^n,l_2^n=0}^{d-1} \prod_{r=1}^{n} \lambda_{l_1^r,l_2^r}
	\bigotimes_{r=1}^{n} \left\vert l_1^r,l_2^r \right\rangle_{2r-1,2r} \notag \\
\Rightarrow &
	\sum_{l_1^1,l_2^1,l_1^2,l_2^2,\dots,l_1^n,l_2^n=0}^{d-1} \prod_{r=1}^{n} \lambda_{l_1^r,l_2^r}
	\left\vert l_1^1,l_2^n \right\rangle_{1,2n} \bigotimes_{r=2}^{n} \left\vert l_1^r,l_2^{r-1} \right\rangle_{2r-1,2r-2}		\notag \\
= &
\frac{\mathcal{P}}{\sqrt{d^{n-1}}}
\sum_{l_1^1,l_2^1,l_2^2,\dots,l_2^n=0}^{d-1}
\sum_{v_1^1,v_2^1,v_1^2,v_2^2,\dots,v_1^{n-1},v_2^{n-1}=0}^{d-1}
\lambda_{l_1^1,l_2^1} \prod_{r=1}^{n-1}  \lambda_{l_2^r \ominus v_2^r,l_2^{r+1}}
\zeta^{-\sum_{r=1}^{n-1} \left( l_2^r - v_2^r \right) v_1^r}
\left\vert l_1^1,l_2^n \right\rangle_{1,2n} \bigotimes_{r=1}^{n-1} \left\vert \phi\left(v_1^r,v_2^r \right)\right\rangle_{2r+1,2r},
\end{align}
where
\begin{gather}
\mathcal{P} = \frac{1}{\sqrt{
\sum_{l_1^1,l_2^1,l_2^2,\dots,l_2^n=0}^{d-1}
\sum_{v_2^1,v_2^2,\dots,v_2^{n-1}=0}^{d-1}
\left\vert \lambda_{l_1^1,l_2^1} \prod_{r=1}^{n-1}  \lambda_{l_2^r \ominus v_2^r,l_2^{r+1}} \right\vert^2
	}}. 				\notag
\end{gather}
It can be seen that when measurement results are
$\left\vert \phi \left( \hat{v}_1^1,\hat{v}_2^1 \right) \right\rangle, \left\vert \phi \left( \hat{v}_1^2,\hat{v}_2^2 \right) \right\rangle,
\dots, \left\vert \phi \left( \hat{v}_1^{n-1},\hat{v}_2^{n-1} \right) \right\rangle$,
the particles (1,2n) collapse onto the state shown in Eq. \ref{generalized-pure-states-ES-chain}.
Likewise, the entanglement swapping chain for general pure states can be characterized as
\begin{align}
&	\bigotimes_{r=1}^{n} \left\vert \mathscr{P}_{\uppercase\expandafter{\romannumeral1}}^r \right\rangle_{2r-1,2r}
\notag \\
\rightarrow &
\frac{\mathcal{P}}{\sqrt{d^{n-1}}}
\sum_{l_1,v_1^1,v_2^1,v_1^2,v_2^2,\dots,v_1^{n-1},v_2^{n-1}=0}^{d-1}
\lambda_{l_1} \prod_{r=1}^{n-1} \lambda_{l_1 \ominus \sum_{i=1}^r \hat{v}_2^i}
\zeta^{- \sum_{r=1}^{n-1} \left( l_1 - \sum_{i=1}^r \hat{v}_2^i \right) \hat{v}_1^r} 
\left\vert l_1, l_1 \ominus \sum_{r=1}^{n-1} \hat{v}_2^r \right\rangle_{1,2n} 
\bigotimes_{r=1}^{n-1} 
\left\vert \phi\left(v_1^r,v_2^r \right)\right\rangle_{2r+1,2r},
\end{align}
where
\begin{gather}
\mathcal{P} = \frac{1}{\sqrt{
\sum_{l_1,v_2^1,v_2^2,\dots,v_2^{n-1}=0}^{d-1}
\left\vert \lambda_{l_1} \prod_{r=1}^{n-1} \lambda_{l_1 \ominus \sum_{i=1}^r \hat{v}_2^i} \right\vert^2
	}}. 				\notag
\end{gather}
Further, we can formulate the entanglement swapping chain for maximally entangled states, which is given by
\begin{align}
& \bigotimes_{r=1}^{n} \left\vert \phi \left(u_1^r,u_2^r,\dots,u_{2m}^r \right) \right\rangle_{2r-1,2r}
\notag \\
\rightarrow
& \frac {1}{d^{n-1}} \sum_{v_1^1,v_1^2,\dots,v_1^{n-1}=0}^{d-1} \sum_{v_{m+1}^1,v_{m+1}^2,\dots,v_{m+1}^{n-1}=0}^{d-1}
\zeta^{\sum_{r=1}^{n-1} \left( u_1^{r+1} - v_1^{r} \right) \sum_{k=1}^{r} \left( u_{m+1}^k - v_{m+1}^k \right)}
\left\vert\phi
\left(
\oplus_{r=1}^n u_1^r \ominus \oplus_{r=1}^{n-1} v_1^r,u_2^1,u_3^1,\dots,u_{m}^1,
\right.\right.			\notag \\
&
\left.\left.
\oplus_{r=1}^{n-1} \left( u_{m+1}^r \ominus v_{m+1}^r \right) \oplus u_{m+1}^n,
\oplus_{r=1}^{n-1}\left( u_{m+1}^r \ominus v_{m+1}^r \right) \oplus u_{m+2}^n,
\dots,\oplus_{r=1}^{n-1} \left( u_{m+1}^r \ominus v_{m+1}^r \right) \oplus u_{2m}^n
\right)
\right\rangle_{1,2n}																	\notag \\
& 
\bigotimes_{r=1}^{n-1} 
\left\vert \phi \left(v_1^r,v_2^r,\dots,v_{2m}^r \right)\right\rangle_{2r+1,2r},
\end{align}
where
\begin{gather}
v_i^r = u_i^{r+1}, \phantom{i} i =2,3,\dots,m, r=1,2,\dots,n-1,
\notag \\
v_j^r = v_{m+1}^r \ominus u_{m+1}^r \oplus u_j^r, \phantom{i} j = m+2,m+3,\dots,2m, r=1,2,\dots,n-1.
\notag
\end{gather}

\subsection{Entanglement swapping chains for mixed states}

Here, we only consider the entanglement swapping chains for mixed Bell states, which can be characterized by
\begin{align}
\label{mixed-Bell-states-ES-chains}
& \bigotimes_{r=1}^{n} \rho \left( u_1^r,u_2^r \right) 
\notag \\
= & \frac{1}{d^n} \sum_{u_1^1,u_2^1,u_1^2,u_2^2,\dots,u_1^n,u_2^n = 0}^{d-1}
\sum_{l_1,l_2,\dots,l_n=0}^{d-1}
\sum_{l_1',l_2',\dots,l_n'=0}^{d-1}
\prod_{r=1}^n \lambda_{u_1^r u_2^r}
\zeta^{\sum_{r=1}^{n} \left( l_r - l_r' \right) u_1^r}
\bigotimes_{r=1}^{n} \left\vert l_r, l_r \oplus u_2^r \right\rangle_{2r-1,2r} \left\langle l_r', l_r' \oplus u_2^r \right\vert
\notag \\
\Rightarrow 
& \frac{1}{d^n}
\sum_{u_1^1,u_2^1,u_1^2,u_2^2,\dots,u_1^n,u_2^n = 0}^{d-1}
\sum_{l_1,l_2,\dots,l_n=0}^{d-1} \sum_{l_1',l_2',\dots,l_n'=0}^{d-1}
\prod_{r=1}^n \lambda_{u_1^r u_2^r}
\zeta^{\sum_{r=1}^{n} \left( l_r - l_r' \right) u_1^r}
\left\vert l_1,l_n \oplus u_2^n \right\rangle_{1,2n} \left\langle l_1',l_n' \oplus u_2^n \right\vert 
\notag \\
&
\bigotimes_{r=2}^n
\left\vert l_r,l_{r-1} \oplus u_2^{n-1} \right\rangle_{2r-1,2r-2} \left\langle l_r',l_{r-1}' \oplus u_2^{n-1} \right\vert
\notag \\
= & \frac {\mathcal{P}}{d^{2n-1}}
\sum_{u_1^1,u_2^1,u_1^2,u_2^2,\dots,u_1^n,u_2^n = 0}^{d-1}
\sum_{v_1^1,v_2^1,v_1^2,v_2^2,\dots,v_1^{n-1},v_2^{n-1}=0}^{d-1}
\sum_{l,l'=0}^{d-1} 
\prod_{r=1}^n \lambda_{u_1^r u_2^r}
\zeta^{\left( \sum_{r=1}^{n} u_1^r - \sum_{r=1}^{n-1} v_1^r \right) \left( l - l' \right)}
\notag \\
&
\left\vert l,l \oplus \sum_{r=1}^{n-1} \left( u_2^r - v_2^r \right) \oplus u_2^n \right\rangle
\left\langle l',l' \oplus \sum_{r=1}^{n-1} \left( u_2^r - v_2^r \right) \oplus u_2^n \right\vert
\bigotimes_{r=1}^{n-1}
\left\vert \phi \left( v_1^r,v_2^r \right)\right\rangle \left\langle \phi \left( v_1^r,v_2^r \right)\right\vert,
\end{align}
where
\begin{gather}
\mathcal{P} = 
\frac{1}
{d
\sum_{u_1^1,u_2^1,u_1^2,u_2^2,\dots,u_1^n,u_2^n = 0}^{d-1} \prod_{r=1}^n \lambda_{u_1^r u_2^r}
}.
\notag
\end{gather}


\section{Concise proof for our recent work}

\noindent
Recently, we have considered some interesting entanglement swapping cases in Ref. \cite{JiZX5852022},
and demonstrated their applications in quantum information processing \cite{JiZX5852022}.
Actually, the conclusion introduced by Ref. \cite{JiZX5852022} can be proved using Eq. \ref{revised-cat-state-ES},
in which the process of proof is similar to that in Ref. \cite{JiZX5852022}.
In what follows, we will provide a more concise proof through Eq. \ref{maximally-ES-contain-first-particle}.
Let us start with the conclusion presented in Ref. \cite{JiZX5852022}, which is as follows:

\begin{theorem}
\label{theorem_generalization}

Suppose that there are $n$ 2-level entangled states,
$\left\vert \phi_{1} \right\rangle,\left\vert \phi_{2} \right\rangle,\dots,\left\vert \phi_{n} \right\rangle$
containing $2m_1,2m_2,\dots,2m_n$ particles each,
are in one of the states $\left\{ \left\vert\phi\left( u_1,u_2,\dots,u_m \right)\right\rangle \right\}_{u_1,u_2,\dots,u_m=0}^{1}$
\textnormal{(see Eq. \ref{2-level-maximally-entangled-states})}.
Suppose that the first or the last $m_1,m_2,\dots,m_n$ particles in 
$\left\vert \phi_{1} \right\rangle,\left\vert \phi_{2} \right\rangle,\dots,\left\vert \phi_{n} \right\rangle$ are selected respectively,
and that the measurement operator $\mathcal{\widetilde{M}}$ is performed on them.
Let us denote the measurement result as $\left\vert \uppsi \right\rangle$,
and mark the joint state of the remaining particles by $\left\vert \upphi \right\rangle$.
If the initial states $\left\vert \phi_{1} \right\rangle,\left\vert \phi_{2} \right\rangle,\dots,\left\vert \phi_{n} \right\rangle$ 
satisfy the following conditions:
\begin{enumerate}

\item
The number of $\left\vert \phi_{1} \right\rangle,\left\vert \phi_{2} \right\rangle,\dots,\left\vert \phi_{n} \right\rangle$
in one of the states $\left\{ \left\vert\phi\left( 1,u_2,\dots,u_m \right)\right\rangle \right\}_{u_1,u_2,\dots,u_m=0}^{1}$ is even,

\item
$\forall r=1,2,\dots,n$, 
\phantom{i} 
$\prod_{i=1}^{m_r} \delta_{u_i^r,u_{m_r+i}^r} + \prod_{i=1}^{m_r} \delta_{u_i^r,\bar{u}_{m_r+i}^r} = 1$,
where the subscripts $i$ represent the selected particles in $\left\vert \phi_{r} \right\rangle$
and the subscripts $m_r+i$ the remaining particles,

\end{enumerate}
then $\left\vert \uppsi \right\rangle$ and $\left\vert \upphi \right\rangle$ are the same.
\end{theorem}

From the entanglement swapping between d-level maximally entangled states, 
the concise proof can be described as follows:
\begin{proof}

Suppose that the first $k_r$ particles in
$\left\vert \phi_r \right\rangle$ for $r =1,2,\dots,t$
and the last $k_r$ particles in $\left\vert \phi_r \right\rangle$ for $r =t+1,t+2,\dots,n$ are selected.
From Eq. \ref{maximally-ES}, after performing the measurement operator $\mathcal{\widetilde{M}}$ on the selected particles, 
we can get
\begin{align}
& \bigotimes_{r=1}^{n} \left\vert \phi \left(u_1^r,u_2^r,\dots,u_{2m_r}^r \right) \right\rangle
\notag \\
\rightarrow 
& \frac {1}{d^{n/2}} \sum_{v_1^1,v_1^2,\dots,v_1^n=0}^{d-1}
\zeta^{\sum_{r=2}^{n} v_1^r u_1^r - \sum_{r=t+1}^{n} u_{m_r+1}^r u_1^r - \left( \sum_{r=1}^n u_1^r -v_1^1 \right) u_{{m_1}+1}^1}
\left\vert \phi \left(v_1^1,v_2^1,\dots,v_{m_1}^1,v_1^2,v_2^2,\dots,v_{m_2}^2,\dots,v_1^n,v_2^n,\dots,v_{m_n}^n \right)\right\rangle \notag \\
& \otimes
\left\vert 
\phi \left(
\tilde{v}_1^1,\tilde{v}_2^1,\dots,\tilde{v}_{m_1}^1,\tilde{v}_1^2,\tilde{v}_2^2,\dots,\tilde{v}_{m_2}^2,\dots,
\tilde{v}_1^n,\tilde{v}_2^n,\dots,\tilde{v}_{m_n}^n 
\right)
\right\rangle
\end{align}
where
\begin{gather}
v_j^1 = u_j^1, \phantom{i} j =2,3,\dots,m_1,
\notag \\
v_j^i = v_1^i \oplus u_j^i, \phantom{i} i =2,3,\dots,t, \phantom{i} j =2,3,\dots,m_i,
\notag \\
v_j^i = v_1^i \ominus u_{m_i+1}^i \oplus u_{m_i + j}^i, \phantom{i} i =t+1,t+2,\dots,n, \phantom{i} j =2,3,\dots,m_i,
\notag \\
\tilde{v}_1^1 = \oplus_{r=1}^{n} u_1^r \ominus v_1^1,
\notag \\ 
\tilde{v}_j^1 = u_{m_1+j}^1 \ominus u_{m_1+1}^1,\phantom{i} j = 2,3,\dots,m_1,
\notag \\
\tilde{v}_j^i = v_1^i \ominus u_{m_1+1}^1 \oplus u_{m_i+j}^1, \phantom{i} i =1,2,\dots,t, \phantom{i} j = 1,2,\dots,m_i,
\notag \\
\tilde{v}_1^i = v_1^i \ominus u_{m_1+1}^1 \ominus u_{m_i+1}^i, \phantom{i} i =t+1,t+2,\dots,n,
\notag \\
\tilde{v}_j^i = v_1^i \ominus u_{m_1+1}^i \ominus u_{m_i+1}^i \oplus u_j^i, \phantom{i} i =t+1,t+2,\dots,n, \phantom{i} j =2,3,\dots,m_i.
\end{gather}

Generating two identical entangled states through entanglement swapping means that the following equations hold:
\begin{subequations}\label{group}
\begin{gather}
\oplus_{r=1}^{n} u_1^r \ominus v_1^1 = v_1^1,
\label{first-condition}
\\
u_{m_1 + j}^1 \ominus u_{m_1 + 1}^1 = u_{j}^1, \phantom{i} j =2,3,\dots,m_1,
\label{second-conditions-1}
\\
v_1^i \ominus u_{m_1 + 1}^1 \oplus u_{m_i + 1}^n = v_1^i, \phantom{i} i =2,3,\dots,t,
\\
v_1^i \ominus u_{m_1 + 1}^1 \oplus u_{m_i + j}^j
= v_1^i \oplus u_j^i, \phantom{i} i =2,3,\dots,t, \phantom{i} j =2,3,\dots,m_i,
\\
v_1^i \ominus u_{m_1 + 1}^1 \ominus u_{m_i + 1}^i = v_1^i, \phantom{i} i =t+1,t+2,\dots,n,
\\
v_1^i \ominus u_{m_1 + 1}^1 \ominus u_{m_i + 1}^i \oplus u_j^i 
= v_1^i \ominus u_{m_i + 1}^i \oplus u_{m_i + j}^i, \phantom{i} i =t+1,t+2,\dots,n, \phantom{i} j =2,3,\dots,m_i.
\label{second-conditions}
\end{gather}
\end{subequations}
From Eq. \ref{first-condition}, we have
\begin{gather}
\oplus_{r=1}^{n} u_1^r = 0,
\end{gather}
which means that the number of $\left\vert \phi_{1} \right\rangle,\left\vert \phi_{2} \right\rangle,\dots,\left\vert \phi_{n} \right\rangle$
in one of the states $\left\{ \left\vert\phi\left( 1,u_2,\dots,u_m \right)\right\rangle \right\}_{u_2,\dots,u_m=0}^{1}$ is even.
From Eqs. \ref{second-conditions-1}-\ref{second-conditions}, we have
\begin{gather}
u_{m_i + j}^i = 
\begin{dcases}
\phantom{i} u_j^i, \quad \mathrm{if} \phantom{i} u_{m_1+1}^1 = 0,
\\
\phantom{i} \bar{u}_j^i, \quad \mathrm{if} \phantom{i} u_{m_1+1}^1 = 1,
\end{dcases}
\quad i=1,2,\dots,n, \phantom{i} j =1,2,\dots,m_i,
\end{gather}
which is consistent with the second condition in Theorem \ref{theorem_generalization}. QED.

\end{proof}

The above conclusion can be further extended to the
entanglement swapping between CAT states, which can be proved through Eq. \ref{revised-cat-state-ES}.
Moreover, it can be extended to general pure states,
but such extensions may not have much significance in finding applications in quantum information processing, 
which leaves us with no motivation to continue exploring in the future.
For two extreme cases, that is, measuring the first half of the particles in each state, or 
the last half of the particles, the proof can be presented similarly.
We would like to give the proof for the latter case here.

\begin{proof}

From Eq. \ref{maximally-ES-without-first-particle}, 
the measurement results are given by
\begin{align}
\label{2m-particle-state-measurement-result}
\left\vert \phi \left(v_1^1,v_2^1,\dots,v_{m_1}^1,v_1^2,v_2^2,\dots,v_{m_2}^2,\dots,v_1^n,v_2^n,\dots,v_{m_n}^n \right)\right\rangle,
\end{align}
where
\begin{gather}
v_j^1 = u_{m_1+j}^1 \ominus u_{m_1+1}^1, \phantom{i} j =2,3,\dots,m_1,			
\notag \\
v_{j}^i = v_1^i \ominus u_{m_i + 1}^i \oplus u_{m_i + j}^i, \phantom{i} i =2,3,\dots,n, \phantom{i} j =2,3,\dots,m_i.
\end{gather}
The remaining particles collapse onto
\begin{align}
\label{2m-particle-state-remaining-result}
& \left\vert \phi \left(\oplus_{r=1}^{n} u_1^r \ominus v_1^1,u_2^1,u_3^1,\dots,u_{m_1}^1,
v_1^2 \oplus u_{m_1+1}^1 \ominus u_{m_2+1}^2,
v_1^2 \oplus u_{m_1+1}^1 \ominus u_{m_2+1}^2 \oplus u_2^2,\dots,
v_1^2 \oplus u_{m_1+1}^1 \ominus u_{m_2+1}^2 \oplus u_{m_2}^2,\dots,
\right.\right.															\notag \\
& \quad \left.\left.
v_1^n \oplus u_{m_1+1}^1 \ominus u_{m_n+1}^n,
v_1^n \oplus u_{m_1+1}^1 \ominus u_{m_n+1}^n \oplus u_2^n,\dots,
v_1^n \oplus u_{m_1+1}^1 \ominus u_{m_n+1}^n \oplus u_{m_n}^n \right) \right\rangle.
\end{align}
Generating two identical entangled states through entanglement swapping implies that 
\begin{gather}
\oplus_{r=1}^{n} u_1^r \ominus v_1^1 = v_1^1,
\label{first-condition-for-second-extreme-case}
\\
u_{m_1 + j}^1 \ominus u_{m_1 + 1}^1 = u_{j}^1, \phantom{i} j =2,3,\dots,m_1,
\notag \\
v_1^i \oplus u_{m_1 + 1}^1 \ominus u_{m_i + 1}^n = v_1^i, \phantom{i} i=2,3\dots,n,
\notag \\
v_1^i \oplus u_{m_1 + 1}^1 \ominus u_{m_i + 1}^i \oplus u_j^i
= v_1^i \ominus u_{m_i + 1}^i \oplus u_{m_i + j}^i, \phantom{i} i=2,3\dots,n, \phantom{i} j =2,3,\dots,m_i.
\label{second-condition-for-second-extreme-case}
\end{gather}
From Eqs. \ref{first-condition-for-second-extreme-case} and \ref{second-condition-for-second-extreme-case}, 
one can derive the equations as Eqs. \ref{first-condition}-\ref{second-conditions}. QED.

\end{proof}


\section{Conclusion}

\noindent
We have studied entanglement swapping theory for pure states and mixed states,
and presented the basic theory for deriving entanglement swapping results.
We have studied the entanglement swapping of 2-level the maximally entangled states
by assuming that all the subsystems are in different bases.
We have studied the entanglement swapping of multiple d-level maximally entangled states 
by assuming that measurement operators are performed on the selected particles 
that contain the first particle in some entangled states
and the ones that do not contain the first particle in the remaining entangled states.
We have considered the generalization of the entanglement swapping between
bipartite general pure states to the multi-system case.
We have provided a proof for the entanglement swapping chains of bipartite general pure states
through mathematical induction.
We have proposed the entanglement swapping chains of multi-particle maximally entangled states,
and proposed the entanglement swapping between general pure states and maximally entangled states.
We have also proposed the entanglement swapping and entanglement swapping chains for mixed states.
We have provided a concise proof for the conclusion introduced in our recent work.
The phenomenon of quantum superposition means that everything is in a superposition of
existence and non-existence, and if we determine that something exists or does not exist, 
it is because we have observed it \cite{JiZX12442025}.
We believe that in the near future, more and more people will turn their attention to philosophy, 
especially ancient Chinese philosophy, in search of ways to live.


\end{document}